\RequirePackage{ifpdf}
\documentclass[hyper,letterpaper]{JHEP3}
\pdfoutput=1
\usepackage{amsmath,amssymb,amsfonts,bm,amscd}
\usepackage{cite}
\usepackage{graphicx}
\usepackage{multirow}
\usepackage{verbatim}
\usepackage{appendix}
\usepackage{fancybox}
\usepackage{array}
\usepackage{url}
\usepackage{float}
\usepackage[mathscr]{euscript}

\newcommand{\bea}{\begin{eqnarray}}
\newcommand{\eea}{\end{eqnarray}}
\newcommand{\be}{\begin{equation}}
\newcommand{\ee}{\end{equation}}

\newcommand{\Z}{{\mathbb Z}}
\newcommand{\R}{{\mathbb R}}
\newcommand{\C}{{\mathbb C}}
\newcommand{\M}{{\cal M}}
\newcommand{\Q}{{\mathbb Q}}

\makeatletter
 \DeclareRobustCommand\widecheck[1]{{\mathpalette\@widecheck{#1}}}
 \def\@widecheck#1#2{%
     \setbox\z@\hbox{\m@th$#1#2$}%
     \setbox\tw@\hbox{\m@th$#1%
        \widehat{%
           \vrule\@width\z@\@height\ht\z@
           \vrule\@height\z@\@width\wd\z@}$}%
     \dp\tw@-\ht\z@
     \@tempdima\ht\z@ \advance\@tempdima2\ht\tw@ \divide\@tempdima\thr@@
     \setbox\tw@\hbox{%
        \raise\@tempdima\hbox{\scalebox{1}[-1]{\lower\@tempdima\box
 \tw@}}}%
     {\ooalign{\box\tw@ \cr \box\z@}}}
\makeatother

\def\Tr{{\rm Tr \,}}

\def\frak{\mathfrak}

\def\tilde{\widetilde}
\def\hat{\widehat}
\def\bar{\overline}

\def\CA{{\mathcal A}}

\def\CC{{\mathcal C}}

\def\CH{{\mathcal H}}
\def\CI{{\mathcal I}}

\def\CM{{\mathcal M}}
\def\CN{{\mathcal N}}
\def\CO{{\mathcal O}}

\def\CQ{{\mathcal Q}}
\def\CR{{\mathcal R}}
\def\CS{{\mathcal S}}
\def\CT{{\mathcal T}}

\newcommand{\cp}{{\mathbb{C}}{\mathbf{P}}}

\renewcommand{\bar}{\overline}
\renewcommand{\hat}{\widehat}

\renewcommand{\d}{\partial}

\def\fD{\mathfrak{D}}
\def\fM{\text{SW}}
\def\Am{{\text{A}}}

\def\pt{\mathrm{pt}}

\def\VW{{\mathrm{VW}}}
\def\RW{{\mathrm{RW}}}
\def\SW{{\mathrm{SW}}}

\def\bn{{\mathbf{n}}}
\def\bm{{\mathbf{m}}}
\def\br{{\mathbf{r}}}

\def\bx{{\mathbf{x}}}

\def\brho{{\boldsymbol\rho}}

\def\red{{\text{(red)}}}

\def\Tor{{\tau}}

\title{Fivebranes and 3-manifold homology}

\author{Sergei Gukov$^{1}$, Pavel Putrov$^{2}$, Cumrun Vafa$^{3}$
\\
$^1$ Walter Burke Institute for Theoretical Physics, California Institute of Technology, Pasadena, CA 91125, USA \\
$^2$ School of Natural Sciences, Institute for Advanced Study, Princeton, NJ 08540, USA \\
$^3$ Jefferson Physical Laboratory, Harvard University, Cambridge, MA 02138, USA}

\abstract{Motivated by physical constructions of homological knot invariants, we study their analogs for closed 3-manifolds.
We show that fivebrane compactifications provide a universal description of various old and new homological invariants of 3-manifolds.
In terms of 3d/3d correspondence, such invariants are given by the $Q$-cohomology of the Hilbert space of partially topologically twisted
3d $\CN=2$ theory $T[M_3]$ on a Riemann surface with defects.
We demonstrate this by concrete and explicit calculations in the case of monopole/Heegaard Floer homology
and a 3-manifold analog of Khovanov-Rozansky link homology.
The latter gives a categorification of Chern-Simons partition function.
Some of the new key elements include the explicit form of the $S$-transform and a novel connection
between categorification and a previously mysterious role of Eichler integrals in Chern-Simons theory.}

\preprint{CALT 2016-004}
\begin{document}
\cornersize{1}


\section{Introduction}

The main goal of this paper is to describe the structural properties and explicit computations of 3-manifold homological
invariant,
\be
\CH_N^{*,*} (M_3)
\label{HN3mfld}
\ee
whose graded Euler characteristic gives quantum $sl(N)$ invariant of $M_3$.
In physics, these spaces will be understood as Hilbert spaces of BPS states or, equivalently, as $Q$-cohomology groups of various systems.

Our study of 3-manifold homologies is largely motivated by and parallels that of knot homologies, which are fairly well understood by now.
The list of new homological invariants of knots and links is constantly growing, and by now there are many examples for knots colored by
various representations of many different groups.
But on the mathematical side the situation was rather different merely a decade ago,
when the only available theories were Khovanov homology categorifying the Jones polynomial \cite{Khovanov}
and the knot Floer homology categorifying the Alexander polynomial \cite{OShfk,RasmussenHFK}.
{}Both of these two theories are extremely concrete and computation-friendly, which immediately led to a number of
surprising observations~\cite{MR2034399,MR2189938}.
For example, while their definition is very different and indicates no direct interrelation,
the total dimension turns out to be equal for many knots,
\be
\dim HFK (K) \; = \; \dim Kh (K) \,,
\label{dimHFKKh}
\ee
including all knots with up to 9 crossings, all alternating knots, {\it etc.}
The discovery of such theories was (and still is) so miraculous that it was not at all clear whether
these two theories, associated to $N=2$ and $N=0$, have cousins for other values of $N$.
In 2004, a considerable hope to the categorification program was given by the seminal work of Khovanov and Rozansky \cite{KR1}
who constructed the entire family of $sl(N)$ knot homologies using matrix factorizations.

\bigskip
\begin{figure}[ht]
\centering
\includegraphics[trim={0 1.5in 0 1.5in},clip,width=5.5in]{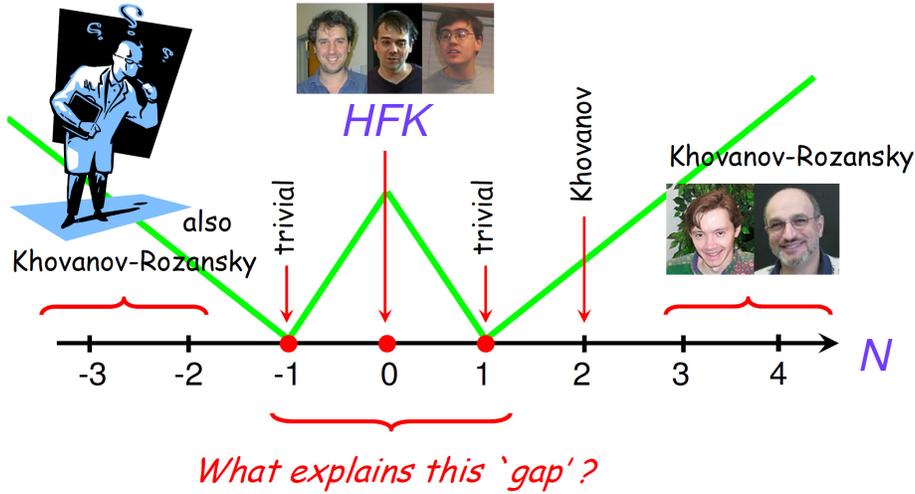}
\caption{The landscape of knot homologies shows peculiar behavior at $N=-1, 0$ and $1$.}
\label{fig:unification}
\end{figure}

This breakthrough, however, led to new questions and more puzzles. Thus, it was not clear why the family of such theories
labeled by $N$ appeared to have an extension to the negative range ($N<0$) where it also gives $sl(N)$ knot homology \cite{Dunfield:2005si}.
Moreover, there were a number of puzzles associated with the behavior at small values of $N$.
For instance, starting from $N \ge 2$ and gradually decreasing its values, one would eventually reach Khovanov homology at $N=2$
and then a ``trivial'' theory for $N=1$. This behavior is not very surprising since decreasing the rank one would expect
to find a simpler theory and the value $N=1$ at the very bottom of this tower corresponds to $sl(1)$ group, which indeed is trivial.
What is surprising, though, is that decreasing $N$ further, to $N=0$ one again finds a very interesting theory, $HFK (K)$,
followed again by a trivial $sl(1)$ theory at $N=-1$.
What explains this peculiar behavior at $N=-1, 0$ and $1$?
And why does the oddball $HFK$, ``sandwiched'' by two trivial theories, have unexpected relations to Khovanov homology {\it a la} \eqref{dimHFKKh}?

The answer to these questions came a bit later, with the advent of the HOMFLY-PT knot homology and its ``colored'' variants,
which came as a surprise \cite{MR2100691}.
They were motivated by independent physics developments where the HOMFLY-PT invariants
were captured by BPS Hilbert spaces associated to knots \cite{Ooguri:1999bv}.
The first connection to knot homologies was then made in \cite{Gukov:2004hz}
which restored the homological grading and led to concrete predictions for many simple knots.
More importantly, it led to new structural properties
that helped to unify knot Floer homology with Khovanov-Rozansky homology \cite{Dunfield:2005si}.
Moreover, these developments helped to explain the extension to negative values of $N$ and the ``gap'' between $N=-1$ and $N=1$
by emphasizing \cite{Gorsky:2013jxa} the role of supergroups $sl(n|m)$ with
\be
N = n-m
\label{Nnm}
\ee
In particular, the ``gap'' at small values of $N$ is best understood by generalizing the theory to colored knot homology,
where it occurs between $N=-$ (the longest row) and $N=$ (the highest column) of the corresponding Young diagram $\lambda$.
For knots colored by the fundamental representation $\lambda = \Box$, one recovers the familiar range $-1 \le N \le 1$.
\FIGURE[h]{
\includegraphics[width=2.2in]{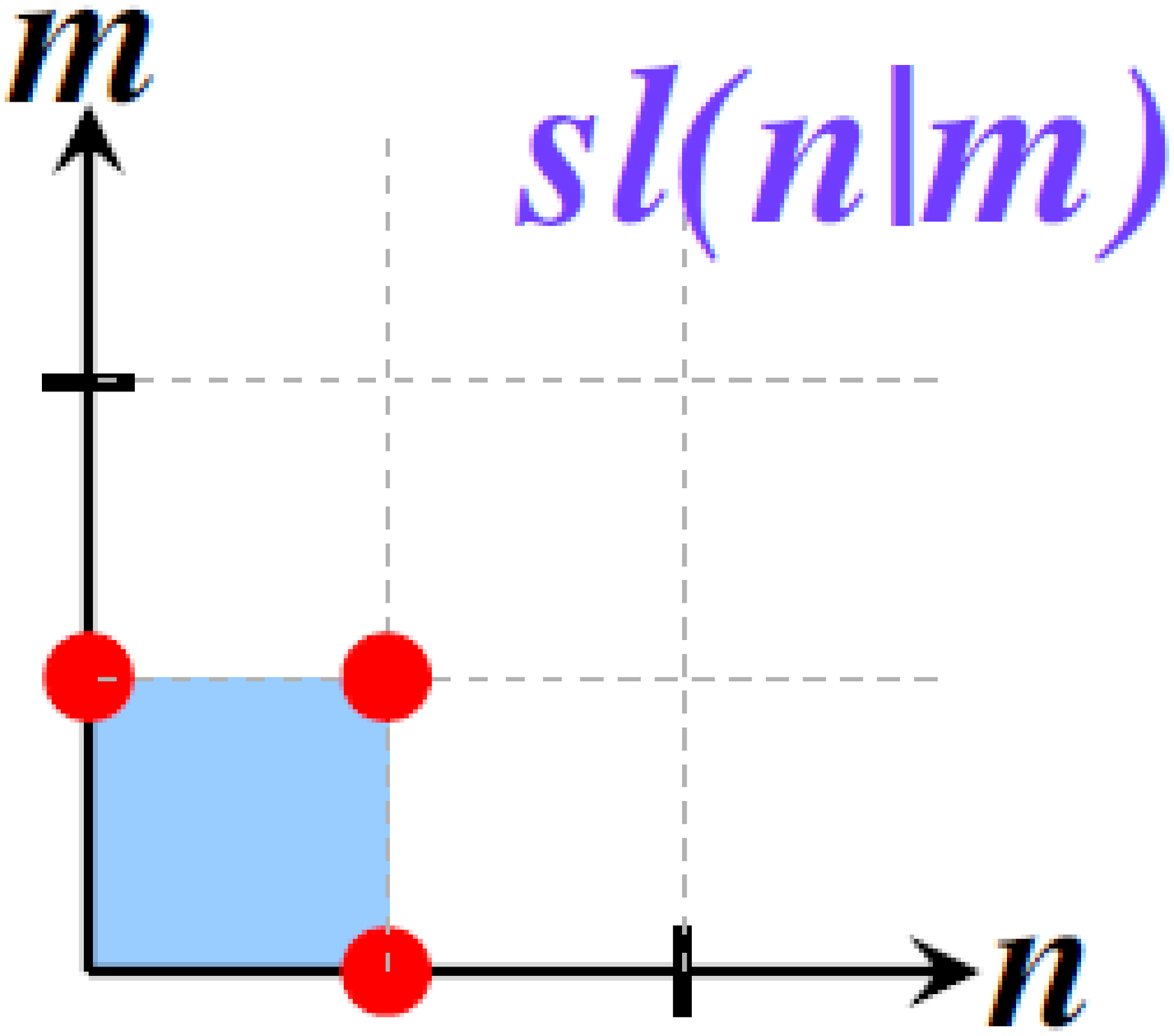}
\caption{\label{fig:super}
The three corners (red $\bullet$) correspond to the three special values of $N=-1,0$ and 1
in $\CH_N^{*,*} (K)$ theory with color $\lambda = \Box$. \vspace{3ex}}
}
A convenient way to visualize the ``landscape'' in Figure~\ref{fig:unification} is to plot $\lambda^t$ in the positive quadrant
of the $(n,m)$ plane, as illustrated in Figure~\ref{fig:super}.
Then, different values of $N$ correspond to boundary points of the positive $(n,m)$-quadrant with $\lambda^t$ excluded.
Traversing the boundary of the unshaded region in  Figure~\ref{fig:super},
one goes through the sequence of homological invariants for $sl(N|0) \cong sl(N)$, $sl(1|1)$, and $sl(0|N)$,
which will also be the list relevant to the present paper.
Indeed, these three classes can be conveniently labeled by $N \in \Z$, such that $N<0$ corresponds to $sl(N|0)$,
$N=0$ corresponds to $sl(1|1)$, and $N>0$ corresponds to $sl(0|N)$.
Of course, the value of the super-rank \eqref{Nnm} does not uniquely specify $sl(n|m)$, but as long as we stay
within these three special cases we can use a more economic notation and label a theory simply by $N$,
which is the notation we adopt in \eqref{HN3mfld} and throughout the paper.\footnote{It would be interesting
to extend this work to computation of quantum (as in \cite{Rozansky:1992td,Chang:1991df,Rozansky:1992zt})
and homological 3-manifold invariants for arbitrary $sl(n|m)$. We hope to return to this problem elsewhere.}

In the physical realization of knot homologies \cite{Gukov:2004hz,Gukov:2007ck,Witten:2011zz,Diaconescu:2011xr},
$N$ enters either as the number of fivebranes or as the K\"ahler modulus (stability parameter) of the conifold $X$:
\be
\begin{array}{rcl}
\multicolumn{3}{c}{~~\text{doubly-graded}~} \\[.1cm]
\hline
{\mbox{\rm space-time:}} && ~~  \R \times T^* S^3 \times TN_4 \\
{\mbox{\rm $n$ M5-branes:}} && ~~  \R \times ~S^3~ \times ~\R^2_q \\
{\mbox{\rm  M5$'$-branes:}} && ~~  \R \times ~L_K~ \times  ~\R^2_q
\end{array}
\quad \xleftarrow[~\text{transition}]{~\text{phase}} \!\!\! \to \quad
\begin{array}{rcl}
\multicolumn{3}{c}{~~~\text{triply-graded}~~} \\[.1cm]
\hline
&& \R \times X \times  TN_4 \\
&& \R \times L_K \times  \R^2_q
\end{array}
\label{conifoldphases}
\ee
where the two systems are related by a geometric transition \cite{Gopakumar:1998ii,Ooguri:1999bv}.
In particular, interpolating from positive to negative values of $N$ on the triply-graded (``resolved'') side
is realized via the flop transition, and the special theory $HFK (K)$ corresponds to the singular limit of $X$.
The systems \eqref{conifoldphases} have been studied from various vantage points and in different duality frames
(see {\it e.g.} \cite{Chun:2015gda} for a recent summary).

In this paper, we will try to replicate some of the successes of this physical framework to explain and predict
the behavior of knot homologies in the world of 3-manifolds.
The theory for $sl(1|1)$ that we label by $N=0$ will again play a very special role;
it is the only value of $N$ for which 3-manifold homology currently admits a rigorous mathematical definition.
In fact, while its cousins for $N \ne 0$ are currently out of reach, the $sl(1|1)$ theory with $N=0$ has
{\it three} equivalent mathematical formulations:

\begin{itemize}

\item via symplectic geometry, ``Heegaard Floer homology'' $HF (M_3)$ \cite{MR2249248},

\item via gauge theory, ``Seiberg-Witten Floer homology'' or ``monopole Floer homology'' $HM (M_3)$ based on Seiberg-Witten equations \cite{MR2388043},

\item  via contact geometry, ``embedded contact homology'' $ECH (M_3)$ \cite{MR3220947,MR2532999}.

\end{itemize}

\noindent
All three are isomorphic\footnote{As we review later in the main text, each of these theories comes in four flavors,
and the isomorphisms hold for the corresponding flavors.} \cite{MR2388043} (see also \cite{MR2873429}):
\be
HF (M_3) \cong HM (M_3) \cong ECH (M_3)
\label{HFHMECH}
\ee

Therefore, in order to develop a picture analogous to Figure~\ref{fig:unification} and to
tackle 3-manifold homologies \eqref{HN3mfld} by a variety of methods that were so successful for knots,
we first need to realize the $N=0$ theory in the physical setup similar to \eqref{conifoldphases}:
\be
\begin{array}{rcl}
\multicolumn{3}{c}{~~\text{doubly-graded}~} \\[.1cm]
\hline
{\mbox{\rm space-time:}} && ~~  \R \times T^* M_3 \times T^* \Sigma \\
{\mbox{\rm $n$ M5-branes:}} && ~~  \R \times ~M_3~ \times ~\Sigma
\end{array}
\quad \xleftarrow[~\text{transition}]{~\text{phase}} \!\!\! \to \quad
\begin{array}{rcl}
\multicolumn{3}{c}{\Z \oplus \Z \oplus H_2 (X;\Z)\text{-graded}} \\[.1cm]
\hline
&& \R \times X \times  T^* \Sigma
\end{array}
\label{M3phases}
\ee
In some ways this setup is simpler than \eqref{conifoldphases};
{\it e.g.} it does not require extra ingredients (branes) associated with knots and links.
But in other ways it is more complicated; one obvious difference is that $S^3$ is replaced by a general 3-manifold $M_3$ in \eqref{M3phases}.
As in the case of knots, analyzing \eqref{M3phases} in various duality frames and from various vantage points will shed light on
different aspects of 3-manifold homologies \eqref{HN3mfld}, which in all duality frames will be realized as $Q$-cohomology (space of BPS states).

To categorify the Chern-Simons partition function $\hat Z_{\text{CS}}^{U(n|m)} (M_3)$ means ``to restore the $t$-dependence''
\be
\hat Z_{\text{CS}}^{U(n|m)} (M_3) \quad \xleftarrow[~]{~~t=-1~~} \quad P_N (q,t) := \sum_{i,j} q^i t^j \dim \CH_N^{*,*} (M_3)
\label{ZCSfromP}
\ee
where $(n|m)=(N|0)$ for $N>0$ and $(n|m)=(1|1)$ for $N=0$.
As we stress throughout the paper --- especially in sections \ref{sec:Amodel} and \ref{sec:higherRank} ---
categorification in \eqref{ZCSfromP} requires writing the CS partition function in a new, slightly unnatural basis,
which is why we put a hat on $Z_{\text{CS}}^{U(n|m)} (M_3)$.
Then, the ``quantum'' variable $q$ (related to the Chern-Simons coupling constant) and the homological variable $t$ can be
interpreted as the equivariant parameters of the Omega-background or, equivalently, as fugacities for the rotation symmetry
\be
\begin{matrix}
U(1)_t ~~~ U(1)_{\Sigma} \\
\ensuremath{\reflectbox{\rotatebox[origin=c]{180}{$\circlearrowleft$}}}
~~\ensuremath{\reflectbox{\rotatebox[origin=c]{180}{$\circlearrowleft$}}} \\
T^* \Sigma
\end{matrix}
\label{UqUtsymmetries}
\ee
Note, in the case of $sl(1|1)$, the variable $q$ corresponds to the Alexander grading of \eqref{HFHMECH},
while $t$ keeps track of the Maslov grading. In this special case, we will be able to give a different
interpretation to the Alexander $q$-grading so that $\Sigma$ is no longer required to enjoy the $U(1)_{\Sigma}$ symmetry.

While the symmetry $U(1)_t$ that gives rise to the homological grading exists for arbitrary $M_3$,
its close cousin $U(1)_{\beta}$ described in section \ref{sec:U1beta} exists only for Seifert $M_3$
and is very handy for practical computations.
Using this symmetry we compute 3-manifold homologies in many concrete examples, sometimes in multiple independent ways.
\begin{table}[h]
	\centering
 \begin{tabular}{|c|c|c|}
	 \hline
3-manifold & $T_G [M_3]$ & $HF^+ (M_3, {\frak s})$ \\
\hline\hline
$M_3 = S^2 \times S^1$
&
\begin{tabular}{c}
	$\CN=2$ vector multiplet\\
	+ adjoint chiral
\end{tabular}
&
\begin{tabular}{cl}
$\CT^+_{-1/2} \oplus \CT^+_{1/2}$ \,, & if ${\frak s} = {\frak s}_0$\\
$0$ \,, & if ${\frak s} \ne {\frak s}_0$
\end{tabular}
\\
\hline
\begin{tabular}{l}
Lens spaces:\\
$M_3 = L(p,1)$ \\
($S^3$ when $p=1$)
\end{tabular}
&
\begin{tabular}{c}
	$\CN=2$ level-$p$ super-CS\\
	+ adjoint chiral
\end{tabular}
&
$\CT^+_0 \,, \quad \forall {\frak s} \in \Z_p$
\\
\hline
\begin{tabular}{c}
the Poincar\'e sphere:\\
$S^3_{-1} (3_1^{\ell}) = \Sigma (2,3,5)$
\end{tabular}
&
see \cite[sec 2.2]{Gadde:2013sca}
&
$\CT^+_2$
\\
\hline
\end{tabular}
\caption{\label{tab:examples}Simple examples. In the last column we use the standard notation
$\CT_k^+$ for the module over the ring $\Z [U]$ abstractly isomorphic to $\Z [U,U^{-1}] / \Z [U]$,
whose lowest degree element is supported in degree $k \in \mathbb{Q}$ \cite{Ellwood:2006vla}.
Note, $HF^+ (\Sigma (2,3,5))$ is isomorphic to $HF^+ (S^3)$ as relatively graded $\Z [U]$ modules,
but the absolute grading distinguishes them.}
\end{table}

The numerical and homological invariants of 3-manifolds that we are interested in
can be naturally calculated in terms of the corresponding 3d $\CN=2$ theory $T_G[M_3]$, {\it cf.} \cite{DGH,FGSS}:
\be
M_3 \quad \leadsto \quad T_G[M_3]
\label{TMMM}
\ee
For a general Lie group $G$ this is the effective theory of 6d $\CN=(0,2)$ SCFT labeled by a corresponding Lie algebra\footnote{Of course, for a given $\mathfrak{g}$ the choice of $G$ is not unique in general. In most of our discussion, the issues related to the global topology and the center of the group will not show up. However, it does not mean these issues can be completely ignored. As will become apparent later in the text, S-duality plays an important role and it should exchange $G$ with its Langlands dual $G^\vee$. Also, we will often omit the explicit dependence on $G$ and write $T[M_3]$ instead of $T_G[M_3]$.} $\mathfrak{g}$ compactified on $M_3$ with a topological twist. In the case when $G=U(N)$ this is the theory describing dynamics of $N=n$ M5 branes in the left hand side of (\ref{M3phases}). The table~\ref{tab:examples} lists basic examples of such correspondence.

\noindent
The rest of the paper is organized as follows:

\begin{itemize}

\item In section \ref{sec:Amodel} we study general features of a 3d $\CN=2$ theory on $\Sigma\times \R$ or $\Sigma\times S^1$ with partial topological twist along $\Sigma$, and its relation to the corresponding A-model on $\Sigma$. Concrete results in this section include the modular transformation of flat $G_\C$-connections on $M_3$ and a proposal for a categorification of the Verlinde algebra associated to 3-manifolds; both will play an important role in the subsequent sections.

\item In section \ref{sec:3d3d} we explore \eqref{M3phases} from the viewpoint of 3d-3d correspondence and compute $HF(M_3)$ (or its Euler characteristic) as the $Q$-cohomology (or, respectively, its index) in 3d $\CN=2$ theory $T[M_3]$ on $\R \times \Sigma$. In particular, we formulate a new way to compute the Seiberg-Witten invariants of 3-manifolds and the homological invariants~\eqref{HFHMECH}. We illustrate the technique in explicit examples of $M_3 = \Sigma' \times S^1$, Lens spaces, and more general plumbed manifolds, always finding an agreement with the known mathematical results. This gives us confidence to move to a mathematically uncharted territory of $sl(N)$ 3-manifold homologies.

\item In section \ref{sec:M3veiw} we reverse the order of compactification and consider the setup \eqref{M3phases} from the viewpoint of the effective 3d $\CN=4$ theory on $M_3$. This vantage point clarifies the connection to Seiberg-Witten invariants and homology groups \eqref{HFHMECH} and also leads to yet another way of computing them, which we call a ``refinement'' of the Rozansky-Witten theory. We illustrate it in a concrete example of $M_3 = S^2 \times S^1$.

\item In section \ref{sec:4mfld} we study the relation to 4-manifold invariants arising from fivebrane compactifications. The goal of this section is twofold: it unifies various twists and vantage points considered in earlier sections and also leads to a new physical interpretation of the ``correction terms'' in the Heegaard Floer homology.

\item In section \ref{sec:higherRank} we propose an analogue of the Khovanov homology for 3-manifolds which categorifies Chern-Simons partition function / quantum group invariant of $M_3$. A key element of this construction is an $S$-transform that, surprisingly, connects categorification with Mock modular forms and somewhat mysterious role of Eichler integrals in Chern-Simons theory. Another surprise is how various terms are grouped into ``homological blocks'' which seem to be labeled only by {\it reducible} $G$-connections on $M_3$. This may be a hint of a deeper relation to Heegaard Floer homology, analogous to the connection between $HFK(K)$ and Khovanov homology. Explicit examples in this section include Lens spaces, and certain more general Brieskorn spheres.

\item In section \ref{sec:Mtheory} we discuss those 3-manifolds for which $T^* M_3$ admits a geometric transition analogous to the conifold,
\be
M_3 \quad \leadsto \quad \text{Calabi-Yau 3-fold}~X
\label{CY3fromM3}
\ee
For this class of 3-manifolds, \eqref{HN3mfld} can be computed as $Q$-cohomology of the right-hand side in \eqref{M3phases}.
The large-$N$ duality that underlies the geometric transition unifies 3-manifold homologies for all $N$ into a bigger Hilbert space,
which is the Hilbert space of a closed string dual and which reduces for each integer $N$ to the 3-manifold homology \eqref{HN3mfld}.
This structural property parallels what was found for knots and links \cite{Dunfield:2005si,Gukov:2004hz,Gukov:2015gmm}.

\end{itemize}

\noindent
Appendices contain useful supplementary material.


\section{Categorification of a 2d A-model}
\label{sec:Amodel}

Even though in this paper we are mostly interested in applications to 3-manifolds,
some of the structure of 3d $\CN=2$ theories topologically twisted along a 2d spatial slice is rather general
and potentially can be useful for refinement and categorification of more general A-models.
Moreover, even for the purpose of studying 3-manifold homology, a purely two-dimensional formulation in terms of A-model
is very illuminating and mathematically a lot more accessible than formulations involving higher-dimensional systems.
In the case of 3d $\CN=2$ theory labeled by a 3-manifold $M_3$, the categorification of the A-model is realized by
the category of the representations of the vertex operator algebra (VOA) associated to a 4-manifold bounded by $M_3$.
A bonus feature of this approach is a concrete description of the modular group action on flat $G_\C$-connections on $M_3$.

\subsection{General A-model and a refinement}
\label{sec:Amodel-cat}

A general 3d $\CN=2$ theory admits a partial topological twist on space-time $\R \times \Sigma$ or $S^1 \times \Sigma$,
where $\Sigma$ can be an arbitrary Riemann surface (possibly with boundary).
The partial topological twist replaces the $SO(2)_{\Sigma}$ little group
in three dimensions with the diagonal subgroup
\be
SO(2)_{\Sigma}' \subset SO(2)_{\Sigma} \times U(1)_t
\ee
where $U(1)_t$ is the R-symmetry of the 3d $\CN=2$ theory.

In the case when space-time is $S^1 \times \Sigma$, compactification on $S^1$ produces an effective 2d $\CN=(2,2)$ theory on $\Sigma$. The partial topologial twist considered above becomes the usual A-model twist in 2d. Let us breifly review basic facts about topologially twisted $\CN=(2,2)$  2d theories. In two-dimensional $\CN=(2,2)$ supersymmetry, the right-moving supercharges are usually denoted as $Q_+$ and $\bar Q_+$,
while the left-moving supercharges are usually denoted as $Q_-$ and $\bar Q_-$ (see {\it e.g.} \cite{Witten:1993yc,Hori:2003ic}).
One also defines $H_{\pm} = (H \pm P)/2$, so that $Q^2_r = H_+$ with $Q_r = (Q_\pm + \bar Q_\pm)/2$. For example, in these conventions, the elliptic genus is
\be
\Tr q^{H_-} e^{i \gamma J_{\ell}} e^{i \pi J_r}
\ee
After the topological twist, which allows to formulate the theory on a general 2-manifold $\Sigma$,
the supercharges $Q_-$ and $\bar Q_+$ have zero spin in the A-model,
while $\bar Q_-$ and $\bar Q_+$ have zero spin in the B-model.
Usually, in either case, one then defines a BRST operator $Q$ to be a sum of these scalar supercharges. The theory becomes effectively topological if one restricts to cohomology of $Q$-operator. The elements of such $Q$-cohomology form a ring $\CR$. For the A-model twist, this is the ring of anti-chiral operators in the left-moving sector and chiral operators in the right-moving sector, the so-called $(a,c)$ ring. Via state-operator correspondence $\CR$, as a vector space, can be identified with the Hilbert space of the topological A-model on a circle.

The standard textbook example of A-model is a twisted sigma-model based on a target space $X$, with
\be
\CR = H^* (X)
\label{Aring}
\ee
such that a product in this $Q$-cohomology ring is the usual cup product in the classical de Rham cohomology of $X$,
$(Q,Q^{\dagger},H) \sim (d,d^*,\Delta)$.
In the large volume limit there are no quantum corrections, but at finite volume the ring $\CR$ gets deformed
into quantum cohomology ring $QH^* (X)$. Here, we will mostly focus on the classical ring / large volume limit.

One of the basic ingredients in $\CN=(2,2)$ 2d theories is a free chiral superfield
\be
\Phi = \phi + \theta^+ \psi_+ + \theta^- \psi_- + \ldots
\ee
Since it appears as a basic building block in many models, it is instructive to consider a slight generalization
where the lowest component $\phi$ carries R-charges $(m,n)$ under $U(1)_V$ and $U(1)_A$ symmetries, respectively.
\be
\begin{array}{l@{\;}|@{\;}c@{\;}|@{\;}c@{\;}|@{\;}c@{\;}|@{\;}c}
\multicolumn{5}{c}{\text{ chiral multiplet}} \\[.1cm]
& U(1)_{\Sigma} & U(1)_V & U(1)_A & m=0 \\\hline
\phi & 0 & m & n & z \\
\psi_- & 1 & m-1 & n+1 & dz \\
\bar \psi_+ & -1 & m+1 & n+1 & d \bar z \\
\bar \psi_- & 1 & m+1 & n-1 & \\
\psi_+ & -1 & m-1 & n-1 &
\end{array}
\notag
\ee
Upon the A-model twist, the generator of $U(1)_{\Sigma}$ ``Lorentz symmetry'' in two dimensions
is replaced by a sum of generators of $U(1)_{\Sigma}$ and $U(1)_V$,
so that the first three fields --- namely, $\phi$, $\psi_-$ and $\bar \psi_+$ --- transform as scalars under $U(1)_{\Sigma}$ when $m=0$.
Note, in this case an observable $\CO \in H^{p,q} (X)$
that corresponds to a cohomology element on $X$ of Hodge degree $(p,q)$ has R-charges $(m,n) = (q-p,q+p)$.

When A-model is obtained by a topological twist of a 2d $\CN=(2,2)$ gauge theory, the R-symmetry $U(1)_V$ acts on the Higgs branch
while the Coulomb branch parametrized by lowest components of twisted chiral superfields can be acted upon by $U(1)_A$.
For future reference, it is helpful to keep in mind that $U(1)_V$ R-symmetry will be identified with R-symmetry $U(1)_t$
that was already introduced in \eqref{UqUtsymmetries}. This symmetry will play a central role throughout the entire paper.
In the present context, its distinguished feature --- compared to $U(1)_A$ --- is that $U(1)_V$ is non-anomalous.
Moreover, it is abelian, which means that $U(1)_V$ remains a symmetry of the A-model even after the topological twist.
This allows to introduce a notion of the ``refined A-model'' where we keep track of the $U(1)_V$ charge
in the partition function on $\Sigma$ and in all other correlation functions.

{}From the point of view of the original 3d $\CN=2$ theory such refinement can be realized as follows. As a result of the partial topological twist we get a theory that associates a vector space $\CH(\Sigma)$ to a 2-manifold $\Sigma$, and a category $\CC$ to a circle $S^1$. The vector space $\CH(\Sigma)$ has a meaning of $Q$-cohomology (now in 3d sense) of the physical Hilbert space of the 3d theory quantized on $\Sigma$, while $\CC$ has a meaning of the category of boundary conditions. In other words, we obtain a 3d theory categorifying the A-model on $\Sigma$:
\be
\begin{aligned}
& \underline{~~~~~~~~~~~~~3d~~~~~~~~~~~~~~} \\
& \Sigma ~\leadsto~ \text{vector space} ~\CH({\Sigma}) \\
& S^1 ~\leadsto~ \text{category} ~\CC
\end{aligned}
\qquad \text{vs.} \qquad
\begin{aligned}
& \underline{~~~~~~~~~~~~~2d~\text{A-model}~~~~~~~~~~~~} \\
& \Sigma ~\leadsto~ \text{number} ~Z (\Sigma) = \chi (\CH({\Sigma})) \\
& S^1 ~\leadsto~ \text{vector space} ~K^0 (\CC)\cong \CR
\end{aligned}
\label{Amodel-functors}
\ee
where $K^0$ denotes the Grothendieck group\footnote{In some theories, the appropriate ``decategorification'' functor
is the Hochschild homology \cite{Gorsky:2013jxa,Chun:2015gda}.}
and $\chi$ denotes the Euler characteristic. The consistency implies $\CH(T^2)=K^0(\CC)$ $\cong \CR$.
Since the R-symmetry of a 3d $\CN=2$ theory is abelian, it survives after the partial topological twist.
Hence, the $Q$-cohomology  $\CH({\Sigma})$ comes equipped with a $\Z$-grading:
\be
\CH({\Sigma}) \; = \; \Z\text{-graded}~Q\text{-cohomology}
\label{HgradedA}
\ee
such that
\be
\chi (\CH({\Sigma})) \; = \; Z_{\text{A-model}} (\Sigma)
\label{EulerAmodel}
\ee
is a partition function of a two-dimensional theory obtained by reduction from 3d to 2d.
In fact, our approach suggests a {\it refined} A-model partition function,
\be
\dim_t \CH({\Sigma}) = \sum_j t^j \dim \CH^j({\Sigma})
\ee
defined as the Poincar\'e polynomial of \eqref{HgradedA} with respect to $t$-grading.
Upon the reduction to 2d, the R-symmetry of 3d $\CN=2$ theory becomes the R-symmetry $U(1)_V$.
Note that the partial topological twist by $U(1)_t$ symmetry has been considered in four dimensions \cite{Johansen:1994aw}
and, more recently, in three dimensions \cite{Cecotti:2013mba,Benini:2015noa},
but without keeping track of the remaining $U(1)_t$ grading.

The category $\CC$ and the vector space $\CH(\Sigma)$ in the left column of (\ref{Amodel-functors}) have additional structures. In particular, the vector space $\CH({\Sigma})$ is equipped with the action of the mapping class group of $\Sigma$. The functor $\Sigma\mapsto \CH(\Sigma)$ should also satisfy particular properties with respect to decomposition of Riemann surfaces $\Sigma=\Sigma_1 \cup \Sigma_2$. All in all, the partial topological twist of a 3d $\CN=2$ theory should provide us with a 2d modular functor (MF). It is known that there is a one-to-one correspondence between 2d modular functors and modular tensor categories (MTC) \cite{Moore:1988qv} (see also \cite{MTClectures} for comprehensive lectures and more references). The category $\CC$ in (\ref{Amodel-functors}) is then the MTC corresponding to this 2d MF. The MTC structure on $\CC$ induces the ring structure and the $SL(2,\Z)$ action on its Grothendieck group $K^0(\CC)$ which will be important later in the text. Note that a 3d (extended) TQFT contains 2d MF/MTC strucutre, but not every MTC defines a TQFT. This is consistent with the fact that 3d theory here defined by a partial topological twist along $\Sigma$ is fully topological (and obeys cutting-and-gluing) along $\Sigma$, but not on a general 3-manifold. In the special case when 3d $\CN=2$ theory is associated to a 3-manifold via 3d/3d correspondence, there is another vantage point on the MTC structure which will be discussed in section~\ref{sec:2bases}.

In general, a topological A-model or, in fact, any 2d TQFT is described by a Frobenius algebra,
{\it i.e.} the data of 2-point functions $\eta_{ij}$ that define the ``metric'' and the 3-point functions $C_{ijk}$
that define the ``structure constants'':
\be
\eta_{ij} = \langle \phi_i \phi_j \rangle_0 = \langle i \vert j \rangle \,, \qquad \phi_i \vert j \rangle = C_{ij}^k \vert k \rangle
\ee
\be
\eta^{ij} \eta_{jk} = \delta^i_k \,, \qquad
C_{ijk} = \langle \phi_i \phi_j \phi_k \rangle_0 = \langle i \vert \phi_j \vert k \rangle = \eta_{il} C_{jk}^l
\ee
where $\phi_i\in \CR$ denotes the operator corresponding to the basis vector $|i\rangle$  in the Hilbert space on the circle and $\langle \ldots \rangle_g$ denotes a correlation function on genus $g$ Riemann surface. Thanks to this structure, inserting a complete set of states $\sum_{ij} \vert i \rangle \eta_{ij} \langle j \vert$
anywhere on $\Sigma$ we get a surgery formula:
\be
\langle \phi_{a_1} \ldots \phi_{a_n} \rangle_g = \sum_{ij}
\langle \phi_{a_1} \ldots \phi_{a_r} \phi_i \rangle_h \eta^{ij}
\langle \phi_j \phi_{a_{r+1}} \ldots \phi_{a_n} \rangle_{g-h}
\ee
which allows to calculate the partition function on any Riemann surface $\Sigma$.

However, to get a nontrivial result for a correlation function one has to ensure cancellation of the ghost number anomaly. In the A-model with a target space $X$, the ghost number anomaly is
\be
(1-g) \dim (X)
\label{ghost-anomaly}
\ee
indicating that low-genus cases are most interesting\footnote{For a non-trivial embedding of the worldsheet Riemann surface $\Sigma\stackrel{\imath}{\rightarrow} X$ there is also $\langle[\imath(\Sigma)],c_1(X)\rangle$ contribution to the ghost number anomaly.}.
In particular, when $g=1$ we have
\be
\langle 1 \rangle_{g=1} = \Tr_{\CR=\CH(T^2)} (-1)^F = \chi (X)
\ee
which, together with \eqref{Aring}, gives an elementary example of a categorification.
In this case, 2d A-model categorifies a one-dimensional QFT (namely, SUSY quantum mechanics), while 2d A-model itself is categorified by a 3-dimensional theory. Starting with section~\ref{sec:3d3d}, we will talk about even more sophisticated
examples of categorification where topologically twisted 3d $\CN=4$ theory or 3d Chern-Simons TQFT is categorified by higher-dimensional structures. The ghost number anomaly (\ref{ghost-anomaly}) also vanishes if $\dim X=0$ which will be relevant for A-models associated to rational homology spheres that we consider later in the text.

In order to be able to categorify a partition function with insertions, it is necessary that the inserted operators lift to line operators in 3d, not local ones. In this case
\begin{equation}
 \langle \phi_1 \ldots \phi_n\rangle_g =\Tr_{\CH(\Sigma_{g,\phi_1,\ldots,\phi_n})} (-1)^F
\end{equation}
where $\CH(\Sigma_{g,\phi_1,\ldots,\phi_n})$ is the ($Q$-cohomology of\footnote{We will omit this clarification later in the text. By default, the Hilbert space will mean the $Q$-cohomology (or, equivalently, BPS part of) of the physical Hilbert space.} the) Hilbert space of 3d theory quantized on genus-$g$ Riemann surface with line operators $\phi_1,\ldots,\phi_n$ supported at points on the Riemann surface (times ``time'').

A good illustration of a 2d topological A-model is a sigma-model with target space $X = \cp^1$.
It has a two-dimensional chiral ring $\CR \cong H^* (\cp^1)$ whose elements we can suggestively denote $1$ and $L$.
They both have degree zero under $U(1)_t = U(1)_V$, but under $U(1)_A$ transform with degree $0$ and $2$, respectively.
Another instructive example is the simplest instance of a vortex moduli space (space of Hecke modifications),
namely $X = \C \times \cp^{N-1}$, whose A-model (and its categorification) is related to HOMFLY-PT knot homology \cite{Gorsky:2013jxa}.
In this case, we have the following non-trivial correlation functions:
$$
\includegraphics[width=3in]{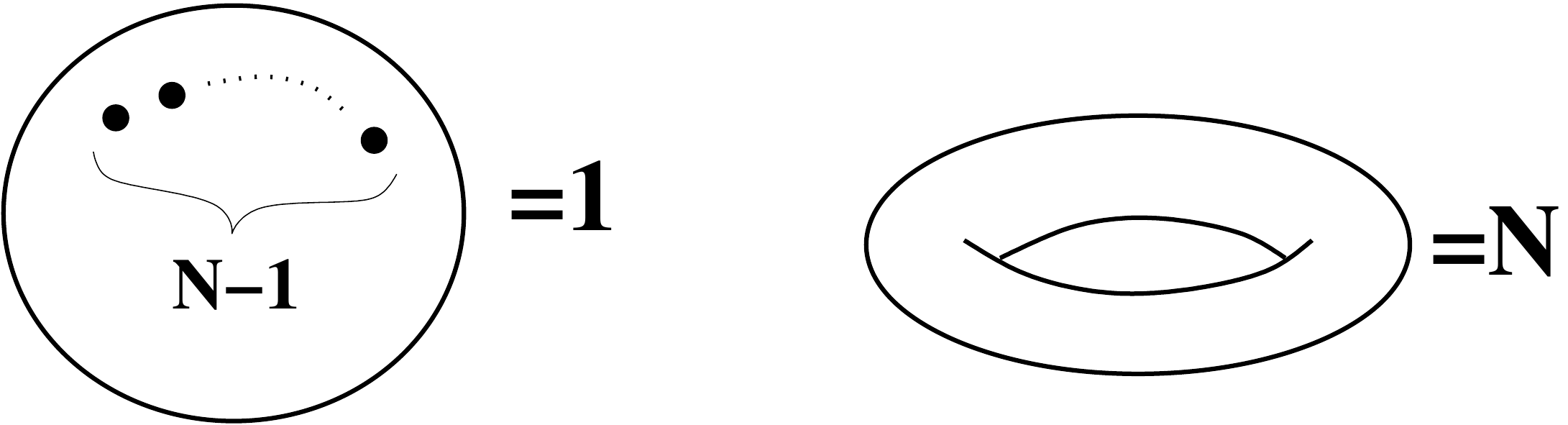}
$$
where a dot denotes insertion of $L$ operator.

\subsection{A-model labeled by 3-manifolds: $T^\Am {[}M_3{]}$}
\label{sec:Amodel-3mfld}

So far, we considered a general 3d $\CN=2$ theory.
Now, let us focus more closely on theories $T_G[M_3]$ labeled by 3-manifolds \eqref{TMMM}.
We will denote the corresponding A-model by $T^\Am_G[M_3]$.

A simple example of such theory is the ``Lens space theory'' listed in Table~\ref{tab:examples}.
It will be one of our working examples throughout the paper, see {\it e.g.} sections \ref{sec:Lp1-example} and \ref{sec:higherRank}.
For the gauge group $G=U(1)$, such 3d theory $T_G[M_3]$ consists of a $U(1)$ super-Chern-Simons theory
at level $p$ and a free chiral multiplet.
Its dimensional reduction to 2d is a theory of a free chiral multiplet and a massive
vector multiplet that can be equivalently described by a twisted chiral multiplet with
the twisted superpotential, {\it cf.} \eqref{LspaceTheory}:
\be
T^\Am_{U(1)} [L(p,1)] \; : \quad \text{2d $\CN=(2,2)$ twisted chiral with $\tilde W = p \sigma^2/2$ and a free chiral}
\ee
Note, the A-model is independent on the superpotential $W$ but depends holomorphically on the twisted superpotential $\tilde W$,
which has charge $(m,n)=(0,2)$ under $U(1)_V \times U(1)_A$.
The $(a,c)$ ring of the Landau-Ginzburg model with the twisted superpotential $\tilde W$ is equal to the Jacobi ring,
which is a mirror version of a more familiar $(c,c)$ ring in the LG B-model.
For $\C^*$-valued fields that describe $U(1)$ gauge multiplets,
the suitable condition is $\exp \left( \frac{\partial \tilde W}{\partial \sigma} \right) = 1$.

If we ignore the trivial free chiral multiplet, the $(a,c)$ ring is given by
\begin{equation}
 \CR=\C[z]/(z^p-1)\cong \C[\Z_p]
\label{Lp1-ring}
\end{equation}
where $z=e^\sigma$. The non-trivial part of the 3d theory is equivalent to the usual $U(1)$ level $p$ bosonic Chern-Simons theory. The elements of (\ref{Lp1-ring}) are lifted to Wilson lines and the multiplication in $\CR$ agrees with the fusion rules.

Suppose we are interested in computing topologically twisted partition function of 3d $\CN=2$ theory $T_G[M_3]$ on $S^1 \times \Sigma$. Such partition function can be interpreted as the partition function of 6d $\CN=(2,0)$ theory on $S^1\times \Sigma\times M_3$ with topological twists along both $\Sigma$ and $M_3$. If we first reduce 6d theory on $\Sigma$ we get an $\CN=2$ 4d theory $T_G[\Sigma]$ on $M_3\times S^1$.
As we explain in detail in sections \ref{sec:twistsM3} and \ref{sec:4mfld},
the topological twist along $M_3$ is equivalent to the Donaldson-Witten topological twist of the 4d theory on $M_4=M_3\times S^1$.
Thus the partition function of $T_G[M_3]$ on $\Sigma\times S^1$ gives us an invariant of $M_3$ categorified by Donaldson-Floer homology associated with 4d $\CN=2$ theory $T_G[\Sigma]$. If, instead, we reduce 3d $\CN=2$ theory $T_G[M_3]$ down to two dimensions we get 2d $\CN=(2,2)$ theory, whose space of vacua is the space of complex $G_{\C}$ connections on $M_3$ \cite{DGH}. The same invariant of $M_3$ is then given by the partition function of the A-model $T^\Am[M_3]$ on $\Sigma$:
$$
\begin{array}{ccccc}
\; & \; & \text{6d $(0,2)$ theory} & \; & \; \\
\; & \; & \text{on $\Sigma \times S^1 \times M_3$} & \; & \; \\
\; & \swarrow & \; & \searrow & \; \\
\text{2d A-model on $\Sigma$ with} & \; & = & \; & \text{4d SW/DW topological twist} \\
\text{target $\CM_{\text{flat}} (G_{\C}, M_3)$} & \; & \; & \; & \text{of $T_G[\Sigma]$ on $S^1 \times M_3$}
\end{array}
$$
In particular, the Seiberg-Witten invariants can be realized either by working with a supergroup $G=U(1|1)$ of (super-)rank $N=0$
or, alternatively, by choosing $G=U(1)$ and $\Sigma=S^2_\fM$, a sphere with particular defects.
This will be explored in more detail in section~\ref{sec:3d3d}.

This fits very well with the analysis of topological twists that will be discussed more fully in sections \ref{sec:twistsM3} and \ref{sec:twistsM4}.
Indeed, the 3d $\CN=4$ theory $T[S^1 \times \Sigma]$ is
twisted on $M_3$ by means of the R-symmetry $SU(2)_R$ which can be lifted to four dimensions (and, hence, categorified)
and under which the scalars in vector multiplet are singlets and scalars in hypermultiplets transform as doublets.
This twist of 3d $\CN=4$ gauge theory was extensively studied by Blau and Thompson \cite{Blau:1996bx,Blau:1997pp,Blau:2000iy},
who showed that in the UV it is precisely the 3d reduction of the SW/DW twist,
while in the IR it gives a RW twist of the 3d $\CN=4$ sigma-model on the Coulomb branch of the theory $T[S^1 \times \Sigma]$.


In the rest of the section we will study various properties of A-model $T^\Am[M_3]$. For general $M_3$, as usual, it should be described in terms of quantum cohomology of the target space, that is
\begin{equation}
 \CR = QH^*(\CM_\text{flat}(G_\C,M_3)).
\end{equation}
In particular
\begin{equation}
 \CR\cong \CH_{T^\Am[M_3]_G}(S^1)\cong H^*(\CM_\text{flat}(G_\C,M_3)).
\end{equation}
as vector spaces over $\C$. In many of our examples, however, $\M_\text{flat}(G_\C,M_3)$ will simply be a discrete set. In particular, this is the case when $M_3$ is a Lens space. When $\M_\text{flat}(G_\C,M_3)$ is a discrete set of points, the Hilbert space of $T^\Am [M_3]$ on $S^1$ is simply a finite dimensional space of complex valued functions on $\M_\text{flat}(G_\C,M_3)$:
\begin{equation}
	\CH_{T^\Am[M_3]}(S^1)=\C[\M_\text{flat}(G_\C,M_3)]
	\label{Hflat}
\end{equation}
equipped with a chiral ring structure.\footnote{To be more precise, it is actually commutative unital algebra over $\C$.} In particular, there is a product map
\begin{equation}
{\,\raisebox{-.8cm}{\includegraphics[width=1.8cm]{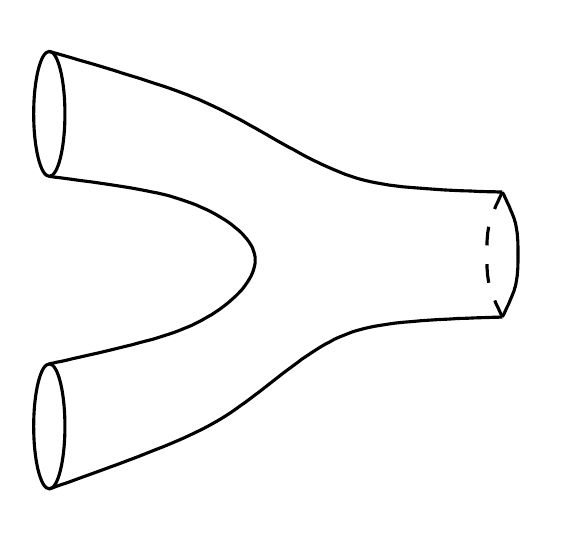}}\,} :
\;\C[\M_\text{flat}(G_\C,M_3)]\otimes \C[\M_\text{flat}(G_\C,M_3)] \;\xrightarrow[~]{~\mu~}\; \C[\M_\text{flat}(G_\C,M_3)]
	\label{Mflatring}
\end{equation}
realized by point-like multiplication of functions on $\CM_\text{flat}$.
Physically the product map $\mu$ is given by the partition function of $T^\Am[M_3]$ on a pair of pants. Together with the scalar product on $\CH_{T^\Am}(S^1)$ (or, equivalently, a unit element) it provides a Frobenius algebra structure. As was already mentioned in section \ref{sec:Amodel-cat} this data is sufficient to calculate the partition function of 2d TQFT $T^\Am[M_3]$ on any Riemann surface with holes (but without any special defects). When the algebra is just the ordinary algebra of functions on $\CM_\text{flat}(G_\C,M_3)$, the result is quite simple and basically provides information about the number of flat connections.
For example, the partition function of the A-model on any closed Riemann surface with positive genus is simply given by
\begin{equation}
Z_{T^\Am[M_3]}[\Sigma] \; = \; \# \CM_\text{flat}(G_\C,M_3) \,.
\label{TAM3counts}
\end{equation}
Many of these statements have a straightforward generalization to the case of arbitrary $M_3$ and $G$.

According to \eqref{Amodel-functors}-\eqref{EulerAmodel}, 3d $\CN=2$ theory $T[M_3]$ on $\R \times \Sigma$
with A-model twist along $\Sigma$ provides a natural categorification of \eqref{TAM3counts}.
In the basic case of $G=U(1)$, {\it i.e.} for a single fivebrane, this theory should be regarded as a physical counterpart
of the ``simplest'' variant of the Heegaard Floer homology $\hat{HF} (M_3)$ that categorifies $|H_1 (M_3)|$:
\be
\chi \left( \hat{HF} (M_3) \right) \; = \; \pm |H_1 (M_3; \Z)|
\label{HFhatEuler}
\ee
where the right-hand side is defined to be zero when $H_1 (M_3; \Z)$ is not finite, {\it i.e.} when $b_1 (M_3) > 0$.
Note, for a 3-manifold with $b_1 (M_3) = 0$, {\it i.e.} for a rational homology sphere,
this implies $\text{rk} \hat{HF} (M_3) \ge |H_1 (M_3)|$
and the equality holds for the so-called {\it L-spaces} that will appear among our examples in section~\ref{sec:3d3d}.

\subsection{$SL(2,\Z)$ action}

Actually, there is an additional structure on (\ref{Hflat}) that contains non-trivial information about $M_3$.
The Hilbert space of the A-model $T^\Am[M_3]$ on a circle can be identified with the Hilbert space of $T[M_3]$ on a 2-torus:
\begin{equation}
	\CH_{T^\Am[M_3]}(S^1) = \CH_{T[M_3]}(T^2)
	\label{HT2}
\end{equation}
Since $T^2$ has a mapping class group $SL(2,\Z)$ it follows that (\ref{HT2}) should be a (projective) representation of $SL(2,\Z)$:
\begin{equation}
	\mathfrak{R}:\;\;SL(2,\Z) \longrightarrow \text{End}(\CH_{T^\Am[M_3]}(S^1))
	\label{HSL2Z}
\end{equation}
which provides us with additional structure on (\ref{Hflat}).
It was discussed in \cite{Cecotti:2013mba} in a slightly different context.
Here, we are going to look more closely at its implications for the A-model $T^\Am[M_3]$
and learn something interesting about the moduli space of complex flat connection on a 3-manifold $M_3$.

There are a few cases when the representation (\ref{HSL2Z}) is well understood. First, let us also assume that the fundamental group $\pi_1(M_3)$ is finite and therefore
\begin{equation}
	\M_\text{flat}(G_\C,M_3)=\M_\text{flat}(G,M_3)
\end{equation}
Consider the case when $G=U(1)$. Denote $H\equiv H_1(M_3)$. Then
\begin{equation}
	\M_\text{flat}(U(1),M_3)\cong \text{Hom}(H,U(1)) \equiv \hat{H}
\end{equation}
where $\hat{H}$ is the Pontryagin dual of $H$. Note that
\begin{equation}
	\CH_{T^A[M_3]}(S^1) =\C[\hat{H}] \cong  \C[H]
	\label{PDualRing}
\end{equation}
as vector spaces. The one-to-one correspondence between the spaces of function $\C[H]$ and $\C[\hat{H}]$ is given by the Fourier transforms:
\begin{equation}
\hat{f}:\;\hat{H}\rightarrow\mathbb{C},\qquad
\hat{f}(q)=\sum_{h\in H}f(h)\,q^{-1}(h)
\label{FourierToHat}
\end{equation}
\begin{equation}
{f}:\;{H}\rightarrow\mathbb{C},\qquad
{f}(h)=\frac{1}{|H|}\sum_{q\in \hat{H}}\hat{f}(q)\,{q}(h)
\label{FourierFromHat}
\end{equation}
The isomorphism (\ref{PDualRing}) also works at the level of rings if we treat $\C[H]$ as the group ring\footnote{Note that group ring structure is not the same as the ring of functions structure, but they get exchanged under the Fourier transform which exchanges point-wise multiplication with convolution.} of the abelian group $H$. Note that if we identify $\hat H$ with $H$ the Fourier transform (up to an overall normalization) plays the role of the $S$-transform acting on $\C[H]$.

\begin{figure}[ht]
\centering
\includegraphics[scale=1]{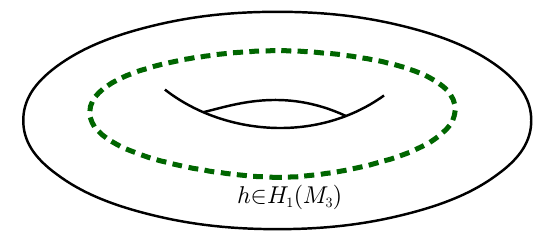}\caption{Introducing a line operator labelled by $h\in H$
along a non-trivial cycle of the solid torus creates a state in $\CH_{T^A[M_3]}(S^1) =  \C[H]$.}
\label{fig:line-op-state}
\end{figure}
Note that the elements of $\CH_{T[M_3]}(T^2)$ are in one-to-one correspondence with (BPS) line operators in $T[M_3]$, as illustrated in Figure~\ref{fig:line-op-state}. The correspondence can be realized by a considering a solid torus with a line operator along a non-trivial cycle of the torus. The 6d theory origin of the line operators are codimension 4 defects wrapping 1-cycles of $M_3$. Therefore, the elements of $H$ play the role of charges. The multiplication on the group ring $\C[H]$ can be then understood as fusion of line operators and is determined by charge conservation. Let us note that for different 3-manifolds with the same $H_1(M_3)$ one obtains the same ring, but the spaces $\CH_{T^\Am[M_3]}(S^1)$ can differ as representations of $SL(2,\Z)$. In other words, the map (\ref{HSL2Z}) can still capture the difference. This happens for example for non-homeomorphic Lens spaces $L(p,q)$ and $L(p,q')$ for both of which $\pi_1(M_3)=\Z_p$.

\subsection{The two bases}
\label{sec:2bases}

In general, the representation (\ref{HSL2Z}) can be constructed in the following way. Consider any 4-manifold $M_4$ such that $\d M_4 =M_3$. Moreover, let us pick a metric on $M_4$ such that it looks like $M_3\times \R$ near the boundary. Consider 6d theory on $M_4 \times T^2 $ with a topological twist mixing the R-symmetry $SO(3)_R$ with the $SU(2)_{\ell}$ subgroup of $SU(2)_{\ell} \times SU(2)_r = SO(4)$ of local rotations. As we explain more fully in section \ref{sec:4mfld}, this will result in Vafa-Witten theory \cite{VafaWitten} with gauge group $G$ on $M_4$. The partition function of the theory will give an element of the Hilbert space of the 4d TQFT associated to $M_3$:
\begin{equation}
	Z_\text{VW}(M_4)(\tau)\;\in \;\CH_{\VW}(M_3)
	\label{HVW}
\end{equation}
where $\tau$ is the modular parameter of the torus, which plays the role of the coupling in Vafa-Witten theory. The Hilbert space of Vafa-Witten theory can be related to the Hilbert space of $T[M_3]$ on $T^2$:
\begin{equation}
	\CH_\text{VW}(M_3)=\CH_{T[M_3\times T^2]}=\CH_{T[M_3]}(T^2).
\end{equation}
Under the action of $SL(2,Z)$ group, the partition function (\ref{HVW}) should transform as
\begin{equation}
	Z_\text{VW}(M_4)\left(\frac{a\tau+b}{c\tau+d}\right)=
	C(\tau;a,b,c,d)\,\mathfrak{R}\left[\left(\begin{array}{cc}
	a & b \\
	c & d
	\end{array}
	\right)\right]Z_\text{VW}(M_4)(\tau)
	\label{VWSL2Z}
\end{equation}
where $C$ is an overall anomaly-related factor and $\mathfrak{R}$ is the same as in (\ref{HSL2Z}). The natural boundary condition for VW theory on $M_4$ requires gauge connection to approach a flat one at the boundary. Therefore it provides a function of $\tau$ for each gauge equivalence class of flat connections on $M_3$:
\begin{equation}
	Z_{\VW}(M_4)_\rho(\tau),\qquad \rho\in \hat{\CM}_\text{flat}(G,M_3)
	\label{VWflat}
\end{equation}
The reason why we put a hat on this set is to formally distinguish it from, of course, an isomorphic set $\CM_\text{flat}(G,M_3)$ that appeared earlier. It will soon become clear why we need such a distinction. The values (\ref{VWflat}) can be understood as components of a vector (\ref{HVW}) in the Hilbert space, expressed in a particular basis:
\begin{equation}
	Z_{\VW}(M_4)\;\in \C[ \hat{\CM}_\text{flat}(G,M_3)] \cong \CH_{\VW}(M_3)=\CH_{T[M_3]}(T^2)
\end{equation}
Moreover, as in \cite{Gadde:2013sca}, the components (\ref{VWflat}) can be interpreted as characters
of modules $\{M_\rho\}$ of the chiral vertex operator algebra $\text{VOA}[M_4]$ of a 2d (0,2) theory $T[M_4]$:
\begin{equation}
	Z_{\VW}(M_4)_\rho(\tau)=\Tr_{M_\rho} (-1)^F q^{L_0}
	\label{ZVWtrace}
\end{equation}
where, as usual, $q=e^{2\pi i\tau}$. From the viewpoint of VW theory, this $q$-series plays the role of a generating function for the Euler characteristic of instanton moduli spaces:
\begin{equation}
 Z_{\VW}(M_4)_\rho(\tau)=\sum_nq^n\,\chi(\CM^\text{inst}_{n,\rho})
\end{equation}
where $n$ is the instanton number, so that one expects that
\begin{equation}
 M_\rho=\bigoplus_n H^*(\CM^\text{inst}_{n,\rho})
\end{equation}
as $\Z$-graded vector spaces. Since modules of $\text{VOA}[M_4]$ are labeled by the elements of $\hat{\CM}_\text{flat}(G,M_3)$ we can expect that as a ring
\begin{equation}
	\CH_{T[M_3]}(T^2) \; \cong \; \C[ \hat{\CM}_\text{flat}(G,M_3)]\; = \;\text{Representation ring of VOA$[M_4]$}
	\label{HVOA}
\end{equation}
The ring structure as well as non-trivial part of $SL(2,\Z)$ representation (\ref{VWSL2Z}) should not depend on a particular choice of the $M_4$ becuase it is fixed by the fact that it can be self-consistently glued with any $M_4'$ such that $\d M_4'=-M_3$. Note that previously we had
\begin{equation}
	\CH_{T[M_3]}(T^2) \; \cong \; \C[ {\CM}_\text{flat}(G,M_3)] \; = \;\text{Ring of functions on $\CM_\text{flat}(G,M_3)$}
	\label{HMflat}
\end{equation}
that is, the multiplication is point-wise in the basis given by ${\CM}_\text{flat}(G,M_3)$. The difference is actually expected since these two bases $\CM_{\text{flat}}(G,M_3)$ and $\hat\CM_{\text{flat}}(G,M_3)$ should be related by $S$-transform. This can be seen for example from the fact that we need to make $S$-transform to get a CS theory with coupling $k=-1/\tau$ on the the boundary of 4d $\CN=4$ SYM with coupling $\tau$ (see {\it e.g.} \cite{Witten:2011zz,Dimofte:2011jd}). The $S$-transform between the two bases can be understood as a Fourier transform translating point-wise multiplication in (\ref{HMflat}) into more non-trivial one in (\ref{HVOA}). The trade-off is that the basis $\hat{\CM}_\text{flat}$ should diagonalize the action of $T$ element of $SL(2,\Z)$. The structure constants in (\ref{HVOA}) are completely fixed by the S-matrix via the Verlinde formula. Note, the relation between $\CM_{\text{flat}}(G,M_3)$ and $\hat\CM_{\text{flat}}(G,M_3)$ generalizes the relation between $\hat{H}$ and $H$ considered previously in the abelian context.

A classic example of this structure is given by Nakajima's result \cite{Nakajima}.
Consider the case where $M_3=L(p,p-1)=-L(p,1)$ and the corresponding 4-manifold is the resolution of $A_{p-1}$ singularity
\begin{equation}
	M_4=\widetilde{\C^2/\Z_p}.
\end{equation}
and $G=U(N)$. The VW partition function is given by the characters of $\hat{su}(p)_N$ affine
algebra\footnote{\label{footnote:VW}A few technical but conceptually not very important clarifications are due here. The explicit computations show that these are not characters of integrable representations of $\hat{su}(p)_N$, but rather products of characters of $\hat{su}(N)_1$ integrable representations \cite{Fujii:2005dk} (see also \cite{Bruzzo:2013daa} and references therein):
\begin{equation}
 Z_{\VW}(M_4)_\rho(\tau)= \prod_{i=1}^N\frac{\chi_{\rho_i}^{\hat{su}(p)_1}(\tau)}{(q;q)_\infty}
\end{equation}
In particular, for trivial flat connection this is the character of so-called Fock space representation of $\hat{su}(p)_N$. Such characters can still be decomposed into characters of $\hat{su}(p)_N$ integrable representations because $\hat{su}(p)_N \subset (\hat{su}(p)_1)^N$. In turn, the latter affine algebra can be embedded into the algebra of $Np$ free chiral fermions $\hat{u}(1)_1^{Np}$. The relation between VW theory on $A_{p-1}$ singularity and free fermions as well as the the physical interpretation of the embedding $\hat{su}(p)_N \subset (\hat{u}(1)_1)^{Np}$ was studied in \cite{Dijkgraaf:2007sw}. It was formally understood as a change from descrete to continous basis in \cite{Gadde:2013sca}. All in all different versions of the partition function corresponding to different steps in the embedding sequence
\begin{equation}
 \hat{su}(p)_N \subset (\hat{su}(p)_1)^N \subset (\hat{u}(1)_1)^{Np}
\end{equation}
differ by inclusion/exclusion of certain degrees of freedom living on the boundary of $M_4$. Note that $(\hat{su}(p)_1)^N$ characters (unlike $\hat{su}(p)_N$ characters) have $q$-expansion of the following form:
\begin{equation}
 q^{S_\text{CS}}(n_1+n_2q+n_3q^2+...),\qquad n_i\in \Z
\end{equation}
where $S_\text{CS}\in\Q$ is the CS action of the corresponding flat connection on $M_3$. As we will see in section \ref{sec:higherRank}, the modular properties of such characters indeed provide us with the correct S-transform of the CS partition function on $M_3$.}:
\begin{equation}
	Z_{\VW}(M_4)_\rho(\tau)=\Tr_{M_\rho} q^{L_0} = \chi_\rho^{\hat{su}(p)_N}(\tau)
\end{equation}
The moduli space of flat connections on $M_3$ is given by
\begin{equation}
	\CM_\text{flat}(U(N),L(p,1))=\text{Hom}(\Z_p,U(N))\,/U(N)=\text{Sym}^N\Z_p
\end{equation}
Then
\begin{equation}
	\C[\CM_\text{flat}(U(N),L(p,1))]=\text{Functions on $\text{Sym}^N\Z_p$}.
\end{equation}
The basis $\hat\CM_{\text{flat}}(G,M_3)$ then can be understood as a particular basis in the ring of functions with elements corresponding to representations of $\hat{su}(p)_N$ so that the product in such basis satisfies the fusion rules\footnote{Such functions can be chosen to be polynomials in $p-1$ variables satisfying $p-1$ polynomial constraints. The solutions the constraint equations are in one-to-one correspondence with elements of $\text{Sym}^N\Z_p$. The constraints can be realized as extremum equations of the so-called fusion potential \cite[\textsection 16.5]{DiFrancesco}.}:
\begin{equation}
	\C[\hat\CM_{\text{flat}}(G,M_3)] = \text{Verlinde algebra of $\hat{su}(p)_N$}.
\end{equation}
This is in agreement with the fact that \cite{Acharya:2001dz,CCV,Gadde:2013sca,Gukov:2015sna,Cecotti:2013mba}:
\be
T_{U(N)}[L(p,1)] \; : \quad \text{3d $\CN=2$ $U(N)$ super-CS at level $p$ with adjoint chiral $\Phi$}
\label{LspaceTheory}
\ee
As in \cite{Acharya:2001dz,Gukov:2015sna}, it is often convenient to give a (real) mass $\beta$ to the adjoint chiral multiplet $\Phi$.
Then, integrating out $\Phi$ shifts the level by $+N$, which precisely compensates the shift $-N$ from
integrating out gluinos. The resulting theory is, therefore, equivalent to bosonic pure Chern-Simons at level $p$
(which, in addition, has the usual renormalization of the level by $+N$).
The non-trivial line operators in this theory, as well as in the parent theory \eqref{LspaceTheory},
obey the fusion rules of $\hat{u}(N)_p$, which is level-rank dual to $\hat{su}(p)_N$ (note that $SL(2,\Z)$ representations associated to 3-manifolds with opposite orientation should be conjugate to each other).


Note, since the setup of 6d $(2,0)$ theory on $M_3\times S^1\times \Sigma$ contains a circle factor,
the observables that we considered in this section can be easily categorified.
For example, from (\ref{ZVWtrace}) it follows that the VW partition function is categorified by the modules $M_\rho$ of $\text{VOA}[M_4]$.
This suggests that the ring (\ref{HVOA}) can be categorified by a category of representation of $\text{VOA}[M_4]$. We expect the categories given by different 4-manifolds $M_4$ with the same boundary $M_3$ to be equivalent, as it was for their Grothendieck rings. The category of representations of a vertex operator algebra has a structure of a modular tensor category (MTC). In particular, it contains the information about $SL(2,\Z)$ representation of its Grothendieck ring $\CH_{T[M_3]}(T^2)$.
This suggests that one can define an MTC-valued invariant of three-manifolds $\text{MTC}[M_3]$.
It is the same as the category $\mathcal{C}$ that appeared in \eqref{Amodel-functors} when the partially twisted 3d $\CN=2$ theory is $T[M_3]$.
This categorification procedure can be summarized in the following diagram:
\begin{equation}
 \begin{CD}
    \begin{array}{c}\text{Modules} \\ \text{of VOA}[M_4]\end{array} @<{Ob}<<
	\begin{array}{c}\text{MTC}[M_3] \\ (=\text{ 2d MF from }T[M_3])\end{array}	 \\
    @VV{\Tr}V @VV{K^0}V \\
     \;\; Z_\VW(M_4)\;\;\in @. \CH_\VW(M_3)=\CH_{T[M_3]}(T^2)
 \end{CD}
\end{equation}

Note, although the basis $\CM_\text{flat}(G,M_3)$ in (\ref{HMflat}) makes the ring structure simple, the categorification is natural in the basis $\hat\CM_\text{flat}(G,M_3)$, {\it cf.} (\ref{HVOA}), that is the elements of $\hat\CM_\text{flat}(G,M_3)$ correspond to simple objects in $\text{MTC}[M_3]$. It would be interesting to compare this description of category $\CC=\text{MTC}[M_3]$ with the one in \cite{haydys2010fukaya}.


\section{Floer homology from $T{[}M_3{]}$}
\label{sec:3d3d}

The so-called 3d/3d correspondence relates topology and geometry of 3-manifolds to physics of supersymmetric
3d $\CN=2$ theories labeled by 3-manifolds.
It can be deduced \cite{DGH} by compactifying a 6d $(0,2)$ theory labeled by a Lie algebra $\mathfrak{g}$ on 3-manifolds and one of its basic features is the relation \eqref{Hflat}--\eqref{HT2} between complex flat
connections on $M_3$ and supersymmetric vacua of $T_G [M_3]$ on a circle, a fact that already played an important role in section \ref{sec:Amodel}.
In later years, the duality was extended to a myriad of sophisticated observables in $T_G [M_3]$, which surprisingly did not include
a much simpler partial topological twist on a Riemann surface $\Sigma$ and its generalizations that, as we show in section \ref{sec:M3veiw},
lead to Seiberg-Witten invariants of $M_3$ and their categorification \eqref{HFHMECH}.
The goal of this section is to extract these 3-manifold invariants directly from $T_G [M_3]$, with suitable choices of the background.
This will lead us to completely new ways of computing the Seiberg-Witten invariants and the Heegaard Floer homology $HF^+ (M_3)$.

\subsection{Twists on $M_3$}
\label{sec:twistsM3}

As was already pointed out in section \ref{sec:Amodel-3mfld}, a large set of numerical 3-manifold invariants allowing natural categorification can be obtained by considering 6d $(0,2)$ theory on
\begin{equation}
\label{choices6d}
	\begin{array}{c}
		\Sigma \times M_4 \\
		\\
		M_4= S^1 \times M_3\qquad \text{or}\qquad \R \times M_3
	\end{array}
\end{equation}
where $\Sigma$ is a 2-manifold (possibly, with punctures or other defects supported at points on $\Sigma$).
In particular, we are going to make contact with \cite{Gukov:2007ck}
where several choices of $\Sigma$ were considered
\be
\Sigma = S^2 \quad , \quad \Sigma = T^2 \quad , \quad \Sigma = \R^2_{q} = \text{``cigar''}
\label{choicesSigma}
\ee
some with partial topological twist along $\Sigma$ and some with Omega-background~\cite{Nekrasov:2002qd}.

In order to preserve a part of supersymmetry one can perform a topological twist along both $\Sigma$ and $M_3$. In the case when $M_3$ and $\Sigma$ are of general holonomy, one can perform topological twisting in the following way. The R-symmetry algebra of the 6d theory is (the universal envelopping of) $SO(5)_R \supset SO(3)_R\times SO(2)_t$. Then one can identify $SO(3)_R$ with $SO(3)$ local rotations of the cotangent bundle of $M_3$ and, similarly, identify $SO(2)_t$ with local rotations of the cotangent bundle of $\Sigma$. After such twist the 6d theory should become independent of metric on both $M_3$ and $\Sigma$. Then, taking them to be small one obtains an effective supersymmetric quantum mechanics along $S^1$. The effective quantum mechanics $T_G[M_3\times \Sigma]$ in general has two supercharges. The partition function of the 6d theory then gives a certain numerical topological invariant of $M_3$ labelled by $G$ and $\Sigma$. Equivalently, it is the partition function of the effective QM on a circe:
\begin{equation}
	Z_{T_G[M_3\times \Sigma]}=\Tr_{\CH_{T_G[M_3\times \Sigma]}}(-1)^F
	\label{ZM_3}
\end{equation}
where $\CH_{T_G[M_3\times \Sigma]}$ is the Hilbert space of the quantum mechanics. By construction $\CH_{T_G[M_3\times \Sigma]}$ provides us with a categorification of the numerical invariant (\ref{ZM_3}). Moreover, it can be extended to the whole functor from the category of 3-manifolds (with cobordisms as morphisms) to the category of vector spaces. Such functor is given by the 4d TQFT obtained by twisting 4d $\CN=2$ theory $T_G[\Sigma]$.

One can also reduce 6d theory on $\Sigma \times S^1 \times M_3$ step-by-step in various ways.
If one first compactifies on $M_3$ one finds a 3d $\CN=2$ theory $T[M_3]$ on $\Sigma\times S^1$ via 3d/3d correspondence \cite{DGH}.
Another possibility is to first compactify on $S^1$. This will give us 5d $\CN=2$ super-Yang-Mills with gauge group $G$ on $M_3 \times \Sigma$. Consider in detail how the topological twist described earlier is realized in terms of the 5d theory. In this background the Euclidean $SO(5)_E$ rotation symmetry is
broken to $SO(3)_{M_3} \times SO(2)_\Sigma$, and we also break the $SO(5)_R$ R-symmetry group accordingly
to $SO(3)_R \times SO(2)_t$ in order to implement the topological twist.
Under this decomposition, the bosons and fermions of the 5d $\CN=2$ super-Yang-Mills transform as
\begin{eqnarray}
& SO(5)_E \times SO(5)_R & \to \quad SU(2)_{M_3} \times SU(2)_R \times U(1)_\Sigma \times U(1)_t  \nonumber \\
\text{bosons:} & ({\bf 5}, {\bf 1} ) \oplus ({\bf 1}, {\bf 5}) & \to \quad
({\bf 3}, {\bf 1})^{(0,0)} \oplus ({\bf 1}, {\bf 3})^{(0,0)} \oplus ({\bf 1} , {\bf 1})^{(\pm 2,0)} \oplus ({\bf 1}, {\bf 1})^{(0, \pm 2)}
\nonumber \\
\text{fermions:} & ({\bf 4}, {\bf 4} ) & \to \quad ({\bf 2}, {\bf 2})^{(\pm 1,\pm 1)} \nonumber
\end{eqnarray}
where all sign combinations have to be considered.
Then, implementing the topological twist along $M_3$ means to replace $SO(3)_{M_3} \cong SU(2)_{M_3}$
with the diagonal subgroup $SU(2)_{M_3}' \subset SU(2)_{M_3} \times SU(2)_R$.
Under the symmetry group $SU(2)_{M_3}' \times U(1)_\Sigma \times U(1)_t$ the fields of the partially twisted
5d $\CN=2$ super-Yang-Mills transform as
\be \begin{array}{r@{\qquad}l}
\text{bosons}: & ({\bf 5}, {\bf 1} ) \oplus ({\bf 1}, {\bf 5}) \to 2 \times {\bf 3}^{(0,0)} \oplus {\bf 1}^{(\pm 2,0)} \oplus {\bf 1}^{(0, \pm 2)} \\
\text{fermions}: & ({\bf 4}, {\bf 4} ) \to {\bf 3}^{(\pm 1,\pm 1)} \oplus {\bf 1}^{(\pm 1,\pm 1)} \,
\end{array}
\label{twistedspectrum}
\ee
Here, one can recognize many familiar facts about 3d-3d correspondence. For instance, two copies of ${\bf 3}^{(0,0)}$
represent adjoint-valued one-forms on $M_3$, which combine into a complex gauge connection $\CA = A + i \phi$. This is the reason for the effective 2d theory on $\Sigma$ to localize on complexified flat connections on $M_3$ \cite{DGH}.

Note, fivebranes wrapped on a general 3-manifold $M_3$ preserve 4 real supercharges (singlets in \eqref{twistedspectrum}),
{\it i.e.} $\CN=2$ in three dimensions.
Turning on Omega-background along $\R^2_q$ or $L(k,1)_b$ as in \eqref{choicesSigma} breaks SUSY by half,
so that the resulting system has only two real supercharges, $\CQ$ and its conjugate $\CQ^{\dagger}$.
It is one of these two supercharges, whose $\CQ$-cohomology gives the desired 3-manifold homology \eqref{HN3mfld}.

\subsection{Orders of compactification}

Starting with 5d $\CN=2$ super-Yang-Mills one can first compactify it on $\Sigma$. This will result in a 3d $\CN=4$ theory on $M_3$ with a topological twist. In the UV such theory usually has a quiver gauge theory description, while in the IR one has a sigma-model description. After topological twisting in the IR the theory becomes Rozansky-Witten theory \cite{Rozansky:1996bq} on the Coulomb branch. This can be illustrated with the following diagram:
$$
\begin{array}{ccccl}
\; & \;\text{6d $(2,0)$ theory} &\text{on}  &\text{$\Sigma \times S^1 \times M_3$}  \; & \; \\
\; & \swarrow \qquad\qquad\qquad& \; &\qquad\qquad\qquad \searrow & \; \\
\text{3d $\CN=2$ theory $T[M_3]$} & \; & \; & \; & \underline{\text{twisted 3d $\CN=4$ theory on $M_3$}} \\
\text{on $S^1 \times \Sigma$} & \; & \; & \; & \text{$\bullet$ UV: gauge theory} \\
 & \; & \; & \; & \text{$\bullet$ IR: Rozansky-Witten}
\end{array}
$$
In the next sections we will consider these two different points of view in greater detail.

In the case $G=U(N)$ the setup can be realized in M-theory as follows~\eqref{M3phases}:
\begin{equation}
 \begin{array}{cccccc}
  \text{space-time:} & S^1 & \times & T^*M_3 & \times & Y_4 \\
   & \Vert &  & \cup &  & \cup \\
N\,\text{M5-branes:} & S^1 & \times & M_3 & \times & \Sigma \\
 \end{array}
\label{Msetup}
\end{equation}
where $Y_4$ is a hyper-K\"ahler four-manifold in which $\Sigma$ is embedded as a calibrated cycle, so that in the neighborhood of $\Sigma$ it looks like $T^*\Sigma$. The ordinary choice is to take just $Y_4=T^*\Sigma$ but we would like to keep the setup more general. The global structure of $Y_4$ and how $\Sigma$ is embedded in it can encode additional information about the 4d $\CN=2$ theory $T_G[\Sigma]$.

One can also introduce the following supersymmetry preserving defects:
\begin{equation}
 \begin{array}{l@{\;}|@{\;}ccccc}
 \text{defect} & \multicolumn{5}{c}{\text{support in}~S^1 \times T^* M_3 \times Y_4} \\[.1cm]\hline
 \text{M5$'$} & S^1 & \times & M_3 & \times & T^*_{\pt\in\Sigma}\Sigma \\
 \text{M2 }\cap\text{ M5} & S^1 & \times & (\text{1-cycle} \in M_3) & \times & (\pt \in \Sigma) \\
 \text{KK-monopole} & S^1 & \times & T^*M_3 & \times & (\pt \in Y_4) \\
 \end{array}
\label{Msetup-defects}
\end{equation}
where the first and the second cases correspond to the usual codimension-2 and -4 defects in 6d $(2,0)$. The third defect can occur if $Y_4$ has a circle fibration structure. The defect is a KK monopole and will modify topology and metric of $Y_4$ in its vicinity. All these defects contribute to the effective quantum mechanics on $S^1$. The insertion of a single M2-brane will lead to 3-manifold invariants, such as {\it e.g.} Seiberg-Witten invariants, labeled by an element of $H_1(M_3)$.

The spectrum of BPS states (or, equivalently, $Q$-cohomology) can be studied from the viewpoint
of the effective 3d $\CN=2$ theory $T[M_3]$ on the fivebrane world-volume after compactification on $M_3$.

In the rest of the paper we will consider particular choices of $\Sigma$ which realize well known 3-manifold invariants
(and their categorification): Chern-Simons partition function and the Seiberg-Witten invariants.


\subsection{SW invariants from $T{[}M_3{]}$}
\label{sec:SWfromTM3}

\begin{figure}[ht]
\centering
\includegraphics[scale=1.2]{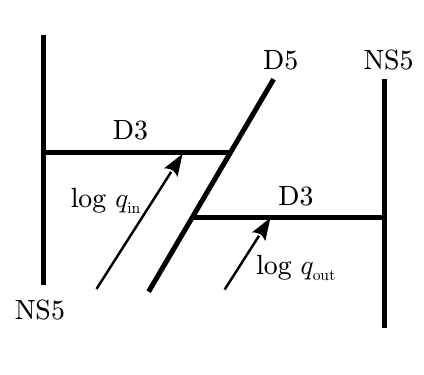}\caption{The brane realization of $\CN=4$ 3d SQED in type IIB string theory.}
\label{fig:SW-branes}
\end{figure}

The Seiberg-Witten invariant of 3-manifolds are computed by topologically twisted 3d $\CN=4$ SQED.
Let us start with the standard Hanany-Witten brane realization of the this theory in flat space:
\begin{equation}
	\begin{array}{rcccc}
		\text{D3:} & ~~123 &  & 7 &  \\
		\text{D5:} & ~~123 & 456 &  & \\
		2\times\text{NS5:} & ~~123 & &  & 890 \\
	\end{array}	
\end{equation}
where the numbers denote space-time directions of type IIB string theory. Denote the positions of the two NS5 branes along directions $890$ by $\log q_\text{in}$ and $\log q_\text{out}$, as depicted in Figure~\ref{fig:SW-branes}. The difference $\log q_\text{in}-\log q_\text{out}$ has the meaning of a FI-parameter (a 3-vector). When it is non-zero the position of NS5 branes along the direction $x^7$ does not really matter and we can pull them to $\pm\infty$.

Now let us introduce a non-trivial three-manifold $M_3$ along directions $123$. After topological twisting the directions $456$ become directions of $T^*M_3$ fibers. Far from D5 ans NS5 branes the theory on the world-volume of D3 brane is topologically twisted 4d $\CN=4$ $U(1)$ super-Yang-Mills on $M_3\times \R$. The path integral of the theory thus localizes on solutions to Vafa-Witten equations on $M_3\times \R$, the gradient flow of complex CS functional on $M_3$. The stationary solutions are given by complex flat connections. Therefore, the boundary conditions for D3 brane at $x^7 \rightarrow \pm \infty$ are given by two elements
\be
q_\text{in},\,q_\text{out} \; \in \; \CM_\text{flat}(U(1)_\C,M_3) \; = \; \hat{H} \,.
\ee
As in the case of $M_3=\R^3$, the partition function should depend only on the ratio $q_\text{out}/q_\text{in}=q$.

\begin{figure}[ht]
\centering
\includegraphics[scale=1]{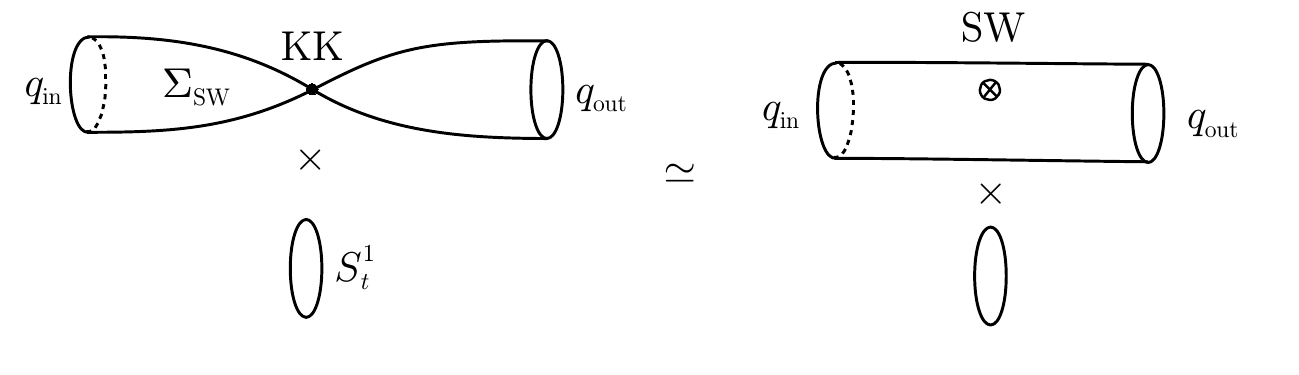}\caption{The space-time of $T[M_3]$: $\Sigma_\fM\times S^1$ where $\Sigma_\fM$ is the Seiberg-Witten curve.}
\label{fig:monopole-curve}
\end{figure}

\begin{figure}[ht]
\centering
\includegraphics[scale=1]{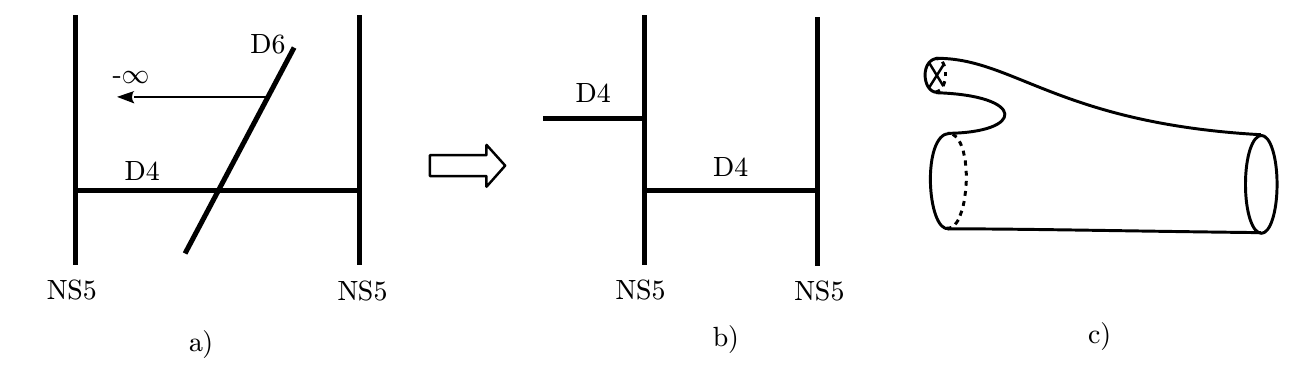}\caption{a) Type IIA brane construction of 4d $\CN=2$ SQED with zero FI-parameter. b) Equivalent description achieved by pulling $D6$ brane to $x^7=-\infty$. c) The curve appearing in M-theory lift.}
\label{fig:monopole-curve-1}
\end{figure}

\begin{figure}[ht]
\centering
\includegraphics[scale=1.2]{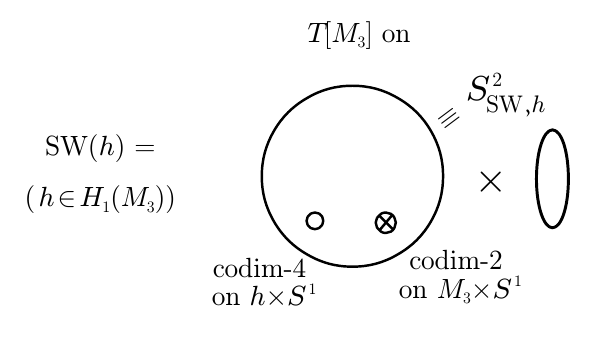}\caption{SW invariants of $M_3$ as the index of $T[M_3]$ on $S^2\times S^1$ with certain line operators inserted. In what follows we will use notation $S^2_{\fM,h}$ (or just $S^2_{\fM}$) to denote sphere with such insertions.}
\label{fig:monopole-curve-3}
\end{figure}

The brane construction can be lifted to M-theory setup of type (\ref{Msetup}) where $Y_4=\text{TN}_4$,
the Taub-NUT space with one center \cite{Witten:2011zz,Gorsky:2013jxa,Mikhaylov:2014aoa}.
The curve $\Sigma=\Sigma_\fM$ embedded into $Y_4$ is a cylinder split into two cigars by a KK monopole,
the Taub-NUT center (see Figure~\ref{fig:monopole-curve}).
Asymptotically, when $x^7 \rightarrow \pm \infty$, we have $T[M_3]$ theory on $T^2\times \R$. The addition of the KK monopole can be understood as insertion of a certain operator, which we denote $\mathscr{SW}$, acting on the Hilbert space of $T[M_3]$ on a torus. The partition function computes the matrix element of this operator:
\begin{equation}
	Z_{T[M_3]}(\Sigma_\fM\times S^1)=\langle q_\text{out}|\mathscr{SW}|q_\text{in}\rangle\equiv \hat{\SW}(q_\text{out}/q_\text{in})
	\label{SW-operator}
\end{equation}
where $\hat{\SW}$ is a function on $\hat{H}$. The action of $\mathscr{SW}$ on $\CH_{T[M_3]}(T^2)\cong \C[H]\cong\C[\hat H]$ can be realized by multiplication by the element $\hat{SW}\in \C[\hat{H}]$:
\begin{equation}
	(\mathscr{SW}\,\hat{f})(q_\text{out})=\hat{(\SW\cdot f)}(q_\text{out})=\sum_{q_\text{in}}\hat{\SW}(q_\text{out}/q_\text{in})\hat f(q_\text{in})
\end{equation}
which means that the cylinder with a KK monopole insertion can be replaced by a cylinder with an extra hole labeled by a state $\SW\in \C[H]$, see Figure~\ref{fig:monopole-curve}. This transition has the following physical meaning. Consider type IIA brane description of the corresponding 4d $\CN=2$ SQED shown in Figure~\ref{fig:monopole-curve-1}a. This can be achieved by  decompactifying $S^1$ into $\R$. One can obtain an equivalent description with a semi-infinite D4-brane by pulling D6-brane to $x^7=-\infty$ (Figure~\ref{fig:monopole-curve-1}b). This brane configuration can be lifted to M-theory with an M5-brane wrapping the curve shown in Figure~\ref{fig:monopole-curve-1}c.
The addition of a semi-infinite D4-brane is equivalent to replacing the left cylindrical end with a pair of pants where one of the in-states corrsponds to insertion of codim-2 defect in 6d theory supported on $M_3\times \R$. It follows that the state SW$\in \C[H]$ as line operator has the following interpretation:
\begin{equation}
	 \CH_{T[M_3]}(T^2) \; \cong \; \C[\{\text{line ops in }T[M_3]\}] \;\;\ni\;\; \SW
	\;=\;
	\text{codim-2 defect on }M_3
\end{equation}
The question of finding SW invariants can then be translated into a question of decomposing
the line operator given by codim-2 defect into basis line operators labelled by $H$ and corresponding to codim-4 defects, {\it cf.} \cite{Frenkel:2015rda}. The coefficients of such decomposition are calculated by the partition function of $T[M_3]$ on a sphere with codim-2 and codim-4 defects inserted (see Figure~\ref{fig:monopole-curve-3}).
The goal of the next few sections is to describe the line operator corresponding to the codim-4 defect purely in terms of $T[M_3]$ for a particular class of 3-manifolds.


\subsubsection{Deformations and spectral sequences}

Here and in what follows, we often identify the Seiberg-Witten curve $\Sigma_\fM$
with a supersymmetric ({\it i.e.} calibrated) submanifold in the Taub-NUT space $Y_4=\text{TN}_4$.
As a complex manifold, the latter, in turn, can be identified with a complex 2-plane,
\be
Y_4 = \text{TN}_4 \cong \C^2
\ee
with complex coordinates $z$ and $w$.
In this complex structure, the Seiberg-Witten curve can be described by the equation
\be
\Sigma_\fM ~: \quad zw = 0
\label{singSWcurve}
\ee
and the two copies of the ``cigar'' illustrated in Figure~\ref{fig:monopole-curve}
correspond to complex lines $z=0$ and $w=0$, respectively.

Note, they meet at a single point $z=w=0$, which is also a fixed point of the
$U(1)_t \times U(1)_{\Sigma}$ symmetry \eqref{UqUtsymmetries}, which in these coordinates
simply acts by phase rotations on variables $z$ and $w$.
It has to be compared with another choice of $\Sigma \cong \R^2_q$ which consists of a single cigar,
say $\Sigma:~ \{ w = 0 \}$, and which also plays an important role in this paper.
As notation $\Sigma \cong \R^2_q$ indicates, in this latter choice the rotation symmetry of the cigar, $U(1)_{\Sigma}$,
gives rise to the $q$-grading on the space of BPS states,
while the rotation symmetry of the complex plane transverse to the fivebranes (parametrized by $w$ in our notations)
is the symmetry we call $U(1)_t$ that gives rise to the homological $t$-grading.
When a fivebrane is supported on $\Sigma_\fM$ which includes both $z=0$ and $w=0$,
both factors in $U(1)_t \times U(1)_{\Sigma}$ act non-trivially on its world-volume.
Moreover, the singular Seiberg-Witten curve \eqref{singSWcurve} has a natural deformation
\be
\Sigma_{\text{def}} ~: \quad zw = \mu
\label{SWmassdeformed}
\ee
which preserves the property that $\Sigma$ is a supersymmetric (calibrated) cycle in $\text{TN}_4 \cong \C^2$.
The parameter $\mu$ can be interpreted as the mass of the monopole field in the Seiberg-Witten theory,
which now in the IR is described by a pure $U(1)$ Maxwell theory (with no charged fields).
Equivalently, turning on $\mu$ can be interpreted as moving onto the Coulomb branch of the original SW theory.
The deformed curve \eqref{SWmassdeformed} is a smooth cylinder, illustrated in Figure~\ref{fig:defSW},
which looks as the right-hand side of Figure~\ref{fig:monopole-curve} but without a KK monopole (codimension-2 defect).
How does $Q$-cohomology of the fivebrane on $M_3 \times \Sigma$ change upon $\mu$-deformation?
\FIGURE[h]{
\includegraphics[width=2.2in]{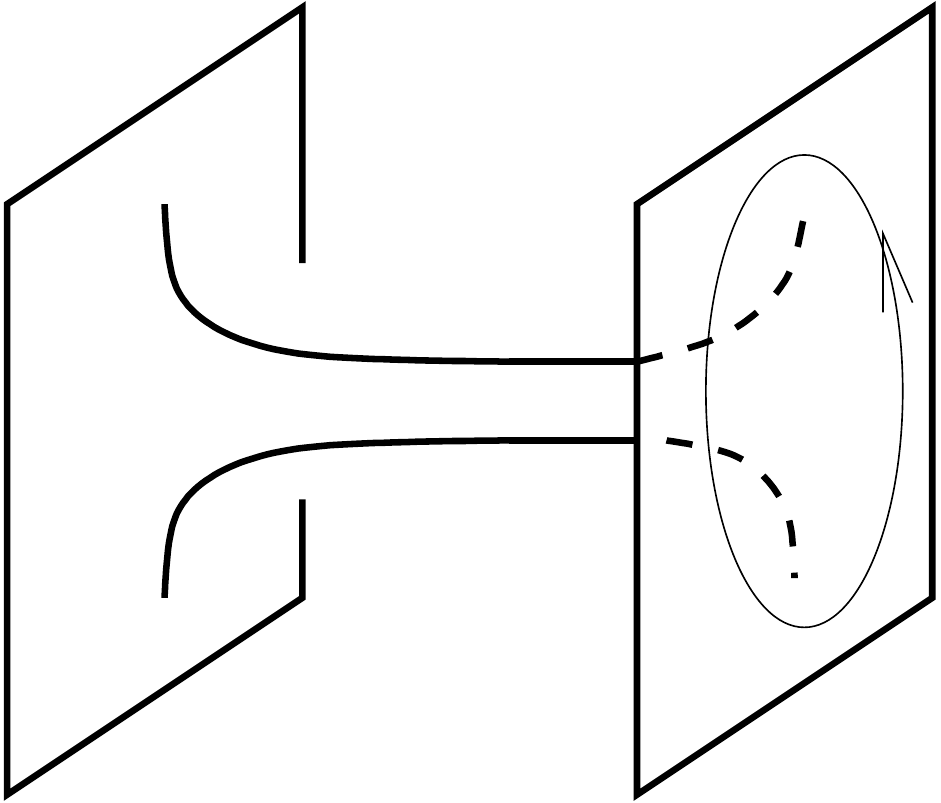}
\caption{\label{fig:defSW}
Deformation of the pinched $\Sigma_\fM$ to a smooth curve that admits a circle action with no fixed points.
\vspace{3ex}}
}

In general, upon deformations the $Q$-cohomology may jump. And, in our present case, the spectrum of BPS states
also changes from the monopole Floer homology at $\mu=0$ to a much simpler homology of $M_3$ associated to $\mu \ne 0$.
To be more precise, these two 3-manifold homology theories (before and after the deformation) are related by a spectral sequence.
Reversing the orders of compactification --- which will be discussed in section \ref{sec:M3veiw} ---
the simpler homology in the final page of the spectral sequence can be identified with the Floer homology
of a 4d TQFT obtained by a topological Donaldson/SW twist of 4d $\CN=2$ Maxwell theory with gauge group $U(1)$.
With the monopole field removed, this theory is almost ``trivial'' and has one-dimensional $Q$-cohomology
in every class $h \in H$.

A curious feature of the deformed theory (that is easy to see in the deformed geometry \eqref{SWmassdeformed})
is the relation between $q$-grading and $t$-grading, which become identified after the deformation.
Indeed, the diagonal subgroup of $U(1)_t \times U(1)_{\Sigma}$ that rotates $z$ and $w$ with opposite phase
is still a symmetry of \eqref{SWmassdeformed},
\be
z \to e^{i \vartheta} z \,, \qquad w \to e^{- i \vartheta} w \,,
\ee
while the ``anti-diagonal'' combination that rotates $z$ and $w$ by the same phase is not.
Furthermore, the deformed curve \eqref{SWmassdeformed} has no fixed points with respect to the action
of the diagonal subgroup of $U(1)_t \times U(1)_{\Sigma}$.
Hence, even the $q$-grading (now identified with the $t$-grading) must be trivial at least for those 3-manifolds
which have $T[M_3]$ with an extra flavor symmetry $U(1)_{\beta}$.
Indeed, as we explain in the next section, one can often extract information about the $Q$-cohomology for such $M_3$
by applying localization techniques to a suitable partition function of $T[M_3]$ with an extra fugacity $y$ for the symmetry $U(1)_{\beta}$.
A version of this argument will be used throughout the paper.
In particular, here it shows that, at least for a class of 3-manifolds whose $T[M_3]$ admits a description with an extra
flavor symmetry $U(1)_{\beta}$, the Floer homology of the deformed theory associated with \eqref{SWmassdeformed}
has no non-trivial grading and does not lead to an interesting concordance invariant.
Since one grading is collapsed in the spectral sequence, it is natural that the original, undeformed theory
has one non-trivial grading and the differential carries the opposite $t$-degree and $q$-degree.

\begin{figure}[ht]
\centering
\includegraphics[scale=0.5]{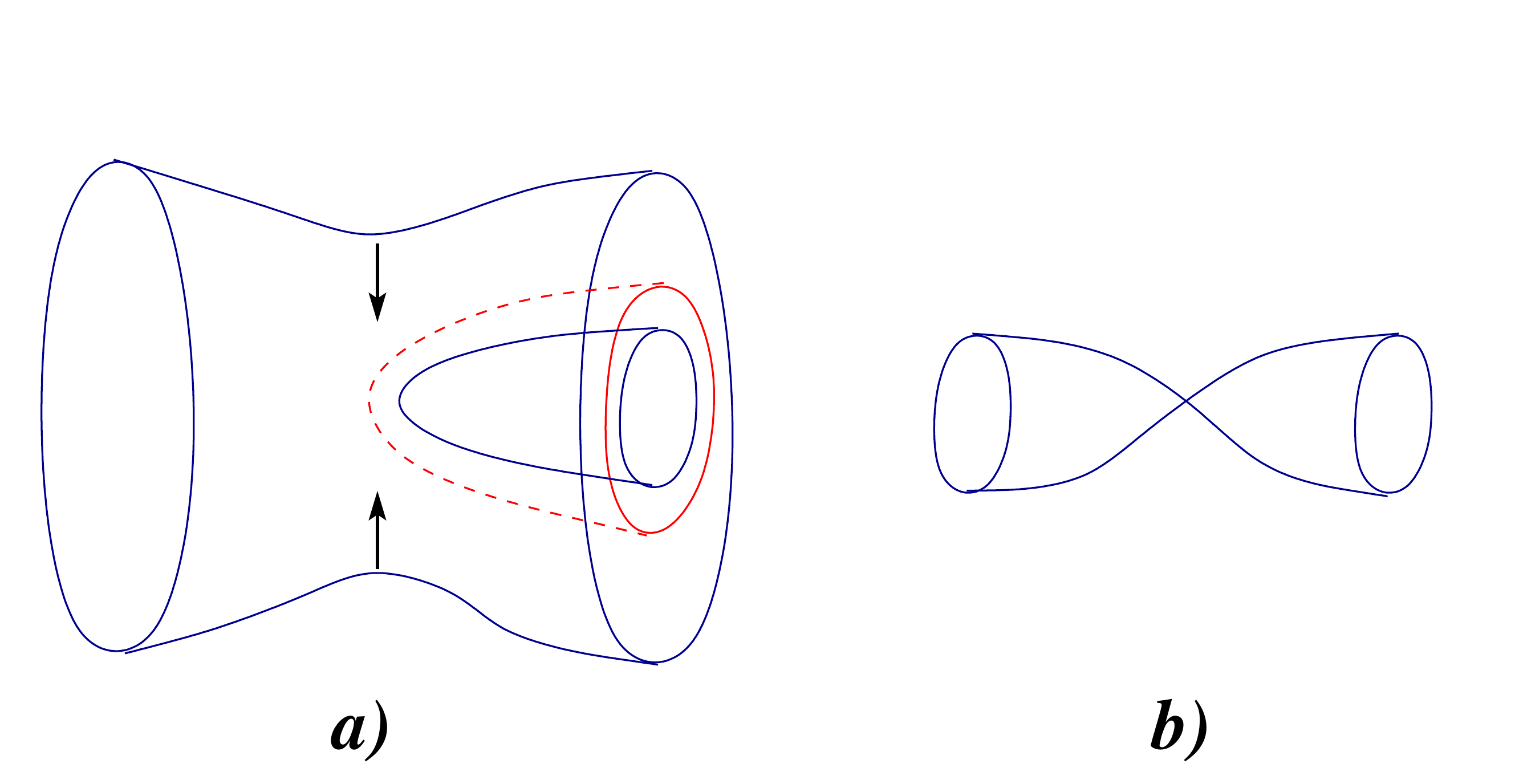}
\caption{In the limit, a deformation of the system of positive and negative branes $(a)$ gives the Seiberg-Witten curve $(b)$.
Solid blue lines represent positive branes, while dashed red stands for negative branes.}
\label{fig:cigar7}
\end{figure}

Let us also briefly mention a relation between the Seiberg-Witten theory discussed here and Chern-Simons theory
with $U(1|1)$ gauge group.\footnote{Although in this paper we do not really talk about Chern-Simons theory
with supergroups as gauge groups --- because $U(N|0)$ and $U(0|N)$ reduce to ordinary Chern-Simons theory with $G=U(N)$,
and even the theory with super-rank $N=0$ only appears in the form of Seiberg-Witten gauge theory rather than Chern-Simons
theory with gauge group $G=U(1|1)$ --- we plan to study quantum and homological $U(n|m)$ invariants of $M_3$ in a future work.}
As in \cite{Vafa:2001qf,DHJV} (see also \cite{Mikhaylov:2014aoa} for a related discussion),
the simplest way to see the connection is to engineer $U(1|1)$ Chern-Simons theory on the world-volume of one ``positive''
and one ``negative'' fivebrane that wrap $M^3$ and one copy of the cigar
(namely, $w=0$ in our complex coordinates on $\text{TN}_4 \cong \C^2$).
If we now imagine adding to this system a positive fivebrane wrapped on \eqref{SWmassdeformed} --- that, as we saw earlier,
gives a very simple homology theory --- and then deforming it by taking the limit $\mu \to 0$ (see Figure~\ref{fig:cigar7}$a$)
in the end of this process it effectively ``annihilates'' the negative brane supported on $w=0$
and adds an extra positive brane supported on the other cigar ($z=0$ in our notations)
resulting precisely in the configuration \eqref{singSWcurve} shown in Figure~\ref{fig:monopole-curve}.
As we saw earlier, one needs to be careful since $Q$-cohomology can change under continuous deformations.
However, modulo spectral sequences (that will be discussed elsewhere),
this suggests that a categorification of $U(1|1)$ Chern-Simons theory on $M_3$ is, roughly speaking,
the monopole Floer homology studied here plus the Floer homology associated with the topological twist of 4d $\CN=2$ super-Maxwell theory which effectively describes the theory on the Coulomb branch. However, one needs to take into account that the Coulomb branch is actually curved in the vicinity of the origin. From the point of Rozansky-Witten theory on $M_3$ the curvature on the Coulomb branch is needed to reproduce Casson-Walker invariant, the mismatch between the torsion (computed by $U(1|1)$ CS) and SW invariants in the case $b_1(M_3)=0$ (\textit{cf}. (\ref{SWresult})).
Note, in the higher-rank version of this argument, the 4d Maxwell theory is replaced
by 4d $\CN=2$ super-Yang-Mills Floer homology.
In such generalizations to $U(n|m)$ homological invariants of $M_3$, the curve \eqref{singSWcurve} should be replaced by
\be
\Sigma ~: \quad z^n w^m = 0
\ee


\subsection{R-symmetry $U(1)_t$ versus flavor symmetry $U(1)_{\beta}$}
\label{sec:U1beta}

Here we wish to emphasize a simple yet important point regarding symmetries of the fivebrane systems \eqref{M3phases}
and the definition of ``refinement'', which in the literature sometimes means slightly different things.
The symmetry $U(1)_t$ that equips the space of BPS states \eqref{HN3mfld} with the homological $t$-grading
is an R-symmetry and acts on supercharges in a non-trivial way, {\it cf.} \eqref{UqUtsymmetries}.

For general $M_3$, the theory $T[M_3]$ has only $U(1)_t$ R-symmetry.
Its close cousin, the symmetry that we call $U(1)_\beta$ exists only for certain 3-manifolds,
but makes life a lot easier as we explain momentarily.
The nature of such symmetry is quite different from that of $U(1)_t$.
In particular, from the viewpoint of the fivebrane theory it is a flavor symmetry, not an R-symmetry.
It can be related, however, to the R-symmetry in a way that also sheds light on the existence of $U(1)_\beta$.
Namely,
\be
U(1)_{\beta} = U(1)_S - U(1)_t
\label{U1symsrel}
\ee
where $U(1)_S$ is the Cartan of $SU(2)_R$ used in topological twist along $M_3$
In particular, the existence of $U(1)_{\beta}$ requires the existence of an extra R-symmetry $U(1)_S$ after the topological twist.
Thus, for $M_3 = \Sigma' \times S^1$ the R-symmetry of $T[M_3]$ is enhanced to $SU(2) \times SU(2)$,
so that $U(1)_t$ and $U(1)_S$ are their diagonals. The case of Seifert $M_3$ is intermediate in
the sense that R-symmetry is $U(1)_t \times U(1)_S$ or, equivalently, $U(1)_t$ R-symmetry plus $U(1)_{\beta}$ flavor symmetry.
In  \cite{Aganagic:2011sg} it was argued that such extra $U(1)_S$ R-symmetry exists for Seifert manifolds. Seifert manifolds have nowhere vanishing vector field associated to semi-free $U(1)$ action on $M_3$. The $U(1)_S$ is the subgroup of $SO(3)_R$ acting on the fibers of $T^*M_3$ which keeps the vector field invariant.

Note, we really need $U(1)_t$ R-symmetry for applications to knot and 3-manifold homologies,
but we can't easily formulate path integral (a partition function) which localizes only to the BPS sector
and keeps track of the $U(1)_t$ grading.
On the other hand, grading by $U(1)_{\beta}$ is easier to implement in the path integral and was heavily explored
in \cite{Aganagic:2011sg,Cherednik:2011nr,Gukov:2015sna}.
For example, one can calculate the index of $T[M_3]$ on $\R^2_q$ refined by $U(1)_\beta$ fugacity $y$:
\begin{equation}
 \CI_{T[M_3]}(q,y)=\Tr (-1)^F q^{L_0-R_t} y^{R_S-R_t}
 \label{index-beta}
\end{equation}
This can be compared to the Poincar\'e polynomial of the space of BPS states \eqref{ZCSfromP}:
\begin{equation}
P (q,t) = \Tr q^{L_0-R_t} t^{R_t}
 \label{Poincare-trace}
\end{equation}
It was shown in \cite{Aganagic:2011sg} that for torus knot complements one can recover (\ref{Poincare-trace}) from (\ref{index-beta}).
This should be possible {\it e.g.} if $R_S$ acts trivially on the BPS spectrum and there are no cancelations
in the index (\ref{index-beta}) due to the $(-1)^F$ factor.
Then, (\ref{index-beta}) will coincide with (\ref{Poincare-trace}) up to some signs in front of coefficients if we replace $y\rightarrow t^{-1}$.

In many concrete examples of $T[M_3]$ that enjoy an extra symmetry $U(1)_\beta$, this symmetry manifests as
a flavor symmetry acting (by phase rotations) on the adjoint chiral multiplet, {\it cf.} Table~\ref{tab:examples}.
Weakly gauging this symmetry leads to a mass deformation of $T[M_3]$ discussed {\it e.g.} below \eqref{LspaceTheory}
in the case of Lens space theory that will be our next topic.

As in the case of knots \cite{Gukov:2011ry}, the basic building blocks of 3-manifold homologies \eqref{HN3mfld}
will be bosonic and fermionic Fock spaces over a single-particle Hilbert space $\CH$:
\begin{subequations}\label{FockoverH}
\be
\text{bosonic:} \quad \bigoplus_{n=0}^{\infty} \text{Sym}^n (\CH)
\ee
\be
\text{fermionic:} \quad \bigoplus_{n=0}^{\infty} \Lambda^n (\CH)
\ee
\end{subequations}
For example, if $\CH$ is generated by a single boson $\phi$, the corresponding Fock space
\be
\CT^+ := 1 \oplus \phi \oplus \phi^2 \oplus \ldots = \text{Sym}^* (\phi)
\label{harmonicH}
\ee
is the Hilbert space of a single harmonic oscillator that we give a special name $\CT^+$.
Similarly, the Fock space of a single fermion $\psi$ is
\be
1 \oplus \psi = \Lambda^* (\psi)
\ee
In the effective quantum mechanics \eqref{ZM_3} obtained by reducing the fivebrane theory on $M_3 \times \Sigma$
or, equivalently, 3d $\CN=2$ theory $T[M_3]$ on the ``cigar'' $\Sigma = \R^2_q$,
the single-particle Hilbert space $\CH$ contains all Fourier modes $\partial^n \phi$ and $\partial^n \psi$,
so that the corresponding Fock spaces graded by the $U(1)_t \times U(1)_{\Sigma}$ symmetry \eqref{UqUtsymmetries} are
\begin{subequations}\label{chiralFock}
\be
\text{boson:} \quad \bigotimes_{n=0}^{\infty} \frac{1}{1 - t^{R_t} q^{n+1/2} x}
\ee
\be
\text{fermion:} \quad \bigotimes_{n=0}^{\infty} (1 \oplus t^{R_t} q^{n + 1/2} x)
\ee
\end{subequations}
In particular, a free chiral multiplet contains a boson with $R_t=1$ and a fermion with $R_t=0$.


\subsection{Example: $M_3 = L(p,1)$}
\label{sec:Lp1-example}

\subsubsection{Turaev torsion}
\label{sec:monopole-defect}

The Lens space $M_3=L(p,1)$ can be understood as an $\mathcal{O}(p)$ circle bundle over $S^2$. Let us consider the Hopf fiber $S^1_\text{Hopf}$ as M-theory circle. In type IIA string theory the information about non-triviality of the Hopf fibration will translate into the fact that there are $p$ units of RR flux through the base. We will denote the base sphere by $S^2_p$ to make the dependence on $p$ explicit and the Hopf fiber by $S^1_\text{Hopf}$ in  order to destinguish it from another $S^1$. The M5-brane in the setup \eqref{Msetup} then becomes a D4-brane on $S_p^2\times S^2_\fM\times S^1$. The presence of RR flux through $S^2_p$ will have two effects. First, as expected, there will be a Chern-Simons term:
\begin{equation}
 \int_{S^2_p\times S^2_\fM\times S^1}C_1\wedge F\wedge F = p\int_{S^2_\fM\times S^1}A\wedge F.
\end{equation}
where $C_1$ is the RR 1-form. Second, the flux of $F$ through $S^2_p$ will be quantized in $\mathbb{Z}_p$ instead of $\mathbb{Z}$. If we formally write $L(p,1)\cong S^2_p\times S^1_\text{Hopf}$, then there should be the following relation which arises from the exchange $S^2_p\leftrightarrow S^2_\fM$ in type IIA where we ``forgot'' about $S^1_\text{Hopf}$:
\begin{equation}
 \CI^\fM_{T[L(p,1)]}\equiv Z_{T[S^2_p\times S^1_\text{Hopf}]}[S^2_\fM\times S^1]=
Z_{T[S^2_\fM\times S^1_\text{Hopf}]}[S^2_p\times S^1]
\label{index-relation}
\end{equation}
and where by $T[S^2_{\fM\text{ or }p}\times S^1_\text{Hopf}]$ we mean the D4-brane theory compactified on $S^2_{\fM\text{ or }p}$.

Consider first the case $p=0$, that is $M_3=S^2\times S^1_\text{Hopf}$.
Using that
\begin{equation}
T_{U(1)}[S^2_\text{SW}\times S^1] \; : \quad \text{3d $\mathcal{N}=4$ $U(1)$ theory with 1 hypermultiplet}
\end{equation}
and we obtain
\begin{multline}
Z_{T[S^2_\fM\times S^1_\text{Hopf}]}[S^2_0\times S^1]= \left(\frac{y^{-1}}{1-y^{-2}}\right) \sum_{h\in \mathbb{Z}}
\int \frac{dz}{2\pi i z}\,\left(\frac{{z^{1/2}y^{1/2}}}{1-zy}\right)^{h}
\,\left(\frac{{z^{-1/2}y^{1/2}}}{1-z^{-1}y}\right)^{-h}(-q)^{-h}=\\
=\frac{1}{y-y^{-1}}\sum_{h\in \mathbb{Z}}
\int \frac{dz}{2\pi i z}\,\left(\frac{y-z}{1-zy}\right)^{h}
q^{-h}
\label{index-S2xS1}
\end{multline}
where we twisted $U(1)$ R-symmetry under which the scalars in the hypermultiplet have charge $1$ and scalars in the vectormultiplet are uncharged ({\it i.e.} this is the correct twist to obtain theory described by SW equations on $M_3$). The fugacity $q\in \hat H=\C^*$ counting different fluxes $h\in \Z$ of the gauge field through $S^2_0$ is the (exponential of) FI parameter that will be discussed in detail in section~\ref{sec:M3veiw}.

The choice of $h\in H=\mathbb{Z}$ corresponds to the choice of Spin$^c$ structure on $M_3=S^1_\text{Hopf}\times S^2_0$. The contribution for a given $h$ is easily calculated using Jeffrey-Kirwan (JK) contour prescription\cite{Benini:2015noa}. Picking "negative" residues we obtain:
\begin{equation}
 \CI^{\fM,h}_{T[S^2\times S^1]}=\left\{
 \begin{array}{cc}
	 0,& h\geq 0\\
 \frac{y^{h}-y^{-h}}{y-y^{-1}}, & h<0
 \end{array}\right.
\label{SWS2S1}
\end{equation}
which agrees with the fact that SW invariants are trivial in this case:
\begin{equation}
	\SW(h)=0
\end{equation}
The result also agrees with the known expression for the Heegaard Floer homology with Spin$^c$ structure $\mathfrak{s}_h$ such that $c_1(\mathfrak{s}_h)=h\neq 0$:
\begin{equation}
\label{HFS2S1triv}
HF^+(S^2\times S^1,\mathfrak{s}_h)=0
\end{equation}
\begin{equation}
HF^-(S^2\times S^1,\mathfrak{s}_h)=HF^\infty(S^2\times S^1,\mathfrak{s}_h)\cong \mathbb{Z}[U]/(U^{|h|}-1)
\end{equation}
if we identify the homological grading with the grading by the flavor symmetry $U(1)_\beta$, for which $y$ is the corresponding fugacity.
When $h=0$ the result is zero, as expected from the Euler characteristic of
\begin{equation}
HF^+(S^2\times S^1,\mathfrak{s}_h) \; \cong \; H_*(S^1)\otimes \CT^+ \,,
\label{HFplusS2S1}
\end{equation}
where $\CT^+$ represents a copy of a bosonic Fock space \eqref{harmonicH}.
Note, one needs to be careful taking the Euler characteristic of the infinite-dimensional space $HF^+ (M_3)$.
For integral homology spheres $HF^+ (M_3)$ decomposes into $\Z [U]$-submodules as
\be
HF^+ (M_3) \simeq \CT^+_{\Delta (M_3)} \oplus HF_{\text{red}} (M_3)
\ee
where $HF_{\text{red}} (M_3)$ is finitely generated and $\CT^+_{\Delta (M_3)}$ is a copy of $\Z [U,U^{-1}] / U \cdot \Z [U]$
with minimal degree $\Delta (M_3)$.
For integral homology sphere, the Heegaard Floer homology categorifies the Casson invariant,
\be
\chi \left( HF_{\text{red}} (M_3) \right) \; = \; \lambda (M_3) + \frac{\Delta (M_3)}{2} \,,
\ee
where $\Delta (M_3)$ is the ``correction term'' \cite{ozsvath2003floer}.
If $M_3$ is an integral homology 3-sphere that bounds a smooth, negative-definite 4-manifold $M_4$, then $\Delta (M_3) \ge 0$.

Similarly, for a 3-manifold $M_3$ with $b_1 (M_3) > 0$ and non-torsion Spin$^c$ structure ${\frak s}$,
\be
\chi \left( HF^+ (M_3, {\frak s}) \right) = \pm \tau (M_3, {\frak s})
\ee
where $\tau : \text{Spin}^c (M_3) \to \Z$ is the Turaev torsion function.
However, the case $b_1 = 1$, closely related to the case of $b_2^+ (M_4) = 1$ in Donaldson theory, is more delicate.
In this case, $\tau ({\frak s})$ should be computed in the ``chamber'' containing $c_1 ({\frak s})$,
{\it i.e.} with respect to the component of $H^2 (M_3; \R) - 0$ containing $c_1 (\frak s)$.
For a 3-manifold with $H_1(M_3;\Z)=\Z$, such as $M_3 = S^2\times S^1$, the set of Spin$^c$ structures is indexed by $\Z$.
Hence, we can write
\be
HF^+ (M_3) \cong \bigoplus_{h \in \Z} HF^+ (M_3,h)
\ee
such that $HF^+ (M_3,h) \cong HF^+ (M_3,-h)$ is endowed with a relative $\Z_{2h}$ grading,
which becomes $\Z$-grading for $h=0$.
For all $h \neq 0$, $HF^+(M_3,h)$ is a finite dimensional vector space,
and it makes sense to take its Euler characteristic (with respect to the $\Z_{2h}$-valued {\it Maslov grading}):
\be
\chi \left( HF^+(M_3,-h) \right) \quad = \quad q^{-h} \text{~coefficient of~} \tau (q) \qquad (h>0)
\ee
For example, if $M_3 = S^1\times S^2$, according to \eqref{HFS2S1triv} we have $HF^+(M_3,h)=0$ for all $h \neq 0$.
Note, when $b_1 (M_3)=1$, the Turaev torsion is not symmetric in $h$, but $HF^+(M_3, h)$ is.
On the other hand, the invariant $HF^-(M_3, h)$ is asymmetric in the same way as the Turaev torsion,
and the relationship holds for both positive and negative $h$:
\be
\chi (HF^- (M_3, \mathfrak{s}_h)) \; = \; \tau_{-\xi} (\mathfrak{s}_h)
\ee
where $\xi$ is the component of $H^2 (M_3 , \Z) - 0$ containing $c_1 (\mathfrak{s}_h) = h$.
The Turaev torsion obeys the wall crossing formula \cite{MR2113020}:
\be
\tau_{-\xi} (M_3, \mathfrak{s}_h) - \tau_{\xi} (M_3, \mathfrak{s}_h) = h
\ee
which, in the case of $M_3 = S^1\times S^2$, relates $\tau (h) = h$ for $h>0$ and $\tau (h) = 0$ for $h<0$, {\it cf.} \eqref{SWS2S1}.
When $h=0$, the group $HF^+(M_3,0)$ is infinitely generated and extra care is needed to define its Euler characteristic.
One can either use twisted coefficient or simply write
\be
\chi \left( HF^+(M_3,0) \right) = \sum_{n >1/2} (-1)^n \dim HF^+_n (M_3,0)+
\label{EulerforTowers}
\ee
$$
{}~~~~~~ + \sum_{n \leq 1/2} (-1)^n (\dim HF^+_n (M_3,0)-1)
$$
rigged so that we get $\chi \left( HF^+(M_3,0) \right) = 0$ for $M_3 = S^1\times S^2$.

In our example of $M_3 = S^1\times S^2$,
one can also consider the total sum\footnote{Note that, while the result for individual $h$ depends on the choice between ``positive'' and ``negative'' poles, the total sum, as a meromorphic function of $q$, does not \cite{Benini:2015noa}.} (\ref{index-S2xS1}) of the invariants \eqref{SWS2S1}. The result is the Turaev-Milnor torsion of $M_3 = S^2\times S^1$ as a function of $q\in \hat H$ and refined by the $U(1)_\beta$ fugacity $y$:
\begin{equation}
\CI^{\fM}_{T[S^2\times S^1]}(q)=\sum_{h\in H}\CI^{\fM,h}_{T[S^2\times S^1]}q^{-h}=\frac{q}{(1-q/y)(1-qy)}
\;\;\stackrel{y\rightarrow 1}{\longrightarrow}\;\;
\frac{q}{(1-q^2)}=\Tor_{S^2\times S^1}(q)
\label{S1S2SWy}
\end{equation}\\

Now let us consider a Lens space with $p>0$. As mentioned earlier, the fluxes $h$ and $h+pm$ for any $m\in \mathbb{Z}$ should be indistinguishable. This is achieved by summing over all $m$, that is (\ref{index-S2xS1}) should be modified as:
\begin{equation}
\CI^{\fM}_{T[L(p,1)]}(q)=Z_{T[S^2_\fM\times S^1_\text{Hopf}]}[S^2_p\times S^1]
=\frac{1}{y-y^{-1}}\sum_{h\in \mathbb{Z}_p}\sum_{m\in\mathbb{Z}}
\int \frac{dz}{2\pi i z}\,\left(\frac{y-z}{1-zy}\right)^{k+mp}
q^{-h}
\end{equation}
where now $h\in H\cong \Z_p$ and $q \in \hat{H}\cong \Z_p$.
Let us make the following change of variables (which, of course, is one-to-one on a complex sphere):
\begin{equation}
x=\frac{y-z}{1-zy}.
\end{equation}
After the change of variables the integral takes the following form:
\begin{equation}
\CI^{\fM}_{T[L(p,1)]}(q)\equiv \sum_{h\in H}\CI^{\fM,h}_{T[L(p,1)]}q^{-h}
=y\sum_{h\in \mathbb{Z}_p}\sum_{m\in\mathbb{Z}}
\int\frac{dx}{2\pi i x} \frac{x^{h+mp}}{(1-xy)(1-y/x)}\,
q^{-h}
\label{lens-space-index}
\end{equation}
The integral has a form of the topologically twisted index of the theory $T_{U(1)}[L(p,1)]$ on $S^2$ with insertion of defects described in Figure~\ref{fig:monopole-curve-3}. The defect contribution reads
\begin{equation}
f^{\fM,h}(x)=\frac{y}{(1-xy)(1-y/x)}\cdot x^h
\label{monopole-defect-vertex}
\end{equation}
where $h$ is the choice of Spin$^c$ structure. One could also obtain this result directly. Indeed, the second factor in (\ref{monopole-defect-vertex}) is the contribution of the basic Wilson line built from a codimension-4 defect. The first factor is the contribution of the codimension-2 defect. After compactification on $S^1_\text{Hopf}$ the codimension-2 defect can be realized by intersection of two D4-branes along $S^1\times S^2_p$. The theory living on the intersection is the theory of a hypermultiplet. Compactifying it further on $S^2_p$ gives a hypermultiplet in the effective quantum mechanics on $S^1$ charged with respect to $U(1)$ gauge symmetry of $T[M_3]$. The first factor in (\ref{monopole-defect-vertex}) is precisely the index of this hypermultiplet.

The result (\ref{monopole-defect-vertex}) will be used in section \ref{sec:plumbed} to write the $S^2_\fM\times S^1$ index of $T[M_3]$ for general plumbed 3-manifolds.


\subsubsection{$HF^+ (M_3)$ from $T{[}M_3{]}$ and the physics of $\CT^+$ towers}
\label{HFLp1}

If all of $HF^+ (M_3)$ is supported in the same mod 2 homological grading,
there are no cancellations in the index and one can try to reconstruct $HF^+(M_3,h)$ from the index $\CI^{\fM,h}_{T[M_3]}$,
especially when the flavor symmetry $U(1)_\beta$ discussed in section~\ref{sec:U1beta} is available.
For example, when $M_3$ is a Seifert homology sphere oriented so that
it bounds a positive definite plumbing (see section~\ref{sec:plumbed} for a definition),
then $HF^+ (M_3)$ is supported in even degrees only.
In particular, in such examples $\chi ( HF_{\text{red}} (M_3)) = \text{rank} HF_{\text{red}} (M_3)$.
Here, we shall apply this principle and walk the reader through the details of the calculation for Lens spaces $M_3=L(p,1)$
in a way that parallels our example $M_3=S^2\times S^1$ considered in section \ref{sec:monopole-defect}.

For the Lens space $M_3=L(p,1)$,
there are $p$ states on a torus labelled by $h\in \mathbb{Z}_p$ with the following wave functions:
\begin{equation}
		\Psi_{h}(x)=\sum_{n\in \Z}q^{\frac{1}{2p}(pn+h)^2}x^{pn+h}
	\label{L21-wave}
\end{equation}
In section \ref{sec:monopole-defect}, we calculated SW invariants and torsion by considering the index of $T[M_3]$ on $S^2_{\fM,h}$, a sphere $S^2$ with the defect $\fM$ and a basic Wilson line labeled by $h\in H$.
Instead, one can consider the partition function on a solid torus
with the defect $\fM$ inserted along a non-contractible cycle and a boundary state $|h\rangle$,
similar to the one illustrated in Figure~\ref{fig:line-op-state}.
The partition function of this system has the following expression:
\begin{equation}
	\CI^{\fM,h}=\int\frac{dx}{2\pi i x}\Psi_{h}(x)\,f^\fM(x)
	\label{L21-index}
\end{equation}
where $\Psi_{h}(x)$ are given by (\ref{L21-wave}) and
\begin{equation}
	f^\fM(x)=\frac{y}{(1-y/x)(1-yx)}
\end{equation}
is the character corresponding to codim-2 defect refined by $U(1)_\beta$ symmetry.
The two factors in the denominator can be interpreted as the contributions of zero-modes of the charged fields $\phi$ and $\tilde \phi$
that compose a hypermultiplet associated with the codim-2 defect; they carry charges $(+1,-1)$ under $U(1)$ gauge symmetry
and charges $(+1,+1)$ under $U(1)_\beta$ flavor symmetry.
Note that when $q$ is equal to 1 we get the same expression as (\ref{lens-space-index}).

If we set $y=1/t$ and $q=t^2$ and calculate (\ref{L21-index}) the result has a surprisingly simple structure:
\begin{equation}
		\CI^{\fM,0}=t^{\frac{(h-p/2)^2}{p}-\frac{p}{4}+1}(1+t^2+t^4+t^6+t^8+\ldots)
		\label{HFLens-Poincare}
\end{equation}
It can be interpreted as the Poincar\'e polynomials of
\be
HF^+(L(p,1),\mathfrak{s}_h) \cong \widecheck{HM}(L(p,1),\mathfrak{s}_h) \cong \CT_0^+
\label{HFplusLp1}
\ee
where $t$ plays the role of fugacity for the homological grading. Up to an overall $h$-independent shift, the gradings coincide with the ones in \cite[Lemma 3.2]{kronheimer2007monopoles}.  The identification of $U(1)_q$ with $U(1)_t$ that we performed corresponds to topological twisting. The identification of $U(1)_\beta$ with $U(1)_t$ (up to normalization of charges) is possible when $U(1)_S$ discussed in (\ref{U1symsrel}) acts trivially.

A general feature of 3-manifold homology that we already encountered in table~\ref{tab:examples}, in eq. \eqref{HFplusS2S1},
and now in \eqref{HFplusLp1} is that it is often infinite-dimensional.\footnote{In Floer theory, the origin of infinite-dimensionality
has to do with reducible solutions, 
while in physics it can often be traced to the Fock space structure of the space of BPS states \cite{Gukov:2011ry}.}
This is a general feature of colored / unreduced knot homology \cite{Gorsky:2013jxa,FGSS} which, as we shall see later,
also persists in ``non-abelian'' variants of 3-manifold homology \eqref{HN3mfld} for $N>1$,
that is for 3-manifold analogues of Khovanov-Rozansky homology.
Moreover, the infinite-dimensional knot homology turns out to be a module over an algebra (of BPS states).
In our present context, $HF^+ (M_3)$ is also a module over the ring $\Z [U]$,
where $U$ lowers degree by two and every element of $HF^+ (M_3)$ is annihilated by a sufficiently large power of $U$.
The simplest such module is the Heegaard Floer homology of the 3-sphere\footnote{We often forget
about the $\Z[U]$-module structure on $HF^+(M_3)$, but still think of it as having a {\it Maslov grading}
with respect to which $U^{-n}$ has degree $2n$.
More generally, for manifolds with $b_1 >0$, $HF^+ (M_3)$ is a module over a larger ring $\Z [U] \otimes_{\Z} \Lambda^* H_1 (M_3;\Z)$,
examples of which will appear {\it e.g.} in section \ref{sec:Sigmaprime} and other places throughout the paper.
We plan to say more about the physical interpretation of the module structure in future work.}
\be
\CT_0^+ \; \cong \; HF^+ (S^3) \; \cong \; \Z [U,U^{-1}] / U \cdot \Z [U] \,,
\ee
which, according to \eqref{harmonicH}, can be identified with a Fock space of a single boson.
Its natural generalization, which often appears as a building block of $HF^+ (M_3)$ for more general 3-manifolds,
is a $\mathbb{Q}$-graded $\Z [U]$-module which is abstractly isomorphic to $\Z [U, U^{-1}] / \Z [U]$,
\be
\CT_k^+ \cong HF^+ (S^3)
\ee
where the bottom-most non-zero homogeneous element has Maslov (homological) degree $k \in \mathbb{Q}$.
In particular, rational homology spheres whose Heegaard Floer homology in every Spin$^c$ structure
has the form \eqref{HFplusLp1} are called {\it L-spaces}.
Such 3-manifolds can be characterized by any of the following conditions \cite{MR2249248}:

$\bullet$ $\hat{HF} (M_3)$ is a free abelian group of rank $|H_1 (M_3; \Z)|$

$\bullet$ $HF^{-} (M_3)$ is a free $\Z [U]$-module of rank $|H_1 (M_3; \Z)|$

$\bullet$ $HF^{\infty} (M_3)$ is a free $\Z [U,U^{-1}]$ module of rank $|H_1 (M_3; \Z)|$, and the map
\be
U : HF^+ (M_3) \xrightarrow{~~~} HF^+ (M_3)
\ee
is surjective.
Since Lens spaces are special examples of $L$-spaces, for every Spin$^c$ structure $\frak s$ we have, {\it cf.} \eqref{HFplusLp1}:
\begin{eqnarray}
\hat{HF} (M_3, \frak s) & = & \Z \,, \nonumber \\
HF^- (M_3, \frak s) & \cong & \CT^- = U \cdot \Z [U] \,, \nonumber \\
HF^+ (M_3, \frak s) & \cong & \CT^+ = \Z [U,U^{-1}] / U \cdot \Z [U] \,, \nonumber \\
HF^{\infty} (M_3, \frak s) & \cong & \CT^{\infty} = \Z [U,U^{-1}] \,, \nonumber \\
HF_{\text{red}} (M_3, \frak s) & = & 0  \nonumber
\end{eqnarray}
More generally, if $M_3$ is a rational homology sphere, there is a spectral sequence starting
at $\hat{HF} (M_3,{\frak s}) \otimes \CT^+$ and converging to $HF^+ (M_3,{\frak s})$.

The $\Z[U]$ module structure on the Heegaard Floer homology has the following physcial meaning. The multiplication by $U$ can be realized as insertion of the  ``meson'' $\phi\tilde\phi$ composed of two fields in the hypermultiplet originating from codimension-2 defect. The meson is uncharged with respect to the gauge field of $T[M_3]$ but carries $U(1)_\beta$ charge $2$. Therefore, if $T[M_3]$ on $\R \times S^2_{\fM,h}$ with defects as in Figure~\ref{fig:monopole-curve-3}
provides a physical realization of $HF^+ (M_3, {\frak s}_h)$,
the same theory on $\R \times S^2_{h\cdot h^{-1}}$ with the codimension-2 defect replaced by a simple codimension 4-defect labeled by $h^{-1}$ should be viewed as a physical counterpart of $\hat{HF} (M_3,{\frak s}_h)$, {\it cf.} \eqref{HFhatEuler}.


\subsection{Invariants for general plumbed 3-manifolds}
\label{sec:plumbed}

\subsubsection{$T{[}M_3{]}$ for plumbed three-manifold}

The wild world of 3-manifolds contains a very tame class of 3-manifolds described by what is called a plumbing graph. Such manifolds generalize the notion of Seifert manifolds. In this section we review some of the results of \cite{Gang:2013sqa,Gadde:2013sca,Chung:2014qpa} about 3d/3d correspondence for 3-manifolds of this type.

A plumbing graph is a graph colored by integer numbers $\{a_i\in \mathbb{Z}\}_{i\in \text{vertices}}$. In general, one can also add extra non-negative integer labels $\{g_i\in \mathbb{Z}_+\}_{i\in \text{vertices}}$ to vertices, but by default $g_i$ are zero and such labels are not shown. Non-zero $g_i$  are depicted by integers in brackets. The vertices and edges correspond to basic building blocks of a 3-manifold $M_3$ glued together. The rules are summarized in the first two columns of the Table~\ref{tab:plumbing}.
\begin{table}[h]
	\centering
 \begin{tabular}{|c|c|c|}
	 \hline
graph & $M_3$ & $T_G[M_3]$ \\
 \hline
$\vcenter{\hbox{\includegraphics[scale=2]{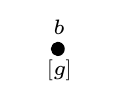}}}$
&
$
 \begin{CD}
    \CO(b)  	 \\
    @V{\pi}VV  \\
    \text{genus $g$ RS}
 \end{CD}
$
&
\begin{tabular}{c}
	$\CN=4$ vector multiplet\\
	w/ level-$b$ Chern-Simons \\
	(breaking to $\CN=2$) \\
	+ $g$ adjoint hypers
\end{tabular}
\\
 \hline
$\vcenter{\hbox{\includegraphics[scale=2]{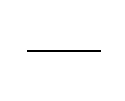}}}$
&
$S^3\setminus$ Hopf link
&
\begin{tabular}{c}
	$T[G]$ theory \\
	(S-duality wall in $\CN=4$ 4d SYM) \\
	w/ $G\times G$ flavor symmetry
\end{tabular}
\\
 \hline
$\vcenter{\hbox{\includegraphics[scale=2]{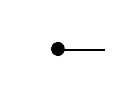}}}$
&
\begin{tabular}{c}
excise $\pi^{-1}(\pt)$, \\
glue along $T^2$ boundary
\end{tabular}
&
gauge $G$ flavor symmetry
\\
\hline
\end{tabular}
\caption{\label{tab:plumbing}The rules for plumbing graphs. $\CO(b)$ denotes a circle bundle with Chern class $b$. }
\end{table}
The third column describes the corresponding 3d theory and how attaching vertices to edges is realized in terms of $T[M_3]$.
\begin{figure}[ht]
\centering
\includegraphics[scale=2]{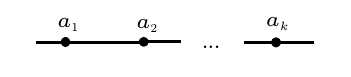}\caption{The plumbing graph realizing $T^2$ mapping cylinder associated to the word $ST^{a_1}ST^{a_2}S\ldots ST^{a_k}S$ in $SL(2,\Z)$.}
\label{fig:SL2Z-word}
\end{figure}
In particular, any linear part of a plumbing graph, such as the one depicted in Figure~\ref{fig:SL2Z-word}, corresponds to a certain element in the mapping class group of the torus, $SL(2,\Z)$, realized as a word of $S$ and $T$ generators. The 3d theory $T[M_3]$ in this case is the corresponding duality wall in $\CN=4$ 4d SYM with gauge group $G$. Let us note that in the case $G=U(1)$ the theory $T[G]$ associated to an edge is just a supersymmetric version of mixed CS interation for two $U(1)$'s. In the case when $G=SU(N)$ it has a quiver description, but for one of the two $SU(N)$'s only its maximal torus is explicitly visible in the UV, which is however enough to calculate the index/sphere partition functions.

\begin{figure}[ht]
\centering
\includegraphics[scale=2]{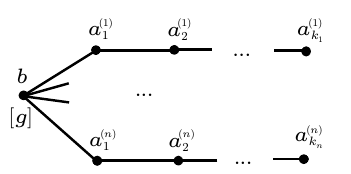}
\caption{A plumbing graph for a Seifert manifold.}
\label{fig:Seifert-plumbing}
\end{figure}

As we already mentioned earlier, the family of plumbed 3-manifolds contains all Seifert 3-manifolds (of orientable type). A Seifert 3-manifold is usually realized as a circle fibration over a genus $g$ Riemann surface, possibly with exceptional fibers. Such fibration can be described by the following data:
\begin{equation}
 (g;b;(q_1,p_1),(q_2,p_2),\cdots,(q_n,p_n))
\end{equation}
where $g$ is the genus of the base, $b\in \mathbb{Z}$ is the ``integer part''\footnote{it can be absorbed into redefinition of $(q_i,p_i)$'s} of the first Chern class of the circle bundle, and $\{(p_i,q_i)\}$ are the pairs of coprime integers charaterizing the exceptional fibers. It can be realized by the plumbing shown in Figure~\ref{fig:Seifert-plumbing}
where the numbers $\{ a_i^{(j)} \}$ should realize continued fraction representation for $p_j/q_j$:
\begin{equation}
\frac{p_j}{q_j}=a^{(j)}_1-\cfrac{1}{a^{(j)}_2-\cfrac{1}{a^{(j)}_3-\cdots}}
\end{equation}

\begin{figure}[ht]
\centering
\includegraphics[scale=1.7]{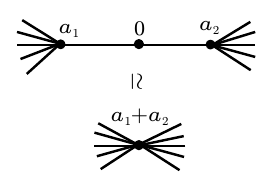}
\includegraphics[scale=1.7]{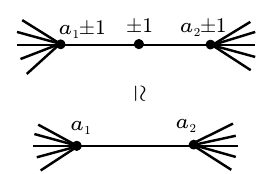}
\includegraphics[scale=1.7]{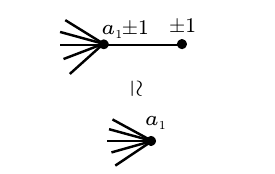}
\caption{3d Kirby moves that relate plumbings giving homeomorphic 3-manifolds.}
\label{fig:moves}
\end{figure}

We will be mostly interested in the case where all extra labels are trivial: $g_i=0,\;\forall i$. In this case one can interpret the plumbing graph as the resolution graph of a complex singularity. The plumbed 3-manifold $M_3$ is then realized as the link of a singularity. Such class of 3-manifolds is a natural home for rational homology spheres. The resolution of the singularity provides us with a smooth 4-manifold $M_4$ such that $\d M_4=M_3$. The plumbing graph is also the plumbing graph of $M_4$. Note that different plumbing graphs can give different $M_4$ but homeomorphic 3-manifold $M_3$. The equivalence relations between plumbing graphs giving three-manifolds of the same homeomorphism and orientation type are given by 3d Kirby moves shown in Figure~\ref{fig:moves}.
Any topological invariant of 3-manifold defined in terms of the plumbing data obviously should be invariant under such moves. Theories constructed using the rules in Table \ref{tab:plumbing} for graphs related by moves should be dual to each other. In particular supersymmetric partition function of $T[M_3]$ on any space provides us with a combinatorial invariant of $M_3$ calculated in terms of the plumbing data. In the next few sections we consider a particular case of such partition function: topologically twisted index on $S^2\times S^1$.

The first homology group $H\equiv H_1(M_3)$ can be easily computed from the graph data. Let the total number of vertices be $b_2$. The plumbing graph defines a bilinear form $Q$ on $\Z[\text{vertices}]\cong\Z^{b_2}$ via its adjacency matrix:
\begin{equation}
	Q^{ij}=\left\{\begin{array}{cc}
	1, & (i,j)\text{ connected by edge}, \\
	a_i, & i=j
	\end{array}
		\right.
\end{equation}
One can associate basis elements of $H_2(M_4)$ with vertices of the plumbing graph. Then $Q$ plays the role of the intersection form on the lattice $\Lambda=H_2(M_4)$. The abelian group $H=H_1(M_3)$ then enters into the following short exact sequence:
\begin{equation}
	0\longrightarrow H_2(M_4) \stackrel{Q}{\longrightarrow} H_2(M_4,M_3) \longrightarrow H_1(M_3)\longrightarrow 0
\end{equation}
where $H_2(M_4,M_3)\cong H^2(M_4)$ can be canonically identified with the dual lattice $\Lambda^*$. Suppose the intersection form is negative definite. Then $H\equiv \Lambda^*/\Lambda\cong \text{Coker}\,Q$ is a finite abelian group and $M_3$ is a rational homology sphere. The non-trivial part of $T[M_3]$ is $U(1)^{b_2}$ CS theory with levels specified by the bilinear form Q and has the following action:
\begin{equation}
	S_{CS}=\frac{1}{2}\sum_{i,j\in \text{vertices}}Q^{ij}\int A_i dA_j
	\label{abelianCS}
\end{equation}
See \cite{Gadde:2013sca} for more details.

\subsubsection{$S^2\times S^1$ topologically twisted index of $T{[}M_3{]}$}

The $S^2\times S^1$ topologically twisted index for general 3d $\CN=2$ gauge theories was considered in detail in \cite{Cecotti:2013mba,Benini:2015noa} and reviewed in Appendix \ref{app:index}. From the rules in Table \ref{tab:plumbing} it follows that for a general group $G$ the index of $T[M_3]$ for plumbed $M_3$ can be constructed from the basic pieces associated to graph elements:
\begin{multline}
 \CI_{T[M_3]}\equiv Z_{T[M_3]}[S^2\times S^1]=\\
\sum_{m_i\in \mathbb{Z}^{\text{rank}G}}\int \prod_{i\in\text{vertices}} \frac{dx_i}{2\pi i x_i}\, \CI_{G_{a_i}}(x_i,m_i)
\prod_{\alpha\in\text{edges}}  \CI_{T[G]}(x_{\alpha_1},m_{\alpha_1};x_{\alpha_2},m_{\alpha_2})
\end{multline}
For each vertex $i$ the integrand has a factor
\begin{equation}
	\CI_{G_{a_i}}(x_i,m_i),
	\label{index-vertex}
\end{equation}
the index of the $\CN=2$ level $a_i$ Chern-Simons theory with adjoint chiral multiplet. It depends on the gauge fugacity $x_i$ (an element of the maximal torus of $G_\C$) and numbers $m_i\in \Z^{\text{rank}\,G}$ (the fluxes through $S^2$). For each edge $\alpha$ connecting vertices $(\alpha_1,\alpha_2)$ there is a factor
\begin{equation}
	\CI_{T[G]}(x_{\alpha_1},m_{\alpha_1};x_{\alpha_2},m_{\alpha_2}),
	\label{index-edge}
\end{equation}
the index of $T[G]$ depending on fugacities and fluxes of $G\times G$ flavor symmetry.

For $G=U(1)$ we have the following simple explicit expressions:
\begin{equation}
 \CI_{U(1)_a}(x,m)=\left(\frac{y^{-1}}{1-y^{-2}}\right)^{-2s-1}\, x^{am},
 \label{index-vertex-u1}
\end{equation}
\begin{equation} \CI_{T[U(1)]}(x_1,m_1;x_2,m_2)=\left(\frac{y^{-1}}{1-y^{-2}}\right)^{2s+1} x_1^{-m_2}x_2^{-m_1},
	\label{index-edge-u1}
\end{equation}
where $y$ is the fugacity for $U(1)_\beta$ flavor symmetry of adjoint chiral multiplet, which decouples in the abelian case. For completeness we included dependence on the integer parameter $s$, the flux of the $U(1)_\beta$ background field through $S^2$. In the above formulae, the $U(1)$ R-symmetry which is used to make the topological twist is the Cartan subgroup of $SU(2)_R$.

For $G=SU(N)$, it is possible to calculate explicitly (\ref{index-vertex}) and (\ref{index-edge}) using their gauge theory descriptions in the UV. The formulae for $G=SU(2)$ are presented in the Appendix~\ref{app:SU2index}.

However if we compute $\CI_{T_{U(1)} [M_3]}$, the result is extremely simple. For any negative definite plumbing graph the result is
\begin{equation}
 \CI_{T_{U(1)}[M_3]}=\left(\frac{y^{-1}}{1-y^{-2}}\right)^{-2s-1}
\end{equation}
Such simple answer is expected from two points of view. First, in the $U(1)$ case the theory is equivalent to topological quiver CS (up to some decoupled free fields) and its Hilbert space on $S^2$ is trivial. The theory only contains non-trivial line operators (which provide states on $T^2$) but no local operators. Second, the corresponding Rozansky-Witten theory is also trivial, becuase there is no Coulomb branch ({\it i.e.} $X_{S^2}=\pt$ in the notations of section \ref{sec:RW}). To get an interesting observable one can insert defects on $S^2$.

\subsubsection{$S^2$ with defects}

Now let us consider $S^2$ with some decoration $\mathfrak{D}$, which can be understood as a collection of defects supported at points on the sphere. In terms of $S^2\times S^1$ these are defects supported at $(\text{pts}) \times S^1 \subset S^2\times S^1$, so they are line operators in the space-time of $T[M_3]$. We will denote such decorated $S^2$ as $S^2_\mathfrak{D}$. We mostly will be interested in the ``monopole'' decoration\footnote{We will often suppress the additional label $h\in H$.} $\mathfrak{D}=\text{SW}$ defined in Figure~\ref{fig:monopole-curve-3} which should provide us with SW invariants of $M_3$.
But for now let us consider some general abstract collection of defects $\mathfrak{D}$. In the case of Chern-Simons theory any collection of line operators can be decomposed into combination of Wilson lines and thus can be encoded by a function of gauge fugacities $f_\fD(x)$. When one computes the index it appears as a factor in the integrand:
\begin{multline}
 \CI_{T[M_3]}^{\fD}\equiv Z_{T[M_3]}[S^2_\fD\times S^1]=\\
\sum_{m_i\in \mathbb{Z}^{\text{rank}G}}\int \prod_{i\in\text{vertices}} \frac{dx_i}{2\pi i x_i}\, \CI_{G_{a_i}}(x_i,m_i)
\prod_{\alpha\in\text{edges}}  \CI_{T[G]}(x_{\alpha_1},m_{\alpha_1};x_{\alpha_2},m_{\alpha_2})f_\fD(x)
\end{multline}
We want $f_\fD(x)$ to have a universal description in terms of plumbing data. Since we can geometrically/physically decompose plumbed $M_3$ into basic building blocks associated to the vertices and edges, the function $f_\fD(x)$ should factorize correspondingly. That is, in the $G=U(1)$ case, we can write
\begin{multline}
 \CI_{T[M_3]}^{\fD}\equiv Z_{T[M_3]}[S^2_\fD\times S^1]=\\
\sum_{m_i\in \mathbb{Z}}\int \prod_{i\in\text{vertices}} \frac{dx_i}{2\pi i x_i}\, \CI_{U(1)_{a_i}}^{\fD}(x_i,m_i)
\prod_{\alpha\in\text{edges}}
\CI_{T[U(1)]}^{\fD}
(x_{\alpha_1},m_{\alpha_1};x_{\alpha_2},m_{\alpha_2})
\label{index-decorated}
\end{multline}
where we introduced defect-modified versions of elementary contributions (\ref{index-vertex-u1}) and (\ref{index-edge-u1}):
\begin{equation}
 \CI_{U(1)_a}^\fD(x,m)=x^{am}\,f_a^\fD(x),
 \label{CSdecorated}
\end{equation}
\begin{equation}
 \CI_{T[U(1)]}^\fD(x_1,m_1;x_2,m_2)=  x_1^{-m_2}x_2^{-m_1} f_S^\fD(x_1,x_2),
\end{equation}
We absorbed constant factors appearing in (\ref{index-vertex-u1}) and (\ref{index-edge-u1}) in the definition of $f^\fD_a(x)$ and $f^\fD_S(x_1,x_2)$. The set of functions $\{f^\fD_a,f^\fD_S\}$ cannot be arbitrary; the index $\CI^\fD_{T[M_3]}$ should be invariant under the moves depicted in Figure~\ref{fig:moves}. For example, it is possible to solve this constraint by the following ansatz:

\begin{equation}
	f^\fD_a(x)=(g^\fD(x))^{-2},\qquad f_S^\fD(x_1,x_2)=g^\fD(x_1)g^\fD(x_2)
\end{equation}

\subsubsection{Torsion of negative definite plumbed 3-manifolds from $T{[}M_3{]}$}
\label{sec:plumbing-torsion}

Instead of understanding directly what is $S^2_\fM$ let us ``bootstrap'' $\{f^\fM_a,f^\fM_S\}$  using their properties described in the previous section and some additional input data.
Consider unrefined case $y=1$. From (\ref{monopole-defect-vertex}) it follows that the contribution of defects on $S^2_\fM$ for each vertex of the plumbing graph reads
\begin{equation}
f_a^{\fM,h}(x)=\frac{x^{h+1}}{(1-x)^2}.
\end{equation}
Where $h$ encodes a choice of Spin$^c$ structure. It is easy to see that the requirement of invariance under moves in Figure~\ref{fig:moves} implies that
\begin{equation}
f_S^{\fM}(x_1,x_2)=(1-x_1)(1-x_2).
\end{equation}
The integral (\ref{index-decorated}) then can be written in the following form:
\begin{equation}
\CI^{\fM,h}_{T[M_3]}=\sum_{m_i\in \mathbb{Z}}\int \prod_{i\in\text{vertices}} \frac{dx_i}{2\pi i x_i}\,x_i^{\sum_j Q^{ij}m_j}\,(1-x_i)^{\delta_i-2} x_i^{h_i}
\label{index-plumbing}
\end{equation}
Where $Q$ is the intersection form associated with the plumbing graph and $\delta_i$ is the degree of the vertex $i$ (the number of adjacent edges). The collection of integers $\{h_i\in \Z\}_{i\in\text{vertices}}$ defines a vector $h=\sum_{i}h_ie^i\in \Lambda^*\cong\Z^{b_2}$ where $e^i$ are basis elements. However the index actually depends only on $h$ modulo the image of $Q$. Therefore, $h$ can be considered as an element of $\text{Coker}\,Q\cong H_1(M_3)\equiv H$ which defines a choice of Spin$^c$ structure on $M_3$.

Taking into account that $Q$ is negative definite and using the JK contour prescription we can sum over fluxes $m_i$:
\begin{equation}
\CI^{\fM,h}_{T[M_3]}=\int \prod_{i\in\text{vertices}} \frac{dx_i}{2\pi i x_i}\,
\frac{1}{1-\prod_j x_j^{-Q^{ji}}}\,
(1-x_i)^{\delta_i-2} x_i^{h_i}
\label{index-plumbing-1}
\end{equation}
The integral (\ref{index-plumbing-1}) can be evaluated by residues:
\begin{equation}
\CI^{\fM,h}_{T[M_3]}=\frac{1}{|\det Q|}\sum_{\prod_j x_j^{Q^{ji}}=1}
(1-x_i)^{\delta_i-2} x_i^{h_i}
\label{index-plumbing-2}
\end{equation}
 It is clear that the whole $H= \Lambda^*/\Lambda$ can be generated by basis elements $e^i$ of $\Lambda^*$. The characters $q\in \hat{H}$ can be naturally identified with the solutions of
\begin{equation}
\prod_j x_j^{Q^{ji}}=1,\;\;\; i=1\ldots b_2
\end{equation}
by
\begin{equation}
q(e_i)=x_i
\end{equation}
Then (\ref{index-plumbing-2}) can be written as
\begin{equation}
\CI^{\fM,h}_{T[M_3]}=\frac{1}{|H|}\sum_{q\in \hat{H}}\prod_i(1-q(e_i))^{\delta_i-2} q(h)
\label{index-torsion}
\end{equation}
which has a form of the Fourier transform (\ref{FourierFromHat}). As in section \ref{sec:monopole-defect} one can consider the partition function as a function of $q\in \hat H$, fugacity corresponding to the FI parameter of SW theory:
\begin{equation}
	\CI^{\fM}_{T[M_3]}(q)=\sum_{h\in H}\CI^{\fM,h}_{T[M_3]}\,q^{-1}(h)=\prod_i(1-q(e_i))^{\delta_i-2}
\end{equation}
This agrees with the known result for torsion of negative definite plumbed 3-manifolds \cite{nemethi2002seiberg}:
\begin{equation}
	\CI^{\fM}_{T[M_3]}(q)=\Tor_{M_3}(q),\qquad q\in \hat H
\end{equation}

In order to get SW invariants for a particular choice of $h$ one needs to perform a Fourier transform of the torsion, that is, to calculate the sum (\ref{index-torsion}). The problem arises because $\Tor_{M_3}(q)$ is singular at $q=1$ and needs regularization. The regularization problem can be translated into regularization of the original integral (\ref{index-plumbing}). The regularized value of $\Tor_{M_3}(1)$ provides an $h$-independent shift of SW invariants and related to Casson invariant $\lambda(M_3)$ \cite{nemethi2002seiberg}:

\begin{equation}
\SW(h)=\CI^{\fM,h}_{T[M_3]}=\frac{1}{|H|}\sum_{q\in \hat{H}}\Tor_{M_3}(q)q(h)
=
-\frac{\lambda{(M_3)}}{|H|}+\frac{1}{|H|}{\sum_{q\in \hat{H}}}'\;\Tor_{M_3}(q)q(h)
\label{SWresult}
\end{equation}
where $\sum'$ denotes the sum with the $q=1$ term omitted.

\subsubsection{Heegaard Floer homology $HF^+ (M_3)$ from $T {[} M_3 {]} $}

Classically, the abelian CS theory is specified by the quadratic action (\ref{abelianCS}). However, as a quantum theory, it is not completely trivial (though relatively simple and has a combinatorial description) and can provide us with interesting information about plumbed three-manifolds $M_3$. abelian CS theory is usually defined in the case when the bilinear form $Q$ is even. In the case when it is not, the quantization of Chern-Simons theory is more involved and requires a choice of Spin structure on the space-time 3-manifold. Such theory is usually called Spin CS and was considered in detail in \cite{Belov:2005ze}.

On the other hand, the Heegaard Floer homology of plumbed 3-manifolds was studied in \cite{ozsvath2003floer} (see also \cite{nemethi2005ozsvath}) where the authors obtained combinatorial description of $HF^+$ in terms of the plumbing data. Since in this case both sides, $T[M_3]$ and $HF^+(M_3)$, have a combinatorial description in terms of plumbing data one can hope to see a relatively simple dictionary between the two. Let us stress, though, that $T[M_3]$ is not the only data one needs in order to reproduce $HF^+$; one also has to provide a distinguished line operator\footnote{A natural question is whether such line operator admits a canonical definition purely in terms of 3d $\CN=2$ theory without referring to $M_3$ itself. That is, if one presents an explicit description of $T[M_3]$ but does not tell us anything about $M_3$ itself, is it possible to define this operator and then calculate $HF^+(M_3)$? The answer is probably ``no'' since the abelian $T[M_3]$ by itself essentially sees only the linking form of $M_3$, which is a very weak invariant.} in $T[M_3]$ that we denote $\fM$. Line operators of $U(1)^{b_2}$ CS theory always can be decomposed into Wilson lines. Therefore, defining a line operator is equivalent to defining a character of $U(1)^{b_2}$ which can be analytically continued to a meromorphic function on $(\C^*)^{b_2}$. In terms of plumbing data, such character was found in section \ref{sec:plumbing-torsion}:
\begin{equation}
	f^\fM(x)=\prod_{i\in\text{vertices}}(1-x_i)^{\delta_i-2}
	\label{monopole-line}
\end{equation}
where $\delta_i$ is the degree of the vertex and $x_i$ belongs to $U(1)$ associated to vertex $i$. The Seiberg-Witten invariants can be understood as the coefficients of (\ref{monopole-line}) decomposed into basic Wilson lines
\begin{equation}
	\prod_{i\in \text{vertices}}x_i^{h_i},\;\; h=h_ie^i\in \Lambda^*
\end{equation}
with the equivalence relation
\begin{equation}
	\prod_{i\in\text{vertices}}x_i^{h_i}\;\sim\; \prod_{i\in\text{vertices}}x_i^{h_i'}\;\;\Leftrightarrow\;\;h-h'\in \text{Im}\,Q
\end{equation}
If we combine together equivalent Wilson lines the expansion will have the following form:
\begin{equation}
	f^\fM(x)=\sum_{\lambda\in \Lambda^*}n_\lambda x^\lambda=\sum_{h\in H}\sum_{\lambda\in \Lambda +h}n_\lambda x^\lambda
\end{equation}
where
\begin{equation}
	x^\lambda\equiv\prod_{i}x_i^{\lambda_i}.
\end{equation}
Note, the expression (\ref{monopole-line}) is singular at $x_i=1$ and, hence, there is an ambiguity in such a decomposition: one can expand either in $|x_i|>1$ or $|x_i|<1$. It follows that
\begin{equation}
	\SW(h)=\sum_{\lambda\in \Lambda +h}n_\lambda.
	\label{SWseries}
\end{equation}
The sum is infinite for rational homology spheres and needs regularization. The zeta-function regularization of such an infinite sum in general gives a rational number (\ref{SWresult}).

The coefficients $n^{(h)}_\lambda$ are integers and can be understood as Euler characteristics of finite dimensional spaces:
\begin{equation}
	n_\lambda=\chi(V_\lambda)
\end{equation}
The Heegaard Floer homology categorifying (\ref{SWseries}) is then given by
\begin{equation}
	HF^{+}(M_3,\mathfrak{s}_h)=\bigoplus\limits_{\lambda\in \Lambda +h}V_\lambda
	\label{HFseries}
\end{equation}
However, without specifying any gradings on $V_\lambda$ the right-hand side of (\ref{HFseries}) is merely an infinite dimensional space for any plumbed 3-manifold and does not provide any non-trivial invariant.

Note that in the case of lens spaces, all $\delta_i\leq 2$ in (\ref{monopole-line}) and there are no cancelations in the expansion,
so it is possible that
\begin{equation}
	n_\lambda=\text{dim}(V_\lambda)
\end{equation}
which is indeed what happened in the case of $M_3=L(p,1)$ as we have seen earlier.


\subsection{A different type of example: $M_3 = \Sigma' \times S^1$}
\label{sec:Sigmaprime}

\subsubsection*{Twists on $M_3 = \Sigma' \times S^1$}

In \eqref{choices6d} we introduced topological twists of the 6d theory on general background of the form
\be
\Sigma \times S^1\times M_3
\label{6donM4}
\ee
and established that we have the A-model twist along $\Sigma$, and the SW = RW twist on $M_3$.
The vantage point of the A-model on $\Sigma$ was the subject of section \ref{sec:Amodel},
while the SW = RW twist on $M_3$ will be discussed in more detail in sections \ref{sec:M3veiw} and~\ref{sec:4mfld}.

If, furthermore, $M_3 = S^1 \times \Sigma'$ for yet another Riemann surface $\Sigma'$ then 3d twist along $M_3$ reduces to the standard A-model twist on $\Sigma'$ and our 6d setup \eqref{choices6d} looks like
\be
\Sigma \times S^1 \times S^1 \times \Sigma'
\label{twoSigmas}
\ee
where $\Sigma$ and $\Sigma'$ now appear on the same footing and can be exchanged.
Indeed, in both cases, we end up with an A-model on one of the Riemann surfaces whose target space is determined by the other
(= Coulomb branch of 3d $\CN=4$ theory).
Yet another vantage point on the system \eqref{twoSigmas} is the 4d $\CN=4$ SYM compactified on a product
of two Riemann surfaces with A-model twist along each one.

\subsubsection*{Torsion refined by $U(1)_\beta$}

It is easy to extend the expression (\ref{lens-space-index}) to the case $M_3=S^1\times \Sigma'$ where $\Sigma'$ is a closed Riemann surface of genus $g$. The theory $T[M_3]$ remains essentially the same as in the case of $M_3=S^2\times S^1$: namely, pure $U(1)$ $\CN=4$ gauge theory, because all $g$ adjoint hypermultiplets are decoupled. The main modification is the index of codimension-2 defect compactified on $M_3=\Sigma'\times S^1$. For general $g$ it will be
\begin{equation}
 \left[\frac{y}{(1-xy)(1-y/x)}\right]^{1-g}
\end{equation}
Therefore, the combined index on $S^2_\fM$ reads
\begin{equation}
\CI^{\fM}_{T[\Sigma'\times S^1]}(q)
=\sum_{h\in \mathbb{Z}}
\int\frac{dx}{2\pi i x} \left[\frac{y}{(1-xy)(1-y/x)}\right]^{1-g}
\,x^h\,q^{-h}
=\left[\frac{y}{(1-yq)(1-y/q)}\right]^{1-g}.
\end{equation}
When $y=1$, the result coincides with the known expression for the torsion of $\Sigma'\times S^1$ where $q$ is the $U(1)_\C$ holonomy along the $S^1$.

This should be compared with the Heegaard Floer homology of $M_3 = \Sigma' \times S^1$,
which is non-trivial (by adjunction inequality) only for Spin$^c$ structures ${\frak s}_h$ with
$c_1 ({\frak s}_h) = 2h [S^1]$, $|h| \le g-1$ (or $k=0$ when $g=0$).
For $h \ne 0$ in this range, $HF^+ (\Sigma' \times S^1, {\frak s}_h)$ is isomorphic \cite{OShfk} to cohomology of the symmetric product of $\Sigma'$,
a fact that has a nice explanation in terms of vortex equations which result from reduction of SW equations on the $S^1$
({\it cf.} appendix \ref{app:vortex}):
\begin{eqnarray}
HF^+ (\Sigma' \times S^1, {\frak s}_h) & = & H^* (\text{Sym}^d (\Sigma_g); \Z) [-g] \\
& \cong & \bigoplus_{i=0}^d \Lambda^i H^1 (\Sigma_g; \Z) \otimes \CT_0 / (U^{i-d-1}) \nonumber
\end{eqnarray}
where $[-g]$ denotes the grading shift down by $g$ units,
the factor $\Lambda^i H^1 (\Sigma_g;\Z)$ is in degree $i-g$,
and $d = g-1-|h|$.

When $h=0$, the calculation of $HF^+ (\Sigma' \times S^1, {\frak s}_0)$
is more subtle \cite{ozsvath2003floer,MR2113019,MR2222356,MR2414320},
and its Euler characteristic needs to be taken as in~\eqref{EulerforTowers}.


\subsection{Triangulations}

In triangulation-based approaches, one tries to circumvent the problem of compactifying 6d $(0,2)$ theory on $M_3$
by guessing the basic building blocks of $T[M_3]$ associated to ideal tetrahedra in a way consistent with gluing and Pachner
moves, see {\it e.g.} \cite{DGG,CCV}.
Usually, this approach leads to a 3d $\CN=2$ theory that does not account for all reducible flat connections on $M_3$
which, as we shall see in section~\ref{sec:higherRank}, are in a sense the most important ones for
categorifying RTW invariants.
(Including reducible flat connections was also important for realizing Khovanov homology and its colored variants
via 3d/3d correspondence \cite{Chung:2014qpa}.)
The advantage of this approach, however, is that refinement or categorification can be easily achieved by passing from
supersymmetric indices to spaces of refined BPS states, where the ``refinement'' means counting with spin
with respect to both the little group in three dimensions that we call $U(1)_{\Sigma}$ and the R-symmetry $U(1)_t$.

In particular, in this approach one usually associates a 3d $\CN=2$ chiral multiplet to an ideal tetrahedron
in a triangulation of $M_3$. Then, various partition functions of a 3d $\CN=2$ chiral multiplet
give the corresponding variants of the quantum dilogarithm, which can be categorified by a vector space
\be
\bigotimes_{n=0}^{\infty} \frac{1 \oplus t^{R_t} q^{n + 1/2} x}{1 - t^{R_t+1} q^{n+1/2} x}
\ee
that simply lists all elements in the cohomology of a supercharge $Q$.
Here, the denominators should be understood as power series expansions in $x$ (or $q$)
representing bosonic Fock spaces, {\it cf.} \eqref{harmonicH}.
Implementing this in the refined 3d/3d correspondence that includes contributions of {\it all} flat connections,
one could try to construct homological invariants of various 3-manifolds {\it e.g.} obtained by surgeries on knots, {\it cf.} \cite{Chung:2014qpa},
and compare the results to the ones here. We plan to explore this elsewhere.


\section{Reversing the order of compactification}
\label{sec:M3veiw}

While 3d/3d correspondence provides novel ways of computing Seiberg-Witten invariants and the Heegaard Floer homology $HF^+ (M_3)$,
it does not give an {\it a priori} explanation why these invariants are encoded in the physics of 3d $\CN=2$ theory $T[M_3]$ the way
we described them in section \ref{sec:3d3d}.
This becomes crystal clear if we look at the system \eqref{M3phases} from the vantage point of the 3-manifold $M_3$
by first compactifying fivebranes on the Riemann surface $\Sigma$; it will directly lead us to more traditional
formulation of Seiberg-Witten invariants, Turaev torsion, and their categorification \eqref{HFHMECH}.

By doing so, we introduce an easily computable ``refinement'' of the Rozansky-Witten theory, which shares many features
with the homological invariants $HF (M_3)$, $HM (M_3)$, $ECH (M_3)$, and in certain cases completely determines the latter.

\subsection{UV: SW invariants and Floer Homology}
\label{sec:N4gauge}

In order to look at the system \eqref{M3phases} from the viewpoint of the 3-manifold $M_3$, one first needs to compactify
the fivebranes on the Riemann surface $\Sigma$ (possibly with defects and punctures).
This gives a 4d $\CN=2$ theory with $SU(2)_{R}\times U(1)_t$ R-symmetry, see {\it e.g.} \cite{Witten:1997sc}.
Keeping track of this R-symmetry helps to describe the topological twist along $M_3$,
under which the R-symmetry $SU(2)_R$ is mixed with the $SO(3)$ group of local rotations on $M^3$.
As a result, three out of five scalars in the fivebrane theory, which are charged under $SU(2)_R$,
no longer transform as scalars on $M_3$ and can instead be thought of as parameterizing the cotangent bundle of $M_3$, {\it cf.} \eqref{M3phases}.
Upon reduction of 4d $\CN=2$ theory on $S^1$, we get a 3d $\CN=4$ theory $T[\Sigma\times S^1]$.
And, after the reduction R-symmetry is enhanced to
\begin{equation}
SU(2)_{R}\times U(1)_t \quad \rightarrow \quad SU(2)_{R}\times SU(2)_{N} \,.
\label{3dRenhanced}
\end{equation}
In three dimensions, $SU(2)_R$ can be distinguished from $SU(2)_N$ by saying that the former acts trivially on the scalars in the vector multiplet. The partition function of the twisted 3d $\CN=4$ theory gives a numerical topological invariant of $M_3$. If instead of $S^1$ we considers a non-compact time direction, $\R$, this will give us a twisted 4d $\CN=2$ theory $T[\Sigma]$ on $M_3\times \R$. Its Hilbert space,
\begin{equation}
	\CH_{T[\Sigma]}(M_3)=\CH_{T[M_3]}(\Sigma),
	\label{HSigma}
\end{equation}
will then provide a categorification of the numerical invariant:
\begin{equation}
	\Tr_{\CH_{T[\Sigma]}(M_3)} (-1)^F \;=\; Z_{\text{twisted}~T[\Sigma\times S^1]}(M_3)
	\label{ZM3}
\end{equation}

Consider a particular case when $G=U(1)$ and $\Sigma=\Sigma_\fM$ such that
\begin{equation}
 \Sigma_\fM:\qquad T_{U(1)}[\Sigma_\text{SW}] = \; \{ \text{4d $\mathcal{N}=2$ $U(1)$ with 1 hypermultiplet} \}
\label{monopole-dec}
\end{equation}
After the topological twist, the scalars of the hypermultiplet become a complex spinor on $M_3$ charged with respect to the $U(1)$ gauged group. To define such spinor one needs to pick a $\text{Spin}^c$ structure ${\frak s}$ on $M_3$. The $U(1)$ gauge bundle then can be identified with $\det ({\frak s})$. The partition function is then expected to give Seiberg-Witten invariants of $M_3$ for a given ${\frak s} \in \text{Spin}^c$:
\begin{equation}
	\text{sw}({\frak s})=Z_{T[\Sigma_\fM\times S^1]}(M_3;{\frak s})
\end{equation}
The space of $\text{Spin}^c$ structures is isomorphic to $H^2(M_3)\cong H_1(M_3)\equiv H$ as an affine space over $H$. By fixing a reference $\text{Spin}^c$ structure ${\frak s}_0$ (the canonical choice is such that $c_1({\frak s}_0)=0$) one can instead define a Seiberg-Witten invariant as a function on $H$:
\begin{equation}
	\SW\in \Q[H]\subset\C[H],\qquad \SW(h)=\text{sw}({\frak s}_0+h).
	\label{SWinvariants}
\end{equation}
It is often also useful to consider a Fourier-transformed SW invariant, a function on the Pontryagin dual group\footnote{Let us note that the space of Spin$^c$ structures is isomorphic to $H_1(M_3)$ for general $M_3$. However only for rational homology spheres it can be identified with flat connections modulo gauge equivalence, {\it i.e.} the space $\CM_{\text{flat}}(U(1)_\C,M_3)$. This happens because when $b_1(M_3)=0$ the first Chern class $c_1({\frak s})$ of any ${\frak s} \in \text{Spin}^c$ is torsion and there is a unique flat connection in $\det({\frak s})$. For $b_1(M_3)>0$ this is not the case. However, the Pontryagin dual $\hat H$, which in this case we define as $\text{Hom}(H_1,\C^*)$, is still the same as $\CM_{\text{flat}}(U(1)_\C,M_3)$. By $\C[\hat H]$ we then understand meromorphic functions on $\hat H$.}:
\begin{equation}
	\hat\SW\in \C[\hat H],\qquad \hat\SW(q)=\sum_{h\in H}\SW(h) q^{-1}(h)
\end{equation}
From the gauge theory point of view, $c_1({\frak s}) \in H^2(M_3)$, ${\frak s} \in \text{Spin}^c$, can be interpreted as the flux of the $U(1)$ gauge field. Therefore the dual variable $q$ has the meaning of (the exponential of) the FI-parameter. This correspondence will be considered in more detail in the next section.

{}From (\ref{ZM3}) it follows that (\ref{HSigma}) should categorify SW invariants and give monopole Floer homology which is isomorphic \cite{MR2388043} to the Heegaard Floer homology (and the corresponding version of the $ECH$ theory, as in \eqref{HFHMECH}):
\begin{equation}
\CH_{T[\Sigma_\text{SW}]}(M_3) \; = \; \widecheck{HM}(M_3)\cong HF^+(M_3) \,.
\end{equation}
This homology naturally splits according to Spin$^c$ structures:
\be
HF^+ (M_3) \cong \oplus_{{\frak s} \in \text{Spin}^c (M_3)} HF^+ (M_3, \frak s)
\ee


\subsection{IR: 3d $\CN=4$ Rozansky-Witten theory}
\label{sec:RW}

In the IR a 3d $\CN=4$ gauge theory $T[\Sigma\times S^1]$ considered in section \ref{sec:N4gauge} has a geometric description in terms of Higgs and Coulomb branches which are both Hyper-K\"ahler and exchanged under 3d mirror symmetry. The $SU(2)_R$ symmetry acts trivially on the scalars parametrizing the Coulomb branch while $SU(2)_N$ acts trivially on the scalars parametrizing the Higgs branch. When $SU(2)_R$ is twisted, the theory becomes a topologically twisted sigma model on the Coulomb branch\footnote{In the sigma model description the R-symmetry should act trivially on the target space itself and only rotate the fermionic fibers.}
\begin{equation}
	X_\Sigma \equiv \CM_\text{Coulomb}(T[\Sigma\times S^1])
\end{equation}
where $X_{\bullet}$ can be considered as a functor from Riemann surfaces to hyper-K\"ahler spaces. This theory is known as the Rozansky-Witten theory \cite{Rozansky:1996bq}. Therefore we have:
\begin{equation}
	Z_{\text{twisted }T[\Sigma\times S^1]}(M_3) \; = \; Z_{\RW[X_\Sigma]}(M_3)
\end{equation}

The Rozansky-Witten (RW) theory computes the same type of perturbative 3-manifold invariants as in the perturbative expansion of CS theory around the trivial flat connection\footnote{To be precise, both CS theory and Rozansky-Witten theory produce a system of weights for LMO universal perturbative invariant \cite{LMO} valued in the algebra $\CA$ of trivalent Feynman diagrams modulo IHX relations. A weight system can be understood as an element of the dual algebra $\CA^*$. The setup arising from 6d theory compactification provides us with a map $\{\text{Riemann surfaces }\Sigma\}\rightarrow \CA^*$.}. However, sometimes it can be possible to consider more refined invariants by turning on certain background fields.

Indeed, suppose $X_\Sigma$ has a $U(1)^b$ tri-holomorphic (that is respecting hyper-K\"ahler structure) symmetry. In the UV description such symmteries can appear as topological $U(1)$ symmetries coupled to $U(1)$ gauge fields. This means that one can couple the sigma-model to $b$ copies of topologically twisted $\CN=4$ $U(1)$ vector multiplets. The scalars of such multiplets transform in a triplet representation of the $SU(2)_R$ symmetry.\footnote{Note, in order to weakly gauge the $U(1)^b$ isometry of $X_{\Sigma}$ it is convenient to realize $X_{\Sigma}$ as a Higgs branch of the mirror 3d $\CN=4$ theory. Since, as we mentioned earlier, 3d mirror symmetry exchanges $SU(2)_R$ and $SU(2)_N$, the twist by $SU(2)_R$ of the original theory in the notations \eqref{3dRenhanced} is equivalent to the twist by $SU(2)_N$ of the mirror 3d $\CN=4$ theory.} In the UV, the scalars in a vector multiplet have meaning of real and complex FI-parameters.
After topological twisting the triplet of scalars together with the vector becomes a complex 1-form on $M_3$.
This, again, can be understood as the Blau-Thompson twist \cite{Blau:1996bx,Blau:1997pp} of the mirror 3d $\CN=4$ theory,
which in the present case localizes on flat $U(1)_\C$ connections.
The partition function refined by the background values of such flat connections depends on
an element of~~$\hat{H}^b \equiv \stackrel{b\text{ copies}}{\times} \CM_\text{flat}(U(1)_\C,M_3)$:
\begin{equation}
	Z_{RW[X_\Sigma]}(M_3)(q_1,\ldots,q_b),\qquad q_i\in \hat{H}
\end{equation}

Consider in more detail the case when $\Sigma=\Sigma_\fM$ as in (\ref{monopole-dec}),
so that $T[\Sigma\times S^1]$ is 4d $\CN=4$ abelian gauge theory with gauge group $U(1)$ and one charged hypermultiplet.
The relation between SW and RW invariants of $M_3$ was proposed in a beautiful
paper of Blau and Thompson \cite{Blau:2000iy} by interpreting these as UV and IR TQFTs, respectively.
A topological twist of the $\CN=4$ SQED with $N_f=1$ hypermultiplet gives the abelian monopole
equations on $M_3$ whose signed count of solutions yields SW invariants of $M_3$.
In the IR this theory flows to a sigma-model whose target space is the Coulomb branch of
$\CN=4$ SQED, namely $\text{TN}_4$, the Taub-NUT space with one center. And the {\it same} kind of topological twist
of this IR theory gives RW invariants of $M_3$:
\begin{equation}
	\begin{array}{c}
\boxed{\phantom{\oint}\text{UV}\phantom{\oint}}
\quad \quad \quad \quad \quad \quad \quad \quad \quad \quad \quad \quad \quad \quad \quad \quad \quad \quad
\boxed{\phantom{\oint}\text{IR}\phantom{\oint}}
		\\
		\\
		\text{SW twist of 3d} ~\CN=4~ \text{SQED} \quad = \quad \text{RW twist of}~\sigma \text{-model with target TN}_4	
	\end{array}
\end{equation}
The relevant twist of both UV and IR theories here involves mixing the Lorentz symmetry group
with the $SU(2)$ subgroup of R-symmetry under which scalars in a vector multiplet are singlets
and scalars in a hypermultiplet transform as ${\bf 2}$.
In type IIB brane setup shown in Figure~\ref{fig:SW-branes} , scalars in vector multiplets correspond to motion of
D3-branes along directions of NS5-branes, that is $x^{8,9,0}$ in our conventions.
Hence, RW and SW theories are obtained by mixing $SO(3)_{123} \cong SU(2)_E$ with $SO(3)_{456} \cong SU(2)_R$.
As a complex manifold, the Coulomb branch is isomorphic to the complex plane:
\begin{equation}
	X_{\Sigma_\fM}=\text{TN}_4 \stackrel{\text{complex}}{\cong}\C^2
	\label{SWCoulomb}
\end{equation}
The UV theory has one topological $U(1)$ symmetry that can be coupled to a $U(1)$ vector multiplet. In the IR, this symmetry rotates two $\C$ factors of (\ref{SWCoulomb}) with opposite charges. The refined RW theory then should give a SW invariant labeled by a corresponding flat connection:
\begin{equation}
	\hat\SW_{M_3}(q)=Z_{\RW[X_{\Sigma_\fM}]}(q),\;\;q\in \hat{H}
\end{equation}

Consider the case when $b_1(M_3)=1$ in more detail.
In this case RW theory calculates the value of a certain characteristic class on $X_\Sigma$ \cite{Habegger:1999yp,Blau:2000iy}:
\begin{equation}
	Z_{\text{RW}[X_\Sigma]}(M_3)=\int_{X_\Sigma} \prod_{i}^{\frac{1}{2}\dim_{\C}X_\Sigma}\lambda_i^2\, \tau_{M_3}(e^{\lambda_i})
	\label{RWcharclass}
\end{equation}
where $(\lambda_1,-\lambda_1,\lambda_2,-\lambda_2,\ldots)$ are the eigenvalues of the holomorphic curvature
and $\tau_{M_3}$ is the Milnor torsion of $M_3$. In the case $b_1(M_3)=1$ the latter has the following form:
\begin{equation}
	\tau_{M_3}(q)=\frac{q\Delta_{M_3}(q)}{(1-q)^2}
\end{equation}
where $\Delta_{M_3}(q)$ is the Alexander polynomial of $M_3$.
A large class of 3-manifolds with $b_1(M_3)=1$ and, moreover, $H=H_1(M_3,\Z)=\Z$ is given by zero-surgeries on knots in $S^3$.

\begin{table}[htb]
\centering
\renewcommand{\arraystretch}{1.3}
\begin{tabular}{|@{\quad}c@{\quad}|@{\quad}c@{\quad}|@{\quad}c@{\quad}|@{\quad}c@{\quad}|}
\hline knot $K$ & signature & Alexander polynomial & $HF^+ (S^3_0 (K))$
\\
\hline
\hline unknot & $\sigma (K) = 0$ & $\Delta_K (q) = 0$ & $\CT^+_{-1/2} \oplus \CT^+_{1/2}$ \\
\hline $3_1^r$ & $\sigma (K) = -2$ & $\Delta_K (q) = - 1 + q^{-1} + q$ & $\CT^+_{-1/2} \oplus \CT^+_{-3/2}$ \\
\hline $4_1$ & $\sigma (K) = 0$ & $\Delta_K (q) = 3 - q^{-1} - q$ & $\Z [- \tfrac{1}{2}] \oplus \CT^+_{-1/2} \oplus \CT^+_{1/2}$ \\
\hline
\end{tabular}
\caption{Simple 0-surgeries on knots. In all these examples $HF^+ (S^3_0 (K), {\frak s}_h) \cong 0$ unless $h=0$.}
\label{tab:surgeries}
\end{table}

In general, a $p/q$ Dehn surgery along a knot $K \subset S^3$ is the operation of removing a tubular neighborhood of $K$
and regluing it back in a way that takes the meridian of the knot to a linear combination of the longitude and the meridian
with coefficients $q$ and $p$, which are assumed to be relatively prime integers.
For example, a Lens space $L(p,q)$ can be realized as a surgery on the unknot with the coefficient $-p/q$
and, in fact, for any knot $M_3 = S^3_{p/q} (K)$ has $H_1 (S^3_{p/q} (K)) = \Z / p \Z$, and
\be
S^3_r (K) \cong - S^3_{-r} (\bar K)
\ee
where $\bar K$ is a mirror of $K$.

In particular, the 0-surgery gives $H_1 (M_3) = \Z$ supplying lots of examples for our discussion here.
The simplest 0-surgery is the surgery along the unknot which, according to the ``Property R'' conjecture,
is believed to be the only knot that produces $S^3_0 (K) = S^2 \times S^1$ which we already discussed earlier.
In this class of examples, $\hat H=\C^*$ which means that there is a continuous deformation
of the background vector multiplet starting from $0$.
In other words, there is only one $q$ parameter and $\Delta_{M_3}(q)=\Delta_K(q)$ is the Alexander polynomial of $K$,
\be
\Delta_K (q) = a_0 + \sum_{h > 0} a_h (q^{h} + q^{-h})
\ee

In general, the (logarithms of the) corresponding values $q_i\in \hat H$ can be understood as equivariant parameters for $U(1)^b$ symmetry acting on $X_\Sigma$. In the presence of such symmetries the equivariant version of the integral (\ref{RWcharclass}) can often be easily calculated using the Atiyah-Bott localization. For example, in the case $\Sigma=\Sigma_\fM$, $X_{\Sigma_\fM}=\C^2$ there is only one fixed point and the result is just the value of the torsion:
\begin{equation}
	Z_{\RW[X_{\Sigma_\fM}]}(q)=\tau_{M_3}(q) \,,
\end{equation}
in perfect agreement with the Meng-Taubes theorem~\cite{meng1996sw}. Note that localization can be easily done also for $G=SU(N)$ when $\Sigma$ is of genus $\geq 1$ with regular punctures, that is when $T[\Sigma]$ is in class $\CS$. This is due to the fact that the Coulomb branch of $T[\Sigma\times S^1]$ has a hyper-K\"ahler quotient description \cite{Benini:2010uu}.

\subsection{A ``refinement'' of the Rozansky-Witten theory}

Let us note that the Coulomb branch \eqref{SWCoulomb} also has a holomorphic symmetry $U(1)$ with respect to which both $\C$ factors have the same charge. This is the ``anti-diagonal'' subgroup $U(1)$ of $SU(2)_R\times SU(2)_N$ R-symmetry. The possibility of turning on the corresponding background 3d $\CN=2$ vector multiplet in a supersymmetric way is equivalent to the existence of an extra symmetry $U(1)_\beta$ discussed in detail in section~\ref{sec:U1beta}.

For example, this extra symmetry exists in the case of $M_3=S^2\times S^1$ that we already encountered in \eqref{S1S2SWy}. Since
\begin{equation}
	\Tor_{S^2\times S^1}(q)=\frac{q}{(1-q)^2}
\end{equation}
the corresponding characteristic class in (\ref{RWcharclass}) is the (complex) $\hat A$-genus\footnote{On a formal level the question about possibility of refinement by holomorphic symmetries seems to be equivalent to the question about extending multiplicative characteristic class appearing in (\ref{RWcharclass}) and a priori defined only for hyper-K\"ahler manifolds to K\"{a}hler manifolds. That is, finding its ``square root'', which means representing it in the following way:
\begin{equation}
 \Tor_{M_3}(e^\lambda)=f(e^{\lambda})f(e^{-\lambda})
\end{equation}
such that $f(q)$ has expansion valued in $\Z[[q]]$. Interestingly enough
such factorization naturally exists for 0-surgeries on slice knots for which there exists Laurent polynomial $g$ such that $\Delta_K(q)=g(q^{-1})g(q)$. Then one can take $f(q)=g(q)/(1-q)$.}:
\begin{equation}
	\Tor_{S^2\times S^1}(e^\lambda)=\frac{\lambda}{2 \sinh (\lambda/2)}\cdot\frac{-\lambda}{2 \sinh (-\lambda/2)}
\end{equation}
Then, the equivariant localization of the integral (\ref{RWcharclass}) gives the following refined torsion of $M_3 = S^2\times S^1$:
\begin{equation}
		Z_{\text{RW}[X_{\Sigma_\fM}]}(M_3) \; = \; \int_{\C^2}\hat A \; = \; \frac{q}{(1-qy)(1-q/y)}
\end{equation}
where $y$ is the fugacity associated to the $U(1)_\beta$ flavor symmetry, {\it cf.} \eqref{index-beta}.
This agrees with the Heegaard Floer homology of $M_3=S^2\times S^1$ discussed in section \ref{sec:monopole-defect}.

%


\section{4-manifold invariants from 5-branes}
\label{sec:4mfld}

In the earlier sections, a few times we already found it useful to answer questions about 3-manifolds by
considering gauge theory on 4-manifolds and the corresponding 2d theory on $\Sigma$, {\it cf.} \eqref{choices6d}.
More generally and more conceptually, the study of homological invariants of knots and 3-manifolds
provides a window into a remarkable world of 4-manifolds with embedded surfaces.
Indeed, every cobordism $M_4$ between 3-manifolds $M_3^-$ and $M_3^+$
induces a map between the corresponding homological invariants, possibly with suitable extra data\footnote{For instance,
if ${\frak s}$ is a Spin$^c$ structure on $M_4$,
we get a map from $HF^+ (M_3^- , {\frak s} \vert_{M_3^-})$ to $HF^+ (M_3^+ , {\frak s} \vert_{M_3^+})$.}:
\be
Z_{T[\Sigma]}(M_4) ~: \qquad \CH_{T[\Sigma]} (M_3^-) \; \longrightarrow \; \CH_{T[\Sigma]} (M_3^+) \,,
\label{TM4functor}
\ee
and similarly for knots. In particular this generalizes relation (\ref{HVW}).
(See also \cite{Gukov:2007ck,Gorsky:2013jxa,Gadde:2013sca} for previous discussion of such cobordisms from physical perspective.)

In this section, our goal will be to revisit the topological twists starting with 6d $(0,2)$ theory on $M_4 \times \Sigma$
and then to explore \eqref{TM4functor} and its variants.
Even at a practical level, 4-manifolds with boundary $M_3$ help us to understand the $SL(2,\Z)$ action on $\CM_{\text{flat}} (G_{\C}, M_3)$
and to calculate the so-called {\it correction terms} in the Heegaard Floer homology $HF^+ (M_3)$.

\subsection{Twists on $M_4$}
\label{sec:twistsM4}

In six-dimensional space-time of the form $M_4 \times \Sigma$, the Euclidean rotation symmetry $SO(6)_E$
of the fivebrane theory decomposes as $SO(4)_E \times SO(2)_{\Sigma}$, and we further identify $SO(4)_E \cong SU(2)_{\ell} \times SU(2)_r$.
The R-symmetry group of the 6d $(0,2)$ theory is $SO(5)_R$, under which self-dual 2-form, scalars and Weyl fermions
transform as ${\bf 1}$, ${\bf 5}$, and ${\bf 4}$, respectively.
The Weyl fermions have positive chirality, {\it i.e.} transform as ${\bf 4}_+$ under $SO(6)_E$,
and obey symplectic reality conditions.
The following branching rules will be useful to us:
\begin{eqnarray}
& SO(6)_E & \to \quad SU(2)_{\ell} \times SU(2)_r \times U(1)_{\Sigma} \nonumber \\
& {\bf 4}_+ & \to \quad ({\bf 2}, {\bf 1})^{+ 1} \oplus ({\bf 1}, {\bf 2})^{-1} \nonumber \\
& {\bf 4}_- & \to \quad ({\bf 2}, {\bf 1})^{- 1} \oplus ({\bf 1}, {\bf 2})^{+1} \nonumber \\
& {\bf 6} & \to \quad ({\bf 2}, {\bf 2})^{0} \oplus ({\bf 1}, {\bf 1})^{+2} \oplus ({\bf 1}, {\bf 1})^{-2} \nonumber
\end{eqnarray}
and
\begin{eqnarray}
& SO(5)_R & \to \quad SU(2)_R \times U(1)_t \nonumber \\
& {\bf 5} & \to \quad {\bf 3}^{0} \oplus {\bf 1}^{\pm 2} \nonumber \\
& {\bf 4} & \to \quad {\bf 2}^{+1} \oplus {\bf 2}^{-1} \nonumber
\end{eqnarray}

The topological twist of 6d $(0,2)$ theory on general background of the form $M_4 \times \Sigma$ that was studied in \cite{Gadde:2013sca}
corresponds to embedding the fivebrane world-volume $M_4 \times \Sigma$ into a product of $G_2$-manifold and a local K3 geometry,
which locally look like
\be
\Lambda^2_+ M_4 \times T^* \Sigma
\ee
For this twist and other applications discussed in the present paper, we decompose the R-symmetry group as $SO(5)_R \to SO(3)_R \times SO(2)_t$.
To summarize this symmetry breaking pattern, let us describe the transformation of the fermions in 6d $(0,2)$ theory:
\begin{eqnarray}
& SO(6)_E \times SO(5)_R & \to \quad SU(2)_{\ell} \times SU(2)_r \times SU(2)_R \times U(1)_{\Sigma} \times U(1)_t  \nonumber \\
\text{fermions:} & ({\bf 4}_+, {\bf 4} ) & \to \quad ({\bf 2}, {\bf 1}, {\bf 2})^{(1,\pm 1)} \oplus
({\bf 1}, {\bf 2}, {\bf 2})^{(-1,\pm 1)} \nonumber
\end{eqnarray}
Now we can consider various topological twists of this system.
Note, when $M_4 = \R \times M_3$ or $M_4 = S^1 \times M_3$,
the rotation symmetry on $M_3$ is a diagonal subgroup $SU(2)_M \subset SU(2)_{\ell} \times SU(2)_r$.

Keeping these facts in mind, we can consider a partial topological twist on general $M_4$
by replacing $SU(2)_r$ with the diagonal subgroup $SU(2)_r' \subset SU(2)_r \times SU(2)_R$.
This gives partial Vafa-Witten twist of the 6d $(0,2)$ theory,
with the new transformation rules:
\begin{eqnarray}
& SO(6)_E \times SO(5)_R & \to \quad SU(2)_{\ell} \times SU(2)_r' \times U(1)_{\Sigma} \times U(1)_t  \nonumber \\
\text{fermions:} & ({\bf 4}_+, {\bf 4} ) & \to \quad ({\bf 2}, {\bf 2})^{(1,\pm 1)} \oplus
({\bf 1}, {\bf 3})^{(-1,\pm 1)} \oplus ({\bf 1}, {\bf 1})^{(-1,\pm 1)} \nonumber
\end{eqnarray}
Note, from the viewpoint of 2d theory on $\Sigma$, the two preserved supercharges are chiral,
which corresponds to 2d $\CN=(0,2)$ supersymmetry along $\Sigma$.
Also note that if $M_4 = \R \times M_3$, then before the twist we have
\begin{eqnarray}
& SO(6)_E \times SO(5)_R & \to \quad SU(2)_M \times SU(2)_R \times U(1)_{\Sigma} \times U(1)_t  \nonumber \\
\text{fermions:} & ({\bf 4}_+, {\bf 4} ) & \to \quad ({\bf 2}, {\bf 2})^{(\pm 1,\pm 1)} \nonumber
\end{eqnarray}
which agrees, as it should, with the transformation rules
of fermions in 5d $\CN=2$ super-Yang-Mills, {\it cf.} section \ref{sec:3d3d}.

Instead of twisting along $M_4$ (or $M_3$) we can start with a partial topological twist along $\Sigma$,
replacing $U(1)_{\Sigma}$ with a diagonal subgroup $U(1)_{\Sigma}' \subset U(1)_{\Sigma} \times U(1)_t$.
Note, since these groups are abelian, $U(1)_t$ is still a symmetry after this twist, so that
\begin{eqnarray}
& SO(6)_E \times SO(5)_R & \to SU(2)_{\ell} \times SU(2)_r \times SU(2)_R \times U(1)_{\Sigma}' \times U(1)_t  \nonumber \\
& ({\bf 4}_+, {\bf 4} ) & \to ({\bf 2}, {\bf 1}, {\bf 2})^{(2,1)}
\oplus \underline{({\bf 2}, {\bf 1}, {\bf 2})^{(0,-1)}} \oplus \underline{({\bf 1}, {\bf 2}, {\bf 2})^{(0,1)}}
\oplus ({\bf 1}, {\bf 2}, {\bf 2})^{(-2,-1)}
\nonumber
\end{eqnarray}
where we underlined the terms which transform as singlets under $U(1)_{\Sigma}'$.
Their significance is that they correspond to unbroken supersymmetries of 6d $(0,2)$ theory partially twisted along $\Sigma$;
they transform precisely as supercharges of 4d $\CN=2$ theory on $M_4$, with the R-symmetry group $SU(2)_R \times U(1)_t$.

As we explain next, a further twist along $M_4$ is the standard Donaldson-Witten (or Seiberg-Witten) twist of this 4d $\CN=2$ theory.
Replacing $SU(2)_r$ with the diagonal subgroup $SU(2)_r' \subset SU(2)_r \times SU(2)_R$, we get
\begin{eqnarray}
& SO(6)_E \times SO(5)_R & \to SU(2)_{\ell} \times SU(2)_r' \times U(1)_{\Sigma}' \times U(1)_t  \nonumber \\
& ({\bf 4}_+, {\bf 4} ) & \to ({\bf 2}, {\bf 2})^{(2,1)}
\oplus ({\bf 2}, {\bf 2})^{(0,-1)} \oplus ({\bf 1}, {\bf 3})^{(0,1)}
\oplus ({\bf 1}, {\bf 1})^{(0,1)}
\oplus ({\bf 1}, {\bf 3})^{(-2,-1)}
\oplus ({\bf 1}, {\bf 1})^{(-2,-1)}
\nonumber
\end{eqnarray}
There is only one supercharge $\CQ$, which is a complete singlet under the symmetries of both $M_4$ and $\Sigma$.
If we denote the generators of $U(1)_{\Sigma}'$ and $U(1)_t$ by $P$ and $R_t$, respectively, then from the above
transformation rules we can easily read off
\be
\CQ^2 = 0
\,, \qquad
[R_t,\CQ] = \CQ
\,, \qquad
[P,\CQ] = 0
\ee
When $M_4 = \R \times M_3$ or $M_4 = S^1 \times M_3$, we have two scalar supercharges;
the second one arises from the decomposition ${\bf 2} \otimes {\bf 2} = {\bf 3} \oplus {\bf 1}$
with respect to $SU(2)_M$ or, equivalently, from implementing the twist along $\Sigma$ in \eqref{twistedspectrum}.


We can compare the above sequence of twists $U(1)_{\Sigma} \to U(1)_{\Sigma}'$ and $SU(2)_r \to SU(2)_r'$
with the standard Donaldson-Witten (or Seiberg-Witten) twist in 4d $\CN=2$ theory\footnote{Sometimes
in the literature the role of $SU(2)_{\ell}$ and $SU(2)_r$ is exchanged or, equivalently,
the sign of the $U(1)_t$ R-charge is flipped.}
\begin{eqnarray}
& SU(2)_{\ell} \times SU(2)_r \times SU(2)_R \times U(1)_t & \to SU(2)_{\ell} \times SU(2)_r' \times U(1)_t \nonumber \\
A_{\mu} & ({\bf 2}, {\bf 2}, {\bf 1})^{0} & \to ({\bf 2}, {\bf 2})^{0} \nonumber \\
\text{scalars:} & ({\bf 1}, {\bf 1}, {\bf 1})^{\pm 2} & \to ({\bf 1}, {\bf 1})^{\pm 2} \nonumber \\
\text{fermions:} & ({\bf 2}, {\bf 1}, {\bf 2})^{-1} \oplus ({\bf 1}, {\bf 2}, {\bf 2})^{1} &
\to ({\bf 2}, {\bf 2})^{-1} \oplus ({\bf 1}, {\bf 3})^{1} \oplus ({\bf 1}, {\bf 1})^{1} \nonumber
\end{eqnarray}
In the last line one can recognize the $U(1)_{\Sigma}'$ invariant fermions of 6d $(0,2)$ theory before and after the $SU(2)_r$ twist.

Finally, we note that the Vafa-Witten and Marcus/GL twists
of 4d $\CN=4$ super-Yang-Mills reduce to the same Blau-Thompson twist in 3d,
a theory that localizes on complex flat connections.
On the other hand, the third twist of 4d $\CN=4$ SYM, namely the theory of adjoint non-abelian monopoles,
reduces to 3d version of DW twist.

\subsection{VW partition function as a CS wave function}

In section \ref{sec:plumbed} we already used the fact (from \cite{Gadde:2013sca}) that $T_{U(1)}[M_3]$ for plumbed $M_3$ has a description in terms of quiver abelian CS theory, so that the role of quiver is played by the plumbing graph. There we also mentioned that the plumbing graph naturally describes not only $M_3$ but a 4-manifold $M_4$ such that $\d M_4=M_3$. In this section we will explore this relation further.

Consider quantization of abelian Chern-Simons theory on $T^2\times \R$. Quantization procedure requires to choose a complex structure $\tau$ on $T^2$ (though the Hilbert spaces for different structures are equivalent). There are $|\text{Coker}\,Q|=|H|$ states on the torus and they correspond to basic Wilson lines of the form
\begin{equation}
	\prod_{i\text{vertices}}x_i^{h_i}\;\in\;\C[x_1,\ldots,x_{b_2}]/\{\prod_j x_j^{Q^{ji}}-1\},\qquad h\in H
\end{equation}
inserted in the solid torus bounded by $T^2$. One can also specify a wave function of such states as a function of $U(1)^{b_2}$ holonomies along one of the cycles. That is, let $|x\rangle\in \CH_{T[M_3]}(T^2)$ be a state with given holonomies and $|h\rangle\in \CH_{T[M_3]}(T^2)$ a state created by a Wilson line. Up to an overall $h$-independent normalization\footnote{The quantization of spin CS on $T^2$ requires to choose Spin structure $\in \left(\frac{1}{2}\Z_2\right)^2$ on a torus. Here we use $(\frac{1}{2},\frac{1}{2})$ choice.}
\begin{equation}
	\langle h|x\rangle \equiv \Psi_{h}(x)=\sum_{\lambda \in \Lambda+h+w_2/2}q^{-\frac{1}{2}(\lambda,\lambda)-b_2/8}x^\lambda
	\label{CS-wave-function}
\end{equation}
where $q= e^{2\pi i\tau}$ and $(\cdot,\cdot)$ is a bilinear form on $\Lambda$ given by $Q$ and extended to $\Lambda^*\subset \Q\otimes_\Z \Lambda$.
The parameter $w_2$ is an extra data one needs to introduce in order to quantize abelian spin CS theory (see \cite{Dijkgraaf:1989pz,Belov:2005ze} for details). The element $w_2\in \Lambda^*$ has to be chosen such that
\begin{equation}
	w_2(\lambda)=(\lambda,\lambda)\;\;\mod 2,\;\;\forall \lambda\in\Lambda
\end{equation}
which fixes the class $[w_2]\in \Lambda^*/2\Lambda^*$.

This is in perfect agreement with what is going on the 3- and 4-manifold side \cite{Gadde:2013sca}. The class $[w_2]\in H^2(M_4,\Z_2)$ is the second Stiefel-Whitney class and reflects the presence of the Freed-Witten anomaly in the $U(1)$ gauge theory on $M_4$. The choice of representative $w_2$ corresponds to a choice of a reference Spin$^c$ structure ${\frak s}_0$ in (\ref{SWinvariants}) which is needed to identify the set of Spin$^c$ structures with $H$ (see e.g. \cite{nemethi2002seiberg}). When $Q$ is even (that is $M_4$ is spin) one can choose $w_2=0$.

The overall factor $q^{-b_2/8}$ is chosen so that the wave function has nice properties with respect to the moves in Figure~\ref{fig:moves}. In particular the $T$ matrix acting on $\CH_{T[M_3]}(T^2)$ is given by
\begin{equation}
	T_{hh'}=\delta_{hh'}\,e^{-\pi i\left[ (h+w_2/2,h+w_2/2)-b_2/4  \right]}
\end{equation}
and is an invariant of $M_3$.

Up to an overall factor (\ref{CS-wave-function}) is equal to the partition function of abelian VW theory on $M_4$ with a boundary condition labeled by $h\in H_1(M_3)$ \cite{Gadde:2013sca}:
\begin{equation}
 Z_\VW[M_4](q;x)_h\propto\sum_{[F/2\pi] \in \Lambda+h+w_2/2}q^{\frac{1}{8\pi^2}\int F\wedge F} x^{\left[{F}/{2\pi}\right]}
\propto\Psi_h(x)
\label{VWwave}
\end{equation}
On the 4-manifold side, the fugacities $x_i$ are the (exponentiated) chemical potentials for the first Chern class of the gauge connection on $M_4$. The label $h$, that is a choice of vacuum of $T[M_3]$ on $T^2$, labels the choice of a flat connection $\rho$ on $M_3$.
As in section \ref{sec:2bases}, on can think of the VW partition function as a vector
\begin{equation}
 Z_\VW[M_4](q;x)\in \CH_\VW(M_3)\equiv \CH_{T[T^2]}(M_3)=\CH_{T[M_3]}(T^2)
\end{equation}
so that
\begin{equation}
 Z_\VW[M_4](q;x)_h=\langle h|Z_\VW[M_4](q;x).
\end{equation}
{}From (\ref{VWwave}) it follows then that
\begin{equation}
 Z_\VW[M_4](q;x)\propto |x\rangle.
\end{equation}

As in the case of $M_3=L(p,1)$ considered in section \ref{HFLp1},
the expressions (\ref{CS-wave-function}) are wave-functions of $T[M_3]$ for states on $T^2$ surrounding codimension-4 defects labeled by $h$ which appear in $S^2_{\fM,h}\times S^1$ (see Figure~\ref{fig:monopole-curve-3}). The partition function (\ref{VWwave}) can be interpreted as the index (partition function on $T^2$) of the 2d (0,2) theory $T[M_4]$. Therefore in the categorified setup, the effective quantum mechanics $T[M_3\times S^2_{\fM,h}]$ (with Hilbert space $\CH_{T[M_3]}(S^2_{\fM,h})\cong HF^+(M_3,\mathfrak{s}_h)$) can be obtained by coupling (via $T[M_3]$ on $S^2$) of boundary CFT $T[M_4]$  on $S^1$ (with states counted by (\ref{VWwave})) and the effective quantum mechanics of the codimension-2 defect compactified on $M_3$ and inserted on $S^2$.

The wave-functions (\ref{CS-wave-function}) have the following $q$-expansions:
\begin{equation}
	\Psi_{h}(x)=q^{-\Delta(h)/2}+\ldots
\end{equation}
The rational numbers $\Delta(h)$ can be interpreted as conformal dimensions of primaries of the boundary CFT, chiral $U(1)^{b_2}$ WZW theory. Their values can be determined in the following way:
\begin{equation}
	\Delta(h)=\max_{\lambda \in \Lambda+h+w_2/2}\left[(\lambda,\lambda)_Q+b_2/4\right]
\end{equation}
This coincides with the formula for ``correction terms'' in the Heegaard Floer homology \cite{ozsvath2003floer}. Note,
\begin{equation}
	\SW(h)=-\frac{\Delta(h)}{2}\;\;\mod\Z
\end{equation}
so that
\begin{equation}
	\SW(h)=\frac{1}{2\pi i}\log T_{hh}\;\;\mod\Z
\end{equation}
and the corresponding operator $\mathscr{SW}$ appeared in section \ref{sec:SWfromTM3} which acts on $\CH_{T[M_3]}(T^2)$ can be identified with
\begin{equation}
	\mathscr{SW}=\frac{1}{2\pi i}\log T\;\;\mod\Z
\end{equation}


\section{Khovanov homology for 3-manifolds}
\label{sec:higherRank}

As in the landscape of knot homologies unified by BPS states, in our study of 3-manifold homologies
the $U(1|1)$ theory plays a central role, literally and figuratively ({\it cf.} Figure~\ref{fig:unification}).
This is the reason why a fair portion of this paper is devoted to a physical realization of this homology
theory in the fivebrane setup \eqref{M3phases} with a topological twist along $\Sigma$, see also \eqref{choices6d}.

Here, we return to another option listed in \eqref{choicesSigma}, namely $\Sigma = \R^2_q$,
which is more relevant to categorification of $U(N)$ (or $SU(N)$) 3-manifold invariants \eqref{HN3mfld} with $N>1$.
Although rather different at first sight, there are many parallels between 3-manifold homologies \eqref{HN3mfld} with $N=0$ and $N>1$.
Thus, as in section~\ref{sec:Amodel}, the $SL(2,\Z)$ action on flat connections will play an important role.
In particular, as we show very concretely, categorification of the Chern-Simons partition function on $M_3$ requires
writing it in a new basis, which is related to a more familiar basis of flat connection on $M_3$ by an $S$-transform.

\subsection{Categorification of $U(N)$ Chern-Simons from $T{[} M_3 {]}$ on $(\text{time}) \times (\text{cigar})$}

As a warm-up example,
consider the partition function of $U(N)_k$ Chern-Simons theory\footnote{Here, $k$ denotes the renormalized level $k=k_0+N$.} on $L(p,1)$.
It can be decomposed into a sum over different flat connections,
\begin{equation}
 Z_{U(N)_k\,\text{CS}}[L(p,1)]=e^{-\frac{2\pi i}{pk}\,\brho^2}\sum_{\bn \in \mathbb{Z}^N/p\mathbb{Z}^N\,/S_N}
e^{2\pi i k\,\frac{\bn^2}{2p}} \;Z_{U(N)_k\,\text{CS}}[L(p,1)]_\bn
\end{equation}
 where the contribution of the flat connection labelled by $\bn \in \mathbb{Z}^N/p\mathbb{Z}^N$ (modulo permutation) can be represented by the following integral \cite{Marino:2002fk}:
\begin{equation}
 Z_{U(N)_k\,\text{CS}}[L(p,1)]_\bn=\frac{1}
{\prod_{k=1}^pN_k^{(\bn)}!}
\int \prod_{i=1}^N \frac{d\sigma_i}{2\pi}\,
e^{\frac{p k i}{4\pi}(\sigma_i-2\pi i n_i/p)^2}\,
\prod_{i\neq j}2\sinh\frac{\sigma_i-\sigma_j}{2}
\end{equation}
where $N_k^{(\bn)}=\#\{j|n_j=k\}$. One can rewrite this expression as
\begin{equation}
 Z_{U(N)_k}[L(p,1)]_\bn=\frac{(\tau)^{N/2}}
{\prod_{k=1}^pN_k^{(\bn)}!}
\sum_{\bm\in\mathbb{Z}^N/p\mathbb{Z}^N}
S^{\mathbb{Z}^N,p}_{\bn,\bm}
\int\limits_{|x_i|=1} \prod_{i=1}^N \frac{dx_i}{2\pi i x_i}\,
\prod_{i\neq j}\left(1-\frac{x_i}{x_j}\right)
\Theta_{\bm}^{\mathbb{Z}^N,p}(\bx;q)
\label{CSLp1}
\end{equation}
where $\bx=\{x_i\}_{i=1}^N\in (\C^*)^N$ is the element of the maximal torus of $U(N)_\C$ and
\begin{equation}
 \tau=\frac{1}{k},\qquad q=e^{2\pi i\tau}
 \label{ktauq}
\end{equation}
\begin{equation}
 \Theta_{\bm}^{\mathbb{Z}^N,p}(\bx;q)=
\sum_{\br\in p\mathbb{Z}^N+\bm}q^{\frac{\br^2}{2p}}\,\bx^{\br},
\end{equation}
\begin{equation}
 S^{\mathbb{Z}^N,p}_{\bn,\bm}=p^{-\frac{N}{2}}e^{2\pi i \bm\cdot\bn/p}
\label{SZNp}
\end{equation}
and $\brho$ is the Weyl vector. Note that $e^{2\pi i k\,\frac{\bn^2}{2p}}$ is the classical contribution. Although the theta-function in the integrand of (\ref{CSLp1}) has an infinite number of terms, only a finite number of them give a non-trivial contribution after integration over $\bx$.
As a function of $\tau$, the contribution of a given flat connection has the following form
\begin{equation}
 q^{-\rho^2/p}e^{2\pi i k\,\frac{\bn^2}{2p}}Z_{U(N)_k\,\text{CS}}[L(p,1)]_\bn = \tau^{N/2}e^{\frac{\pi i \bn^2}{p\tau}}\;P_{\bn}(q^\frac{1}{p})
\label{Znoncat}
\end{equation}
where $P_\bn$ is a Laurent polynomial whose coefficients are \textit{algebraic} numbers. In particular,
\begin{equation}
 Z_{U(N)_k\,\text{CS}}[L(p,1)]_\mathbf{0} = Z_{U(N)_k\,\text{CS}}[S^3]_\mathbf{0}|_{q\rightarrow q^\frac{1}{p}}
\end{equation}

It is not quite clear how to categorify this quantity directly since, unlike the Jones polynomial,
it is not a polynomial in $q$ with integer coefficients.
However, using modular properties of the theta-function, one can rewrite it in a more promising way:
\begin{multline}
 e^{2\pi i k\,\frac{\bn^2}{2p}}\;Z_{U(N)_k\,\text{CS}}[L(p,1)]_\bn=\\
\frac{1}{\prod_{k=1}^pN_k^{(\bn)}!}
\int\limits_{|x_i|=1} \prod_{i=1}^N \frac{dx_i}{2\pi i x_i}\,
\prod_{i\neq j}\left(1-\frac{x_i}{x_j}\right)e^{-\frac{pi(\log \bx)^2}{4\pi\tau}}
\,\Theta_{\bn}^{\mathbb{Z}^N,p}(\tilde\bx;\tilde q)\,\tilde{q}^{-\frac{\bn^2}{2p}}
\label{CSside}
\end{multline}
where, as usual,
\begin{equation}
 \tilde{q}=q^{2\pi i\tilde{\tau}},\qquad \tilde\tau=-\frac{1}{\tau},
\end{equation}
\begin{equation}
 \tilde{x_i}=e^{2\pi i \xi_i/\tau},\qquad {x_i}=e^{2\pi i\xi_i}.
\end{equation}
Note that the classical factors and $\tau^{N/2}$ are all absorbed into a nice $\tilde q$-series by S-transform of the theta-functions.


Using the 3d/3d correspondence, we can formulate the partition function of
analytically continued $U(N)$ Chern-Simons theory on $M_3$ as the partition function of the fivebrane system \eqref{Msetup}
on $S^1 \times \R^2_q \times M_3$ where $\Sigma = \R^2_q$ is a ``cigar'' embedded in $Y_4 = TN_4 \cong \R^2_q\times \R^2_t$,
and $\tau$ plays the role of the equivariant parameter rotating the cigar.
Equivalently, it can be expressed as the so-called vortex partition function \cite{DGH} of the 3d $\CN=2$ theory $T[M_3]$
on $\R^2_q$ times a time circle, which sometimes is denoted by $D^2\times_q S^1$,
\be
S^1 \times \R^2_q \cong D^2\times_q S^1
\ee
where the disk $D^2$ is rotated by the angle $\mathrm{Re}\,\tau$ as one goes around $S^1$,
and with a certain boundary condition labelled by $\bn$.
The latter way of writing the vortex partition function is sometimes called the ``half-index'' since
it computes roughly half of the index on $S^2 \times S^1$ (and since the space-time itself is roughly half of $S^2 \times S^1$).

We wish to apply this to $M_3 = L(p,1)$ and to express $Z_{U(N)_k\,\text{CS}}[L(p,1)]$ as the the partition function of the ``Lens space theory'' \eqref{LspaceTheory}. The latter looks like \cite{Yoshida:2014ssa} (see also \cite{02branes}):
\begin{equation}
Z_{T[L(p,1)]}[D^2\times_q S^1]_\bn=\frac{1}{\prod_{k=1}^pN_k^{(\bn)}!}
\int\limits_{|x_i|=1} \prod_{i=1}^N \frac{dx_i}{2\pi i x_i} Z_\text{3d}(\bx) Z_\text{2d}^\bn(\bx)
\label{TM3side}
\end{equation}
where
\begin{equation}
 Z_\text{3d}=\frac{(q;q)_\infty^N}{(qy^2;q)_\infty^N}\prod_{i\neq j}\frac{(x_i/x_j;q)_\infty}{(x_i/x_jy^2q;q)_\infty}e^{-\frac{pi(\log \bx)^2}{4\pi\tau}}
\label{bulk}
\end{equation}
combines the bulk one-loop contributions of the vector multiplet, of the adjoint chiral multiplet with $R$-charge $2$ (we pick only the traceless part) and a classical contribution of the CS action. The extra parameter $y$ is the $U(1)_\beta$ fugacity, as in section \ref{sec:U1beta}.
The boundary contribution should be chosen in a way that cancels the anomaly inflow. This condition can be reformulated as a requirement for $Z_\text{2d}(\bx)$ to satisfy a certain difference equations with respect to $\bx$. Equivalently, $Z_\text{2d}(\bx)$ should be a section of $p$-th power of the prequantum line bundle over $\CM_\text{flat}(T^2,U(N)_\C)$. The theta-functions or form a natural basis of such sections:
\begin{equation}
 Z_\text{2d}^\bn(\bx)=\frac{\Theta_{\bn}^{\mathbb{Z}^N,p}(\tilde\bx;\tilde q)}{(q;q)_\infty^N}
\label{boundary}
\end{equation}
The partition function (\ref{TM3side}) with $y=1$ then matches with (\ref{CSside}) up to a universal simple factor:
\begin{equation}
  Z_{T[L(p,1)]}[D^2\times_q S^1]_\bn=\frac{Z_{U(N)_k\,\text{CS}}[L(p,1)]_\bn}{(q;q)_\infty^N}.
\end{equation}
As will be seen later it is often convenient to use ``reduced'' index of $T[M_3]$ with contribution of the Cartan part of the chiral multiplet in the adjoint representation factored out:
\begin{equation}
	 Z_\text{3d}^\red\equiv{(q;q)_\infty^N}\prod_{i\neq j}\frac{(x_i/x_j;q)_\infty}{(x_i/x_jy^2q;q)_\infty}e^{-\frac{pi(\log \bx)^2}{4\pi\tau}}
	\label{bulk}
\end{equation}
\begin{equation}
Z_{T[L(p,1)]}^\red[D^2\times_q S^1]_\bn\equiv\frac{1}{\prod_{k=1}^pN_k^{(\bn)}!}
\int\limits_{|x_i|=1} \prod_{i=1}^N \frac{dx_i}{2\pi i x_i} Z^\text{red}_\text{3d}(\bx) Z_\text{2d}^\bn(\bx)
\label{TM3side}
\end{equation}
Turning on $y\neq 1$ gives the refined Chern-Simons theory on $L(p,1)$.
For example, if we take $L(1,1)=S^3$, and factor out the contribution of the Cartan part of the chiral multiplet in the adjoint represenation:
\begin{equation}
 Z^\red_{T[L(1,1)]}[D^2\times_q S^1]_\mathbf{0} \propto  \prod_{j=1}^{N-1}\left(\frac{((qy^2)^j;q)_\infty}{((qy^2)^{j+1};q)_\infty}\right)^{N-j}=\frac{(qy^2;q)_\infty^{N-1}}{\prod_{k=2}^N((qy^2)^k;q)_\infty}.
 \label{ZS3-refined}
\end{equation}
The result coincides with the refined CS partition function from \cite{Aganagic:2011sg}. In particular, if $y^2=q^{\beta-1}$ with $\beta\in \mathbb{Z}_+$ (as in loc. cit.) we get
and
\begin{equation}
 Z^\red_{T[L(1,1)]}[D^2\times_q S^1]_\mathbf{0} \propto \prod_{k=0}^{\beta-1} \prod_{j=1}^{N-1}(1-(qy^2)^jq^k)^{N-j}
\end{equation}
Note, the factor
\begin{equation}
  \,\prod_{i=1}^N \frac{dx_i}{2\pi i x_i}\;\prod_{i\neq j}\frac{(x_i/x_j;q)_\infty}{(x_i/x_jy^2q;q)_\infty}
\end{equation}
in (\ref{TM3side}) can be understood as the measure orthogonalized by the MacDonald polynomials.

\subsection{Mock modularity and homological blocks}

Up to the anomalous 2d factor, the bulk contribution (\ref{bulk}) can be understood as the trace over the Hilbert space of 3d theory $T[M_3]$ on the ``cigar'' $D^2$ (with Neumann boundary conditions for the adjoint chiral), where the time direction runs along the $S^1$ factor in $D^2\times_q S^1$. On the other hand, the boundary contribution (\ref{boundary}) can be understood as the trace (over the integrable representation of $U(1)^N_p$) only if we treat $\tilde{S}^1\equiv \d D^2$ as the time circle, not the $S^1$. This is the source of difficulties with categorifying (\ref{Znoncat}) directly.

Instead, one can perform an $S$-transformation ({\it i.e.} exchange $S^1$ and $\tilde{S}^1$) of the boundary piece only.
This will lead us to different quantities
\begin{multline}
 \hat Z_{U(N)_k\,\text{CS}}[L(p,1)]_\bn
  \equiv\hat{Z}^\red_{T[L(p,1)]}[D^2\times_q S^1]_\bn\equiv\\
\int\limits_{|x_i|=1} \prod_{i=1}^N \frac{dx_i}{2\pi i x_i} Z^\red_\text{3d}(x) \hat{Z}_\text{2d}^\bn(x)\equiv
\int\limits_{|x_i|=1} \prod_{i=1}^N \frac{dx_i}{2\pi i x_i}
\prod_{i\neq j}\frac{(x_i/x_j;q)_\infty}{(x_i/x_jy^2q;q)_\infty}
\,\Theta_{\bn}^{\mathbb{Z}^N,p}(\bx;q)
\label{TM3mod}
\end{multline}
related to the original ones in (\ref{CSLp1}) via a linear transformation,
whose $(\bn,\bm)$ matrix elements appear in front of the integral (\ref{CSLp1}). Namely,
\begin{equation}
 Z_{U(N)_k\,\text{CS}}[L(p,1)]_\bn =
\frac{\tau^{N/2}}
{\prod_{k=1}^pN_k^{(\bn)}!}
\sum_{\bm\in\mathbb{Z}^N/p\mathbb{Z}^N}
S^{\mathbb{Z}^N,p}_{\bn,\bm}
\hat Z_{U(N)_k}[L(p,1)]_\bm
\label{Lp1-Stransform}
\end{equation}
The $S$-matrix defined in (\ref{SZNp}) is the same as the $S$-transform of $\hat{u}(1)_p^N$ characters. Note, here we consider the set of representations $\cong \Z_p^N$ modulo $S_N$ permutations. Since $\hat{u}(1)_p^N$ is level-rank dual to $\hat{su}(p)_1^N$, the $S$-matrices transforming their characters are inverse of each other. Therefore the $S$-transform (\ref{Lp1-Stransform}) is consistent with modular properties of VW partition function on the resolution of $A_{p-1}$ singularity whose boundary is $\d M_4=L(p,p-1)=-L(p,1)$, {\it cf.} footnote \ref{footnote:VW}.

The $S$-transform in (\ref{Lp1-Stransform}) has also the following physical interpretation. Take $V\in U(N),\; V^p=1$ to be a representative corresponding to $\bn \in \Z_p^N/S_N$. That is $V$ is the holonomy along the generator $H_1(M_3)$ in a particular flat connection background. Denote by $\mathrm{R}$ a linear representation of $U(N)$. Then, by analogy with \cite{Ooguri:1999bv} one expects the following decomposition of the Chern-Simons partition function in a particular background:
\begin{equation}
 Z_{U(N)_k\,\text{CS}}(V)
\propto \sum_{\mathrm{R}}\Tr_{\mathrm{R}}(V)
\hat Z_\mathrm{R}(q)
\end{equation}
where $Z_\mathrm{R}$ is a contribution of multi-particle states produced by M2-branes in representation $\mathrm{R}$ wrapping a non-trivial cycle in $M_3$ (that is $\mathrm{R}$ is the ``total'' effective representation of the multiparticle state). When $V$ is unconstrained the characters $\Tr_{\mathrm{R}}(V)$ can be realized as symmetric polynomials in $\{v_i\}_{i=1}^N$, eigenvalues of $V$. The condition $V^p=1$ implies $v_i^p=1$. The ring of such polynomials is
\begin{equation}
	\CH_{T[M_3]}[T^2]\cong \C[v_1,\ldots,v_N]/\{v_i^p=1\}_{i=1}^N\cong \C[\Z_p^N/S_N]
\end{equation}
so they can be labelled by $\bm\in \Z_p^N/S_N$ so that their values coincide with the values of the $S$-matrix in (\ref{Lp1-Stransform}) given $v_i=e^{2\pi in_i/p}$.

As usual, the integral over $x_i$ in the expressions like \eqref{TM3mod} or \eqref{ZhatyL31}
corresponds to taking gauge-invariant combinations of the operators accounted by the integrand.
The latter, in turn, contains $q$-Pochhammer factors
\be
(x;q)_{\infty} = \prod_{n=0}^{\infty} (1 - x q^n)
\ee
that correspond to bosonic and fermionic Fock spaces (\ref{chiralFock})
in 3d $\CN=2$ theory $T[M_3]$, depending on whether they appear in the denominator or numerator.
Finally, the theta-functions in the integrals like \eqref{TM3mod} or \eqref{ZhatyL31}
correspond to 2d degrees of freedom at the boundary of the 3d space-time $D^2\times_q S^1$, {\it cf.} \cite{Gadde:2013wq}.

The new expressions (\ref{TM3mod}) have the advantage that now they can be interpreted as traces over certain vector spaces.
In particular, unlike (\ref{Znoncat}), the new expression \eqref{TM3mod}
has expansion in $y$ and $q$ with \textit{integer powers} (up to an overall rational power of $q$) and \textit{integer coefficients}. In the unrefined case ($y=1$) we have:
\begin{equation}
 \hat Z_{U(N)_k\,\text{CS}}[L(p,1)]_\bn \;\in q^{\frac{\bn^2}{2p}}\Z[q]
\label{hatZLp1}
\end{equation}
For example:
\begin{equation}
 \hat{Z}_{T[L(p,1)]}[D^2\times_q S^1]_\mathbf{0}=
N!\prod^{N-p}_\text{\footnotesize $\begin{array}{c} j=1 \\ j=N\mod p \end{array}$}
(1-q^{N-j})^j
\end{equation}
One can also consider corresponding ``unreduced'' quantities produced by (\ref{TM3mod}) but with kept contribution of the Cartan part of the chiral multiplet in adjoint representation:
\begin{equation}
	\hat Z^\text{(unred)}_{U(N)_k\,\text{CS}}[L(p,1)]_\bn\equiv \frac{1}{(qy^2;q)^N_\infty}\,\hat Z_{U(N)_k\,\text{CS}}[L(p,1)]_\bn
\end{equation}

At the end of this section we will give explicit examples of how integrality can be restored in CS partition function for various 3-manifolds.
In order to do this, we now wish to formulate a general proposal for what should be the $S$-transform for arbitrary rational homology spheres. We verify in various examples that it produces ``categorifiable'' quantities similar to (\ref{hatZLp1}) from the original ``non-categorifiable'' CS partition function. Along the way, we put in a natural context (of the categorification program) various explicit expressions for the CS partition function obtained previously in the literature and their intriguing properties, including connection to Mock modular forms and Eichler integrals.

Specifically, in \cite{hikami2011decomposition} it was proposed (based on many examples from \textit{loc. cit.} and \cite{Hikami,hikami2006quantum,lawrence1999modular}) that for rational homology spheres the $SU(2)$ Chern-Simons partition function (a.k.a. RTW invariant\footnote{The $SU(2)$ CS parition that usually appears in the physics literature and that we use here has a different normalization compared to RTW invariant $\tau_k(M_3)$ which usually apears in the math literature:
\begin{equation}
 Z_{SU(2)_k\,\text{CS}}(M_3)=\frac{\tau_k(M_3)}{\tau_k(S^2\times S^1)}
\end{equation}
where
\begin{equation}
 \tau_k(S^2\times S^1)=\sqrt\frac{2}{k}\frac{1}{\sin\frac{\pi}{k}}
\end{equation}
  }) has the following decomposition:
\begin{equation}
 Z_{SU(2)_k\,\text{CS}}[M_3]=\frac{q^{\Delta}}{\sqrt{8k}}\sum_{a\in H_1(M_3)/\Z_2}e^{2\pi i k S_a} Z_a(q)|_{\tau \searrow 1/k}
\label{ZSU2red}
\end{equation}
where $S_a=\lambda_{M_3}(a,a)$ are the diagonal values of the linking form\footnote{For general $M_3$ the linking form is a bilinear part on the torsion part of $H_1(M_3,\Z)$:
\begin{equation}
	\lambda_{M_3}:\text{Tors}\,H_1(M_3)\otimes \text{Tors}\,H_1(M_3) \rightarrow \Q/\Z
\end{equation}
For a rational homology sphere $M_3$ it can be defined as follows. Consider $a\in \text{Tors}\,H_1(M_3)$ and $s \in \Z$, such that $sa=0$. Then, there exists a 2-chain $B$ such that $\d B=sa$. The value of the linking form between $a$ and $a'\in \text{Tors}\,H_1(M_3)$ then equals
\begin{equation}
	\lambda_{M_3}(a,a')=\frac{\#(B\cap a')}{s}.
\end{equation}
For plumbed rational homology spheres considered in section \ref{sec:plumbed}, the linking form on $H_1(M_3)=\Lambda^*/\Lambda$ is given by extension of the intersection form on lattice $\Lambda$ to $\Lambda^*\subset \Q\otimes_\Z \Lambda$.}
on $H_1(M_3)$, $Z_a(q)$ are $q$-series convergent for $\tau$ in the upper-half plane,
and $\tau \searrow 1/k$ means taking a limit from above.
Note, that the values $S_a$ can also be interpreted as values of Chern-Simons action for reducible flat connections. Reducible flat connections correspond to representations of $\pi_1(M_3)$ into $U(1)$ subgroup of $SU(2)$. They are labeled by elements of $H_1(M_3)$ modulo $\Z_2$ Weyl symmetry action. The latter will result in technical differences between the cases where $2$ is a divisor of $0$ in $H_1(M_3)$ or not. For the sake of simplicity in what follows we consider the case when it is not\footnote{In practical, $2$ is not a divisor of zero \text{iff} $H_1(M_3)\cong \Z_{p_1} \oplus \Z_{p_2} \oplus \ldots$ where all $p_i$ are odd.}. The quantities $Z_a(q)$ in general are contributions to CS path integral from the corresponding reducible flat connection \textit{plus} the contributions of irreducible flat connections. This structure will be discussed in more detail at the end of this section.

In general, {\it i.e.} when $M_3$ is not an integer homology sphere, the $q$-series $Z_a(q)$ have algebraic coefficients and contain powers of $q$ with exponents which differ by rational numbers and, therefore, still not suitable for categorification. However, we want to make the following conjecture: there exists a $k$-independent $S$-matrix such that:
\begin{equation}
 Z_a(q)=\sum_{a}S_{ab} \hat Z_a(q)
 \label{SZhat}
\end{equation}
where\footnote{When $2$ is a divisor of $0$ in $H_1(M_3)$ we expect
\begin{equation}
 \hat Z_a(q)\in q^{\Delta_a}\Z[[q^{1/2}]],\qquad \Delta_a\in \mathbb{Q}
\end{equation}
}
\begin{equation}
 \hat Z_a(q)\in q^{\Delta_a}\Z[[q]],\qquad \Delta_a\in \mathbb{Q}
\label{ZSU2integrality}
\end{equation}
and, moreover, the $S$-matrix depends only on $H_1(M_3)$.

For example, in the case when $H_1(M_3)=\Z_p$ and $p$ is odd, the $S$-matrix given by the appropriate $\Z_2$ reduction of the $\hat{u}(1)_p$ $S$-matrix already appeared earlier in the text:
\begin{equation}
 S_{ab}=\frac{e^{\frac{2\pi i ab}{p}}+e^{-\frac{2\pi i ab}{p}}}{1+\delta_{a,0}}=\frac{2\cos\frac{2\pi ab}{p}}{1+\delta_{a,0}},\qquad a,b\in 0,\ldots,(p-1)/2
\label{SSU2}
\end{equation}
It is easy to check that the matrix (\ref{SSU2}) squares to the identity. In general, $H_1(M_3)$ is a product of cyclic groups and the $S$-matrix is the corresponding tensor product of matrices (\ref{SSU2}). Similarly to the case of lens spaces considered in the previous section on can introduce unreduced quantities, which may be more appropriate for categorificarion:
\begin{equation}
	\hat Z^\text{(unred)}_a(q)\equiv \frac{\hat Z_a(q)}{(q;q)_\infty}\in q^{\Delta_a}\Z[[q]]
\end{equation}

We support this conjecture by verifying it explicitly in various examples of spherical Seifert fibered 3-manifolds. For many such examples, $M_3$ can be realized as a link of a singularity $f(x,y,z)=0$, and the corresponding expression for $Z_{SU(2)_k \,\text{CS}}$ is explicitly written in \cite{hikami2011decomposition,hikami2006quantum}. Using these expressions we find a perfect agreement with our general conjecture.

The fact that the structure of the decomposition (\ref{ZSU2red}) and the $S$-transform (\ref{SZhat}) is essentially dictated only by $H_1(M_3)$ is quite mysterious but might have the following physical interpretation. When one considers multiparticle states from several M2 branes wrapping non-trivial cycles in $M_3$ one expects them to be labelled by elements of $H_1(M_3)$, not (conjugacy classes of) $\pi_1(M_3)$, since there is no natural order between different particles.

Before we proceed to examples, let us introduce the following $q$-series convergent for $\tau$ in the upper-half plane:
\begin{equation}
 \Tilde \Psi_p^{(a)}(q)=\sum_{n=0}^\infty \psi^{(a)}_{2p}(n)q^{\frac{n^2}{4p}}\qquad \in q^\frac{a^2}{4p}\,\Z[[q]]
\label{Psis}
\end{equation}
where
\begin{equation}
 \psi^{(a)}_{2p}(n)=\left\{
\begin{array}{cl}
 \pm 1, & n\equiv \pm a\mod 2p \\
0, & \text{otherwise}
\end{array}\right.
\end{equation}
These $q$-series are Eichler integrals of certain weight-$3/2$ modular forms
\begin{equation}
   \Psi_p^{(a)}(q)=\frac{1}{2}\sum_{n\in \Z} n\psi^{(a)}_{2p}(n)q^{\frac{n^2}{4p}}
\end{equation}
and exhibit Mock modular properties (see {\it e.g.} \cite{hikami2006quantum}).
For the 3-manifolds that we are going to consider, the $q$-series $Z_a(q)$ in
(\ref{ZSU2red}) can be expressed as a linear combinations (with algebraic coefficients)
of $\tilde{\Psi}_p^{(a)}(q)$ for a fixed $p$.
In what follows, we use the same notation for 3-manifolds and the corresponding surface singularities.

We first list examples with $H_1(M_3)=\Z_3$. The $S$-matrix (\ref{SSU2}) reads
\begin{equation}
 S=\frac{1}{\sqrt{3}}\left(
\begin{array}{cc}
 1 & 1 \\
 2 & -1 \\
\end{array}
\right)
\end{equation}

\begin{enumerate}

\item
The simplest example of a 3-manifold with $H_1(M_3)=\Z_3$ is a Lens space $M_3=L(3,1)$,
which can be thought of as the link of a surface singularity $f(x,y,z)=x^3+yz$.
The $S$-transform of the Chern-Simons partition function
\begin{equation}
 \sqrt{2k}\,Z_{SU(2)_k \,\text{CS}}[L(3,1)]=\frac{1}{\sqrt{3}}(q^{1/3}-1)+\frac{e^{2\pi ik/3}}{\sqrt{3}}(-2-q^{1/3})
\end{equation}
gives
\begin{equation}
 \hat Z_0=-2,\qquad \hat Z_1=2q^{1/3}
\label{ZaunrefL31}
\end{equation}
The corresponding quantities refined by the $U(1)_\beta$ fugacity $y$ can be obtained from the $SU(2)$ analogue of the formula (\ref{TM3mod}):
\begin{equation}
\label{ZhatyL31}
\hat Z_a=
	-\int\limits_{|x|=1}  \frac{dx}{2\pi i x}
	\frac{(x^2;q)_\infty}{(x^2y^2q;q)_\infty}\frac{(x^{-2};q)_\infty}{(x^{-2}y^2q;q)_\infty}
	\,\sum_{n\in\Z}q^{\frac{(3n+a)^2}{3}}\,x^{2(3n+a)}
\end{equation}
To obtain ``naive Poincar\'e polynomials'' of the Khovanov homology \eqref{Poincare-trace},
it is convenient to write these $q$-expansions in terms of the variable $t = - y^2$,
which can be interpreted as a fugacity for the $U(1)_t$ R-symmetry, {\it cf.} section \ref{sec:U1beta}.
For the Lens space $L(3,1)$ we find
\begin{equation}
	 \hat Z_0 = -2(1+q(1+t)+q^2(2+3t+t^2)+\ldots)
\label{ZarefL31}
\end{equation}
$$
\hat Z_1=2q^{1/3}(1+q(2+2t)+q^2(2+4t+2t^2)+\ldots)
$$
As this example ideally illustrates, the refined ``homological blocks'' \eqref{ZarefL31} have a lot more
terms compared to their unrefined cousins \eqref{ZaunrefL31}. The fact that all coefficients are positive integers supports hypothesis that they indeed can give us Poincar\'e polynomials. However, to be more conservative, one should only say that they provide lower bounds on the homology dimensions for a given $U(1)_q$ grading.  The way the specialization to $t=-1$ works is
that many terms cancel in pairs {\it a la} \cite{Gukov:2015gmm}:
\be
\hat Z_a  \; = \; \hat Z_a\vert_{t=-1} + (1+t) \cdot P_{a,+} (q,t)
\ee
where $P_{a,+} (q,t) \in \Z_+ [[q,t]]$.
This seems to be a part of the general structure of categorification \eqref{ZCSfromP}.

One can also keep the contribution of the Cartan part of the chiral in adjoint representationconsider corresponding unreduced refined ``blocks'' defined as
\begin{equation}
	\hat Z^\text{(unred)}_a=\frac{\hat Z_a}{(-qt;q)_\infty}
	\label{ZCSunred}
\end{equation}
These quantites are more natural from the index of $T_{SU(2)}[M_3]$ and, as we will see in the next section, from the M-theory point of view after geometric transition. However their unrefined limit reproduces blocks $Z_a$ in the CS partition function with an extra factor:
\begin{equation}
	\hat Z^\text{(unred)}_a|_{t=-1}=\frac{\hat Z_a}{(q;q)_\infty}
\end{equation}
The denominator can be interpreted as the contribution of point-like instantons in twisted $\CN=4$ SYM on $\R_+\times M_3$ realizing CS on its boundary. This extra factor in (\ref{ZCSunred}) will not affect positivity of the coefficients in the $q$-expansion:
\begin{equation}
	 \hat Z_0^\text{(unred)} = -2(1+q+q^2(2+t+t^2)+q^3(3+2t+2t^2)+\ldots)
\end{equation}
$$
\hat Z_1^\text{(unred)}=2q^{1/3}(1+q(2+t)+q^2(2+t+t^2)+\ldots)
$$
and therefore can be naively interpreted as Poincar\'e polynomials. The same holds for any $M_3=L(p,1)$ in the case $G=SU(2)$. Further research should answer questions about mathematical meaning of these two versions, reduced or unreduced one.

\item
The Brieskorn sphere
$M_3=\Sigma (2,3,4)$ corresponds to $f(x,y,z)=x^4+y^3+z^2$ and has the following set of invariants
\begin{multline}
 \sqrt{2k}\,q^{25/24}\,Z_{SU(2)_k \,\text{CS}}[\Sigma (2,3,4)]=
\frac{1+2e^{2\pi ik/3}}{\sqrt{3}}\,q^{1/24}\\-
\frac{1+2e^{2\pi ik/3}}{2\sqrt{3}}(\tilde\Psi^{(1)}_6(q)+\tilde\Psi^{(5)}_6(q))
-\frac{1-e^{2\pi ik/3}}{\sqrt{3}}\,\tilde\Psi^{(3)}_6(q)
\end{multline}
\begin{equation}
 \hat Z_0=2q^{1/24}-\tilde\Psi^{(1)}_6(q)+\tilde\Psi^{(5)}_6(q)\;\in q^{1/24}\Z[[q]],
\qquad \hat Z_1=-2\tilde\Psi^{(3)}_6(q)\;\in q^{3/8}\Z[[q]]
\end{equation}

\item
Another Brieskorn sphere with $H_1(M_3)=\Z_3$ is $M_3=\Sigma (2,3,8)$, the link of a surface singularity $f(x,y,z)=x^8+y^3+z^2$:
\begin{multline}
 \sqrt{2k}\,q^{-1/48}\,Z_{SU(2)_k\,\text{CS}}[\Sigma (2,3,8)]=
\frac{1}{2\sqrt{3}}
(\tilde\Psi^{(1)}_{12}(q)-\tilde\Psi^{(7)}_{12}(q)-2\tilde\Psi^{(9)}_{12}(q))
\\
+\frac{e^{2\pi ik/3}}{\sqrt{3}}
(\tilde\Psi^{(1)}_{12}(q)-\tilde\Psi^{(7)}_{12}(q)+\tilde\Psi^{(9)}_{12}(q))
\end{multline}
\begin{equation}
 \hat Z_0=\tilde\Psi^{(1)}_{12}(q)-\tilde\Psi^{(7)}_{12}(q)\;\in q^{1/48}\Z[[q]],
\qquad \hat Z_1=-2\tilde\Psi^{(9)}_{12}(q)\;\in q^{11/16}\Z[[q]]
\end{equation}

\item
The link of a surface singularity $f(x,y,z)=x^4+y^3+xz^2$ is $M_3=Q_{10}$, with the following invariants:
\begin{multline}
 \sqrt{2k}\,q^{71/72}\,Z_{SU(2)_k\,\text{CS}}[Q_{10}]=
\frac{1}{2\sqrt{3}}
(\tilde\Psi^{(1)}_{18}(q)-\tilde\Psi^{(5)}_{18}(q)-\tilde\Psi^{(13)}_{18}(q)+\tilde\Psi^{(17)}_{18}(q))
\\
+\frac{e^{-2\pi ik/3}}{2\sqrt{3}}
(2\tilde\Psi^{(1)}_{18}(q)+\tilde\Psi^{(5)}_{18}(q)+\tilde\Psi^{(13)}_{18}(q)+2\tilde\Psi^{(17)}_{18}(q))
\end{multline}
\begin{equation}
 \hat Z_0=\tilde\Psi^{(1)}_{18}(q)+\tilde\Psi^{(17)}_{18}(q)\;\in q^{1/72}\Z[[q]],
\qquad \hat Z_1=-\tilde\Psi^{(5)}_{18}(q)-\tilde\Psi^{(13)}_{18}(q)\;\in q^{25/72}\Z[[q]]
\end{equation}

\noindent
We now consider examples with $H_1(M_3)=\Z_5$:
\begin{equation}
 S=\frac{1}{\sqrt{5}}
\left(
\begin{array}{ccc}
 1 & 1 & 1 \\
 2 & \frac{1}{2} \left(-\sqrt{5}-1\right) & \frac{1}{2} \left(\sqrt{5}-1\right) \\
 2 & \frac{1}{2} \left(\sqrt{5}-1\right) & \frac{1}{2} \left(-\sqrt{5}-1\right) \\
\end{array}
\right)
\end{equation}

\item
The simplest example of a 3-manifold with $H_1(M_3)=\Z_5$ is the Lens space $M_3=L(5,1)$,
which can be realized as a link of a surface singularity $f(x,y,z)=x^5+yz$.
It has
\begin{multline}
 \sqrt{2k}\,q^{-1/2}\,Z_{SU(2)_k\,\text{CS}}[L(5,1)]=\\
\frac{1}{\sqrt{5}}(q^{1/5}-1)+\frac{e^{2\pi i k/5}}{2\sqrt{5}}((-1-\sqrt{5})q^{1/5}-4)+\frac{e^{-2\pi i k/5}}{2\sqrt{5}}((-1+\sqrt{5})q^{1/5}-4)
\end{multline}
\begin{equation}
 \hat Z_0=-2,
\qquad \hat Z_1=2q^{1/5},
\qquad \hat Z_2=0
\end{equation}
The corresponding refined quantities read
\begin{equation}
    \hat Z_0=-2(1+q(1+t)+q^2(2+3t+t^2)+ \ldots),
\end{equation}
\begin{equation}
    \hat Z_1=2q^{1/5}(1+q(2+3t)+q^2(3+5t+2t^2)+\ldots),
\end{equation}
\begin{equation}
    \hat Z_2=-2q^{4/5}(0+q(1+t)+q^2(1+3t+2t^2)+\ldots),
\end{equation}
while their unreduced versions are
\begin{equation}
    \hat Z_0^\text{(unred)}=-2(1+q+q^2(2+t+t^2)+q^3(3+2t+2t^2)+ \ldots),
\end{equation}
\begin{equation}
    \hat Z_1^\text{(unred)}=2q^{1/5}(1+q(2+t)+q^2(3+2t+t^2)+\ldots),
\end{equation}
\begin{equation}
    \hat Z_2^\text{(unred)}=-2q^{4/5}(0+q(1+t)+q^2(1+2t+t^2)+\ldots),
\end{equation}

\item
The Brieskorn sphere $M_3 = \Sigma (2,4,5)$ corresponds to $f(x,y,z)=x^5+y^4+z^2$ and has quantum invariants
\begin{equation}
 \sqrt{8k}\,q^{19/40}\,Z_{SU(2)_k\,\text{CS}}[\Sigma (2,4,5)]=
Z_0(q)+e^{4\pi i k/5}Z_1(q)+e^{-4\pi i k/5}Z_2(q)
\end{equation}
\begin{equation}
 \hat Z_0=\tilde\Psi^{(1)}_{10}(q)+\tilde\Psi^{(9)}_{10}(q)\;\in q^{1/40}\Z[[q]],
\qquad \hat Z_1=0,
\qquad \hat Z_2=-2\tilde\Psi^{(5)}_{10}(q)\;\in q^{5/8}\Z[[q]]
\end{equation}

\end{enumerate}

Now let us make a few comments about contribution of irreducible flat connections. As was exploited in \cite{lawrence1999modular,Hikami,hikami2006quantum,hikami2011decomposition}, the large-$k$ asymptotics of the CS partition function can be easily found starting from decomposition (\ref{ZSU2red}) and using mock-modular properties of functions (\ref{Psis}):
\begin{equation}
 \tilde\Psi^{(a)}_p(q)=
-\sqrt\frac{k}{i}\sum_{b=1}^{p-1}M_{ab}\tilde\Psi^{(b)}_p(e^{-2\pi i k})
+\sum_{n=0}^\infty \frac{L(-2n,\psi^{(a)}_{2p})}{n!}\left(\frac{\pi i}{2pk}\right)^n
\label{Psis-S}
\end{equation}
where
\begin{equation}
 L(-n,\psi^{(a)}_{2p})=-\frac{(2p)^n}{n+1}\sum_{m=1}^{2p}\psi^{(a)}_{2p}B_{n+1}\left(\frac{n}{2p}\right),
\end{equation}
\begin{equation}
 M_{ab}=\sqrt\frac{2}{p}\sin\frac{\pi ab}{p}
\end{equation}
and
\begin{equation}
 \tilde\Psi^{(a)}_p(e^{-2\pi i k})=(1-a/p)\,e^{\frac{\pi i k a^2}{2p}}
\end{equation}
for integer $k$. It follows that $\hat Z_a(q)$, their linear combinations, will have asymptotics $k\rightarrow \infty$ of the following form:
\begin{equation}
 \hat Z_a(q)=\sqrt\frac{k}{i} M'_{ab}\sum_b  e^{2\pi i k C_a}+\hat W_a(q)
\end{equation}
where $M'_{ab}$ are algebraic $k$-independent numbers, $C_a\in\Q$ and
\begin{equation}
  \hat W_a(q)\in q^{\Delta_a}\Q[[q-1]]
\end{equation}
Therefore the asymptotics of the total partition function has the following form:
\begin{equation}
  \frac{\sqrt{8k}}{q^{\Delta}}\,Z_{SU(2)_k\,\text{CS}}[M_3]=\sum_{a\in H_1(M_3)/\Z_2}e^{2\pi i k S_a}\,\left(\sum_{b\in H_1(M_3)/\Z_2} S_{ab}\hat W_b(q)\right)+
\sqrt\frac{k}{i}\sum_{c}e^{2\pi i k S_c} A_c
\label{WRTasympt}
\end{equation}
where the sum over $a$ correspond to the sum over reducible flat connections and the sum over $c$ is the sum over irreducible flat connections. Again, $A_c$ are algebraic $k$-independent numbers.

The contribution from the trivial flat connection (the $a=0$ term) is an element of $\Q[[q-1]]$ and is known as Ohtsuki series (perturbative RTW invariant) in the math literature (see {\it e.g.} \cite{ohtsuki1996polynomial}). For integer homology spheres, it is an element of $\Z[[q-1]]$ \cite{rozansky1998p,habiro2002quantum}. Let us stress an important difference between decompositions (\ref{WRTasympt}) and (\ref{ZSU2red}) of the Chern-Simons partition function. The right-hand side of (\ref{WRTasympt}) is an asymptotic expression in the $k\rightarrow\infty$ limit with coefficients in front of each $e^{2\pi ikS_a}$ being asymptotic series in $1/k$, or, equivalently in $q-1$, while the right-side of (\ref{ZSU2red}) is an exact expression with coefficients being the convergent series in $q$.

As an example, consider the Poincar\'e sphere $M_3=\Sigma (2,3,5)$ which has only one reducible flat connection (the trivial one) and two irreducible flat connections. The partition function is given by \cite{lawrence1999modular}:
\begin{multline}
 \sqrt{2k}\,q^{181/120}\,Z_{SU(2)_k\,\text{CS}}[\Sigma (2,3,5)]=
q^{1/120}-\\-\frac{1}{2}(\tilde\Psi^{(1)}_{30}(q)+\tilde\Psi^{(11)}_{30}(q)+\tilde\Psi^{(19)}_{30}(q)+\tilde\Psi^{(29)}_{30}(q))\;\in\frac{1}{2}\,q^{1/120}\Z[[q]]
\end{multline}
and has the following asymptotic expansion:
\begin{equation}
 \sqrt{2k}\,q^{1/2}\,Z_{SU(2)_k\,\text{CS}}[\Sigma (2,3,5)]=\sum_{n=0}^\infty q^n(q^n)_n+S'_{11}e^{-\pi ik/60}+S'_{21}e^{-49\pi ik/60 }
\label{E8asympt}
\end{equation}
Where $(x)_n\equiv(1-x)\ldots (1-xq^{n-1})$. The first term in (\ref{E8asympt}) is the contribution of the trivial flat connection and $S'$ is the $S$-matrix
\begin{equation}
 S'=\sqrt\frac{k}{i}\,\frac{2}{\sqrt{5}}\left(
\begin{array}{cc}
 \sin\frac{\pi}{5} & \sin\frac{2\pi}{5} \\
 \sin\frac{2\pi}{5} & -\sin\frac{\pi}{5} \\
\end{array}
\right)
\end{equation}
acting on the vector
\begin{equation}
 \left(\begin{array}{c}
 \tilde\Psi^{(1)}_{30}(q)+\tilde\Psi^{(11)}_{30}(q)+\tilde\Psi^{(19)}_{30}(q)+\tilde\Psi^{(29)}_{30}(q)  \\
 \tilde\Psi^{(7)}_{30}(q)+\tilde\Psi^{(13)}_{30}(q)+\tilde\Psi^{(17)}_{30}(q)+\tilde\Psi^{(23)}_{30}(q)
\end{array}\right)
\end{equation}
according to (\ref{Psis-S}). That is the $S$-transform of the contributions from irreducible flat connections with the matrix $S'$ gives integer valued vector $(1,0)$. This property is general for Brieskorn integer homology spheres \cite{Hikami}.

\subsection{Back to Heegaard Floer homology}

In this section we show that similarly to the case of knots it might be possible to reconstruct Heegaard Floer homology starting from Khovanov homology. Let us start with the (reduced version of) proposed Poincar\'e polynomial for $S^3$:
\begin{equation}
	\hat Z_{SU(2)_k\,\text{CS}}[S^3]_0=\frac{(-qt;q)_\infty}{(q^2t^2;q)_\infty}=\frac{(-qt;q)_\infty}{(at^2;q)_\infty}
	\label{ZS3-a-refined}
\end{equation}
where we naively restored an extra fugacity $a=q^N$ with $N=2$ which corresponds to the choice of $SU(2)$ gauge group. This is the fugacity for $U(1)_a$ grading similar to the extra $U(1)_a$ grading that appears in the triply graded HOMFLY-PT homology for knots. As will be evident in the next section this way to resore the extra grading, or equivalently, the way $a$ appears in the right hand side of (\ref{ZS3-a-refined}) agrees with counting of BPS states of M-theory in the resolved conifold background.

First let us consider a spectral sequence from $SU(2)$ Khovanov-Rozansky homology to its $SU(1)$ version, which should be a trivial one. It can be realized by deforming a differential (or $Q$ operator in the physical language) by the following one:
\begin{equation}
	d_1:\;\;(-1,1,-1)_{a,q,t}
\end{equation}
where 3 numbers on the right denote its $U(1)_a$, $U(1)_q$ and $U(1)_t$ gradings correspondingly. In case of Khovanov homology for knots, the analogous differential to $N=1$ theory was constructed by Lee \cite{lee2005endomorphism}. Since
\begin{equation}
	\hat Z_{SU(2)_k\,\text{CS}}[S^3]_0|_{a=-q/t}=1
\end{equation}
The Poincar\'e polynomial of the new homology w.r.t. to the deformed differential is then given by the first term in the right hand side of the following decomposition formula and indeed corresponds to a trivial space:
\begin{equation}
\hat Z_{SU(2)_k\,\text{CS}}[S^3]_0=1+(1+at/q)\,(\ldots)
\end{equation}
where $(\ldots)$ has expansion with positive coefficients.

Now consider deformation by the following differential:
\begin{equation}
	d_0:\;\;(-1,0,-1)_{a,q,t}	
\end{equation}
One can similarly decompose (\ref{ZS3-a-refined}) as follows:
\begin{equation}
	\hat Z_{SU(2)_k\,\text{CS}}[S^3]_0=\frac{1}{1-at^2}+(1+at)(\ldots)
\end{equation}
The first term agrees with the Poincar\'e polynomial of $HF^+(S^3)$ (cf. formula (\ref{HFLens-Poincare})). It would be interesting to a similar spectral sequence that relates categorification of WRT invariant with $HF^+$ for some other rational homology spheres.


\section{Homology / BPS spectrum from M-theory}
\label{sec:Mtheory}

In this section, we approach the space of BPS states \eqref{HN3mfld} in the setup \eqref{M3phases} from
the vantage point of the Calabi-Yau 3-fold.
Since (closed) refined BPS invariants now have a rigorous mathematical definition \cite{Kontsevich:2008fj},
this gives an opportunity to define mathematically 3-manifold homologies, at least on the resolved / triply-graded side,
as motivic (= refined \cite{Dimofte:2009bv}) invariants of the Calabi-Yau 3-fold \eqref{CY3fromM3} labeled by $M_3$.
Of course, as we already mentioned in the introduction, the map \eqref{CY3fromM3} is expected to exist
only for a certain class of 3-manifolds and identifying this class will be one of the goals in the present section.

Studying the BPS Hilbert space associated for each manifold $M^3$ for each fixed $N$ would be cumbersome.
Instead it would be helfpul if there is some regularity property that the existence of the large $N$ dual predicts for this. This has beed used effectivly in the context
of knot homology invariants \cite{Gukov:2004hz}. It would be natural to ask for its extension to the BPS Hilbert space associated with closed 3-manifold.


Consider the simplest case of closed 3-manifold $M_3=S^3$. It is known \cite{Gopakumar:1998ki} that large number of A-branes wrapped around $S^3$, is dual to the topological string on the resolved
condifold $O(-1)+O(-1)\rightarrow \mathbb{P}^1$.  This duality which was orignially studied in the context of large $N$ Chern-Simons theory
continues to hold for refined version as well \cite{Aganagic:2011sg}.  Moreover, this duality can be viewed as the statement of equality even
for fixed finite $N$ in the refined case \cite{AganagicPC}.

In what follows we review this connection and try to generalize it.  We also discuss its implications for the structural properties of
the Hilbert space associated with the refined Chern-Simons theory, by the existence of the large $N$ dual theory.

\subsection{BPS spectrum for resolved conifold}

Consider the case when Calabi-Yau threefold $X$ is resolved conifold in more detail. We are interested in studying BPS spectra of M-theory in $\R\times \text{TN}_4 \times X$ background. Denote by $q_{1,2}$ fugacities\footnote{Compared to the notations of  \cite{Aganagic:2011sg} $q_1=t_\text{there}$, $q_1=q_\text{there}$} for  $U(1)_{q_1}\times U(1)_{q_2}$ symmetry acting on $\text{TN}_4\cong \C_{q_1}\times \C_{q_2}$. They are related as follows to the fugacities appeared in section \ref{sec:higherRank}:
\begin{equation}
	q_1=q\qquad q_2=-tq=y^2q
\end{equation}

The BPS states are realized as M2-branes or, via reduction to type IIA, as D0-D2 bound states. The contribution from D2 branes to the partition function of refined top strings on resolved conifold (in the large volume chamber) is given by
\begin{equation}
	Z_\text{top}^{\text{D2}}=\exp\left\{-\sum_{k=1}^\infty \frac{Q^k}{k(q_1^{k/2}-q_1^{-k/2})(q_2^{k/2}-q_2^{-k/2})}\right\}
\end{equation}
where $\log Q$ is complexified K\"ahler modulus of the base $\mathbb{P}^1$. The full partition function which takes into account the contribution of D0 branes then reads:
\begin{equation}
	Z_\text{top}=\frac{Z_\text{top}^{\text{D2}}}{Z_\text{top}^{\text{D2}}|_{a=q_2^{1/2}q_1^{-1/2}}}.
	\label{Ztoprefined}
\end{equation}
After introducing a rescaled fugacity $a=Q\,q_1^{1/2}\,q_2^{-1/2}$ the numerator can be rewritten as follows:
\begin{equation}
	Z_\text{top}^\text{D2}=\exp\left\{-\sum_{k=1,i}^\infty \frac{i}{k}\,\frac{a^k\,q_2^{-ik}(1-q_2^k)}{(1-q_1^k)}\right\}=
	\prod_{i=1}^{\infty}\left(\frac{(aq_2^{-i};q_1)_\infty}{(aq_2^{-i+1};q_1)_\infty}\right)^i=\prod_{i=0}^{\infty}\frac{1}{(aq_2^{-i};q_1)_\infty}
\end{equation}
The full topological string partition function (\ref{Ztoprefined}) then reads
\begin{equation}
	Z_\text{top}(a,q_1,q_2)=\prod_{i=0}^{\infty}\frac{(q_2^{-i};q_1)_\infty}{(aq_2^{-i};q_1)_\infty}
	\label{Ztoprefinedfull}
\end{equation}
Now, if one sets $a=q_2^N$ where $N$ is a positive integer it agrees with the ``unreduced'' version of the refined CS partition function on $S^3$ (\ref{ZS3-refined}):
\begin{equation}
	 Z_\text{top}|_{a=q_2^N}=\prod_{i=0}^{\infty}\frac{(q_2^{-i};q_1)_\infty}{(q_2^{N-i};q_1)_\infty}=\prod_{i=0}^{N-1}\frac{1}{(q_2^{N-i};q_1)_\infty}=\frac{\hat Z_{U(N) \text{ CS}}[S^3]_\mathbf{0}}{(q_2;q_1)^N_\infty}\equiv \hat Z^\text{(unred)}_{U(N) \text{ CS}}[S^3]_\mathbf{0}
	\label{ZS3-unreduced}
\end{equation}
The fact that it reproduced the unreduced version is expected since the latter is the answer given by the index of the full, unreduced $T[M_3]$ on $D^2\times_q S^1$ (see section \ref{sec:higherRank}). Then both sides of this equality are direct M-theory calculations, one of which is done before geometric transition, another is after. At the same time the precise (that is including all normalization factors) relation to the CS partition function on $M_3$ a priori is not so clear.

Note that the expression in the middle of (\ref{ZS3-unreduced}) can be interpreted as the index of $N$ free chiral multiplets on $D^2\times_q S^1$ with Neumann boundary condition imposed on the boundary. This is not surprising since $\CN=2$ 3d $U(N)$ Chern-Simons theory with level one is conjectured to be dual to a theory of $N$ free chiral multiplets \cite{Kapustin:2011vz,Jafferis:2011ns,Pei:2015jsa}. Therefore the 3-manifold analog of $U(N)$ Khovanov-Rozansky homology for $S^3$ is conjecturelly given by a Fock space of $N$ bosons (and all their rotational modes). Since there are no cancellation in the refined partition function (\ref{ZS3-unreduced}) it can be understood as the corresponding Poincer\'e polynomial. The $SU(N)$ version has one less boson and the corresponding refined partition function has $(q_2;q_1)^{-1}$ factor removed.

Let us describe how the formula
\begin{equation}
	\left.\prod_{i=0}^{\infty}\frac{(q_2^{-i};q_1)_\infty}{(aq_2^{-i};q_1)_\infty}
	\right|_{a=q_2^N} =\prod_{i=0}^{N-1}\frac{1}{(q_2^{N-i};q_1)_\infty}
	\label{SSindex}
\end{equation}
can be understood on the level of Hilbert spaces. It has meaning analogous to the spectral sequence from HOMFLY-PT knot homology to $U(N)$ Khovanov-Rozansky knot homology. The space of BPS states in M-theory is $\Z_2\oplus \Z^3$-graded. Two of three $\Z$-gradings correspond to charges with respect of $U(1)_{q_1,q_2}$ rotational symmetries. The third grading counts the number of times that M2-branes wrap the non-trivial two-cycle $\mathbb{P}^1$ in the resolved conifold. It has a meaning of $U(1)_a$ gauge symmetry charge in terms of the effective 5d theory on $\R\times \C^2_{q_1,q_2}$ (see section \ref{sec:5d3d}). The refined topological string partition function then counts BPS states weighted with $\pm 1$ according to the $\Z_2$-grading. In the particular case of conifold (\ref{Ztoprefinedfull}) counts multiparticle states generated by the following single particle states (see fig. \ref{fig:HBPS-S3}):
\begin{figure}[ht]
\centering
\includegraphics[scale=1.3]{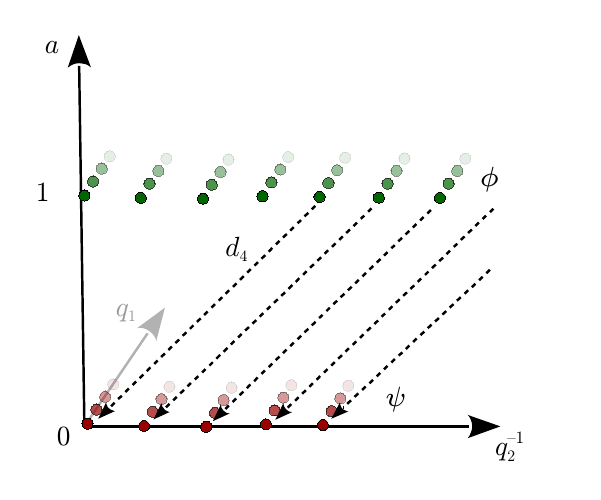}
\caption{A visualization of the single-particle states in $\CH_\text{BPS}$ for the resolved conifold. Three axes correspond to $\Z^3$ gradings corresponding to fugacities $q_1,q_2$ and $a$. One particle states $\phi_{i,j}$ and $\psi_{i,j}$ span two quadrants in $(q_1,q_2)$ plane shifted by $1$ along $a$ axis. Dashed arrows show the action of $d_4$ differential which has $U(1)_{q_1}$ charge zero and maps $\Z$-graded towers to $\Z$-graded towers.}
\label{fig:HBPS-S3}
\end{figure}
\begin{equation}
	\left.\begin{array}{cc}
	\phi_{i,j}\,: & \;\;(0,1,i,-j)_{\Z_2,a,q_1,q_2}, \\
	\psi_{i,j}\,: & \;\;(1,0,i,-j)_{\Z_2,a,q_1,q_2},	
	\end{array}
	\right.\qquad i, j\geq 0
\end{equation}
where the numbers in parentheses denote their $\Z_2\oplus \Z^3$ gradings. That is
\begin{equation}
	\CH_\text{BPS}=\text{Sym}^*\C[\{\phi_{i,j}\}_{i,j\geq 0}]\otimes \Lambda^*\C[\{\psi_{i,j}\}_{i,j\geq 0}]
\end{equation}
The formula (\ref{SSindex}) at the level of then lifts to the following relation
\begin{equation}
	H_*(\CH_\text{BPS},d_N)=\CH_N=\text{Sym}^*\C[\{\phi_{i,j}\}_{i\geq 0,\,N-1\geq j\geq 0}]
\end{equation}
where $\CH_N$ in the right-hand side is $U(N)$ Khovanov-Rozansky homology for $S^3$ and the left-hand side is homology of $\CH_\text{BPS}$ with respect to a differential acting as follows:
\begin{equation}
	d_N \phi_{i,j}=\psi_{i,j-N}.
\end{equation}
It has the following charges:
\begin{equation}
	d_N\,: \;\;(1,-1,0,N)_{\Z_2,a,q_1,q_2}
\end{equation}
We expect that the similar structure will hold for homological invariant of other 3-manifolds that admit large $N$ duals.

\subsection{Generalizations}

The large-$N$ duality for $N$ fivebranes wrapping $M_3 \times \R \times \Sigma$ should have a generalization to geometric
transition for positively curved $M_3$.
When $M_3$ is not positively curved, such as hyperbolic 3-manifold, we still expect a large-$N$ transition,
but as we will discuss in the end of this section we expect the transition in the $\R \times \Sigma$.
Most of this section will be devoted to the geometric transition for positively curved $M_3$.

It has already been argued, and checked \cite{Aganagic:2002wv,Halmagyi:2003ze,Brini:2008ik,Borot:2015fxa} that this large $N$ duality extends at least to the case where we quotient both
sides of the duality by a discrete group, which preserves $S^3$.  In particular if we choose a discrete subgroup
$\Gamma \subset SO(4)$ and mod out, we can get an arbitrary spherical 3-manifold
\begin{equation}
	M_3=S^3/\Gamma
	\label{M3orb}
\end{equation}
instead of $S^3$ and on the right hand side
we get a resolved geometry with more K\"ahler classes.

Let us discuss how this works. Consider for example modding $S^3/\Z_p$ where $\Z_p$ acts on the
conifold geometry $xy-zt=\mu$ by
\begin{equation}
	(x,y,z,t)\rightarrow ( \omega x, \omega^{-1}y,\omega z,\omega^{-1} t),\qquad \omega^p=1
\end{equation}
Passing this through duality one finds that one can choose the action so that $\mathbb{P}^1$ is fixed but the fiber of the resolved conifold $(O(-1)+O(-1))|_\text{pt}\cong \C^2$ is modded
out by $\Z_p$ giving $\mathbb{P}^1$ as the locus of $A_{p-1}$ singularity.  We thus obtain a singular geometry on the resolved side which
can be blown up giving $p-1$ additional moduli, leading to the closed string side with a total of $p$ K\"ahler classes.
These classes were interpreted from the open Chern-Simons perspective as follows.
Consider $U(N)$ Chern-Simons theory on $L(p,1)=S^3/\Z_p$.  The saddle points for CS theory are captured by choices
of the holonomy of flat $U(N)$ connection, which amounts to picking a homomorphism
\begin{equation}
	\rho: \qquad \pi_1(S^3/\Z_p)=\Z_p \rightarrow U(N)
\end{equation}
This in turn can be labeled by the decomposition of $\rho$ into  irreducible representations which are one dimensional and corrspond to a choice of $p$-th
root of unity:
\begin{equation}
	e^{2\pi in_i/p},\;\;i=1\ldots N
\end{equation}
The corresponding vector $\bn \in \Z_p^N$ (modulo permutations) has already appeared in section \ref{sec:higherRank} as the label for $U(N)$ flat connections on $L(p,1)$. As in section \ref{sec:higherRank} let us denote $N_k^{(\bn)}=\#\{j|n_j=k\}$,
the number of times the $k$-th representation appears.
Then $\sum_{k=1}^p N_k^{(\bn)}= N$. We expect the following generalization of (\ref{ZS3-unreduced}):
\begin{equation}
	\left. Z_\text{top}(a_1,\ldots,a_p;q_1,q_2)\right|_{a_i=q_2^{N_i^{(\bn)}}}=\frac{\hat Z_{U(N) \text{ CS}}[L(p,1)]_\mathbf{\bn}}{(q_2;q_1)^N_\infty}
	\label{ZLp1-unreduced}
\end{equation}
where the right-hand side is computed by (\ref{TM3mod}) and $a_i$ are fugacities for $H_2(X)$ gradings (with appropriate rescaling by powers of $q_{1,2}$). The equality makes sense since both sides have epansions with integer coefficients. It would be interesting to verify this explicitly. The left-hand side can be calculated using refined topological vertex or K-theoretic Nekrasov partition function of pure $U(p)$ super-Yang-Mills.

One can express the right-hand side of (\ref{ZLp1-unreduced}) via contributions to CS partition function corresponding to a particular flat connection by inverting (\ref{Lp1-Stransform}):
\begin{equation}
	Z_\text{top}(a_1,\ldots,a_p;q_1,q_2)|_{a_i=q_2^{N_i^{(\bn)}}}\propto
	\sum_{\bm\in\mathbb{Z}_p^N/\,S_N}
	S^{\mathbb{Z}^N,p}_{\bn,\bm}
	 Z_{U(N)_k}[L(p,1)]_\bm
	\label{ZLp1-unhat}
\end{equation}
This formula can be interpreted as the discrete analog of the Fourier-like transform relating topological string partition function and Chern-Simons partition function in the large $N$ limit. Let us review this point briefly. Consider unrefined case $q_1=q_2=q$ for simplicity and denote $g_s=\log q=2\pi i/k$, $T_i=\log a_i$ (K\"ahler parameters of $X$) and $t_i=g_sN_i^{(\bn)}$ (partial 't Hooft parameters in Chern-Simons gauge theory). In the large $N$/large volume limit one can consider both topological string partition function and CS partition function for a given flat connection as asymptotic series in string coupling constant $g_s$:
\begin{equation}
	\log Z_\text{top}=\sum_{g\geq 0}g_s^{2g-2}F_\text{top}^{(g)}(T_1,\ldots,T_p),
\end{equation}
\begin{equation}
	\log Z_\text{CS}=\sum_{g\geq 0}g_s^{2g-2}F_\text{CS}^{(g)}(t_1,\ldots,t_p).
\end{equation}
Both $Z_\text{CS}$ and $Z_\text{top}$ can be interpreted as wave-functions on the K\"ahler moduli space so that $\{T_i\}$ and $\{t_i\}$ are  natural coordinates near large volume and conifold points respectively. The relation between asymptotic expansions in $g_s$ then should be given by a Fourier-like transform \cite{Aganagic:2006wq} (which follows from \cite{Bershadsky:1993ta,Witten:1993ed}):
\begin{equation}
	Z_\text{top}(T_1,\ldots,T_p;g_s)=\int e^{\frac{1}{g_s^2}S(t,T)}Z_\text{CS}(t_1,\ldots,t_p;g_s)
\end{equation}
where $S(t,T)$ is a quadratic function. In genus zero it reduces to a symplectic transform between periods of mirror Calabi-Yau threefold (that is solutions of the Picard-Fuchs equations) and the integral formula above realizes its unambiguous lift to the quantum level. In the case $M_3=L(2,1)$ the explicit form of such transform was found in \cite{Drukker:2010nc,Marino:2011eh}

Suppose we consider the Chern-Simons theory on general $M_3$ and ask under what conditions we expect there to be a large
$N$ dual.  The above examples show that if there is a large $N$ dual the number of its K\"ahler classes should be given by
the number of inequivalent representations of $\pi_1(M^3)$.  The reason for this is that from the CS side the relevant
saddle points are given by flat connections which are in turn captured by homomorphisms\footnote{However, observations made in section \ref{sec:higherRank} in the case of $SU(2)$ gauge group suggest that one might be able to use decomposition of the CS partition function into a smaller set of terms. This will be explored further elsewhere.}
\begin{equation}
	\rho:\qquad \pi_1(M^3)\rightarrow U(M)
\end{equation}
and this map is characterized by the number of times $N_i$ where the $R_i$ irreducible representation of $\pi_1(M_3)$
occurs in this map. Moreover we have the relation
\begin{equation}
	\sum_i N_i \,{\rm dim}R_i=N
\end{equation}
Again we would identify $t_i=N_i g_s$ with coordinates on the K\"ahler moduli space of the closed string dual. Since the number
of irreducible representations are related to the number of conjugacy classes of $\pi_1(M_3)$ we thus predict that
\begin{equation}
	h^{1,1}=  |\text{Conj}(\pi_1(M_3))|
\end{equation}
If $\pi_1(M_3)$ has infinitely many conjugacy classes, then we would not expect there to exist a meaningful large $N$ dual
because that would have suggested that the local CY dual would have infinitely many K\"ahler classes.
In particular $M_3$ would need to be a rational homology 3-sphere. The number of conjugacy classes is automatically finite if the fundamental group $\pi_1(M_3)$ is finite. In this case the 3-manifold can always be realized as (\ref{M3orb}) so that $\Gamma=\pi_1(M_3)$. It is hard to imagine a 3-manifold with infinite fundamental group that have a finite number of conjugacy classes. Although there are examples of infinite groups with finitely many conjugacy classes, there are no known examples of finitely presented infinite groups with such property. And from any surgery-like construction one is always bound to obtain a 3-manifold with a finitely presented fundamental group. This is in agreement with general experience with geometric transitions
which strongly suggests that we must have a positively curved three manifolds if it were to shrink.

Let us point out that the large $N$ behavior is very different for hyperbolic and spherical 3-manifolds. In the case when $M_3$ hyperbolic the backreaction happens not along the $M_3$ directions of fivebranes, but on the other part which becomes a boundary of $AdS_4$ (see e.g. \cite{Gauntlett:2006ux}). It would be interesting to study in detail large $N$ limit for three-manifolds with ``intermediate'' geometry, for example the ones which are modeled on hyperbolic plane times $\R$ or even with more exotic nil- or solv- geometries.


\subsection{BPS states and grid diagrams}

In the five-brane system \eqref{M3phases}, the $N=0$ case corresponds to the most singular limit of the conifold.
Whether we start on resolved or deformed side, setting $N=0$ means taking the size of $\mathbb{P}^1$ to zero.
While this limit looks pretty singular, it actually is precisely the kind of orbifold / conifold limit where
one often finds an alternative description --- typically, algebraic, in terms of quivers --- of the space of BPS states.
And the space of BPS states is precisely what we are after:
\be
\CH_{N=0} = HF^+ (M_3)
\label{HFgridconif}
\ee
A toy version of such relation was considered at the end of section \ref{sec:higherRank}. So, just from knowing the standard five-brane system on the conifold and the fact that $HF^+$ theory corresponds to $N=0$,
we could {\it predict} from physics that $HF (M_3)$ should have a combinatorial description in terms of quiver representations or something similar,
where a quiver would be associated to a given CY singularity (obtained by shrinking $M_3$ inside $T^* M_3$ or its resolved version,
if we start on the resolved side).

Let us compare this ``prediction'' with mathematics. It comes extremely close!
Namely, there is a nice formulation of the Heegaard Floer homology in terms of the so-called grid diagrams~\cite{HFgrid},
which is very reminiscent of dimer tilings and similar models that appear in the description of quiver representations / BPS states
at Calabi-Yau singularities (see {\it e.g.} \cite{Yamazaki:2008bt} for a review).

Moreover, as we pointed out earlier, the volume on the resolved side is related to the super-rank \eqref{Nnm},
so that the singular conifold limit corresponds to the entire collection of $U(n|n)$ theories, of which \eqref{HFgridconif} is only a special case.
It would be interesting to understand the BPS spectra for different $U(n|n)$ and their relation to grid diagrams.


\subsection{Surface operators in 5d theories labeled by 3-manifolds}
\label{sec:5d3d}

The goal of this section is to mention that there is yet another physical way to look at 3-manifold invariants produced by fivebrane compactifications.
Compactification of eleven-dimensional M-theory on $T^* M_3$ or $X$ (when the latter exists) from setup (\ref{M3phases}) leads to a 5d $\CN=2$ theory that captures all
the relevant physics, in particular, the spectrum of BPS states we are interested in.
The two phases of the CY 3-fold geometry, {\it deformed} and {\it resolved}, are realized in this 5d effective low-energy theory as
different branches. For instance, in the basic case of $M_3 = S^3$ the effective 5d theory is $\CN=2$ SQED,
namely abelian $\CN=2$ vector multiplet coupled to a hypermultiplet and deformed (resp. resolved) phase
of the conifold geometry corresponds to the Higgs (resp. Coulomb) phase of 5d $\CN=2$ SQED \cite{Strominger:1995qi}. On the Coulomb branch the v.e.v. of the scalar in the vector multiplet gives mass to the hypermutliplet and the 5d theory can be effectively described as pure $U(1)$ 5d SYM.

Incorporating $N$ fivebranes wrapped on $M_3 \subset T^* M_3$ then gives rise to 3-dimensional defect
{\it a la} codimension-2 surface operator in 5d $\CN=2$ theory.
The theory on this codimension-2 defect should flow to 3d $\CN=2$ SCFT $T[M_3]$.
Note, in particular, that even when 5d theory is non-conformal the 5d/3d coupled system enjoys
3d $\CN=2$ superconformal symmetry, whose bosonic subgroup is $SO(3,2) \times U(1)_R$. The corresponding homological invariants 3-manifolds than can be computed by studying the spectrum of this 5d/3d coupled system.

\begin{figure}[ht]
\centering
\includegraphics[scale=1]{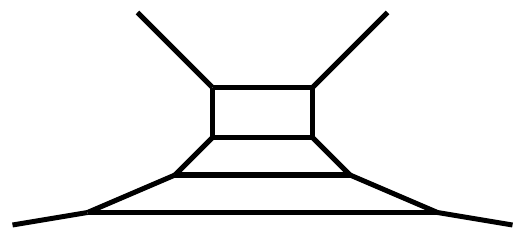}
\caption{A toric diagram for Calabi-Yau threefold $X$ obtained by geometric transition from $T^*L(4,1)$. It can be intepreted as a $(p,q)$ brane web relizing effective 5d $\CN=2$ $U(4)$ SYM.}
\label{fig:Lp1-toric}
\end{figure}

Consider for example the case of $M_3=L(p,1)$. In the resolved phase the 5d theory can be effectively described as 5d $\CN=2$ $U(p)$ SYM \cite{Aganagic:2002wv}. This is easy to see from the toric diagram of the Calabi-Yau threefold $X$, which can be dualized into a web of $(p,q)$-branes in type IIB string theory (see Fig. \ref{fig:Lp1-toric}). Apart from the $U(1)_{q,t}$ gradings the spectrum of BPS states will have $p$ extra gradings corresponding to $p$ generators of $H_2(X)$. From the point of view of 5d theory these gradings can be interpreted as charges with respect to the $U(1)^p$ diagonal subgroup of $U(p)$.


\acknowledgments{We would like to thank M.~Aganagic, A.~Gadde, M.~Khovanov, C.~Manolescu, S.~Nawata, P.~Ozsvath, J.~Rasmussen, M.~Romo, L.~Rozansky,
and E.~Witten for useful comments and discussions.
The work of S.G. is funded in part by the DOE Grant DE-SC0011632 and the Walter Burke Institute for Theoretical Physics. P.P. gratefully acknowledges support from the Institute for Advanced Study and also would like to thank Caltech and UT Austin theory groups for hospitality during the final stage of the project.
The work of C.V. is supported in part by NSF grant PHY-1067976. C.V. would like to thank KITP for hospitality. Opinions and conclusions expressed here are those of the authors and do not necessarily reflect the views of funding agencies.
}


\appendix


\section{Heegaard Floer homology of $M_3 = \Sigma' \times S^1$ and vortices}
\label{app:vortex}

In section \ref{sec:Sigmaprime}, we encountered a natural module over the ring $\Z [U] \otimes_{\Z} \Lambda^* H_1 (M_3;\Z)$
in the case of $M_3 = \Sigma' \times S^1$, namely \cite{OShfk}:
\begin{eqnarray}
HF^+ (\Sigma' \times S^1, {\frak s}_{h \ne 0}) & \cong & H^* (\text{Sym}^d (\Sigma_g)) \\
& \cong & \bigoplus_{i=0}^d \Lambda^i H^1 (\Sigma_g) \otimes \CT_0 / (U^{i-d-1}) \nonumber
\end{eqnarray}
where, as usual, $U$ has degree $-2$, and ${\frak s}_h$ denotes the Spin$^c$ structure with
\be
\langle c_1 ({\frak s}_h) , [\Sigma'] \rangle = 2h \qquad \text{and} \qquad \langle c_1 ({\frak s}_h) , \gamma \times S^1 \rangle = 0
\ee
for all closed curves $\gamma$ on the genus-$g$ Riemann surface $\Sigma'$.
This result has a simple and direct derivation via reduction of the Seiberg-Witten theory to two dimensions.

On $M_3 = \R \times \Sigma'$ the spinor bundle $W$ splits
into two eigen-bundles, $W \cong E \oplus F$, which
correspond to spinors of positive and negaitve chirality,
respectively.
In fact, we have $E = (K_{\Sigma'} \otimes L)^{1/2}$
and $F = K_{\Sigma'}^{-1} \otimes E$.
Correspondingly, we decompose
$\Psi = (\phi, \rho) \in \Gamma (E) \times \Gamma (F)$,
on which the Dirac operator ${\slash\!\!\!\! D}_A$ acts as
$$
{\slash\!\!\!\! D}_A \Psi =
\begin{pmatrix}
- i \frac{\partial}{\partial t} & \bar \partial^*_B \\
\bar \partial_B & i \frac{\partial}{\partial t}
\end{pmatrix}
\begin{pmatrix}
\phi \\
\rho
\end{pmatrix}
$$
Here, $t$ is the coordinate on $\R$ and $B$ is a connection on $E$
(equal to a linear combination of the connection $A$ on $L$
and the natural connection on the canonical bundle $K_{\Sigma'}$).
Under this decomposition, the three-dimensional
Seiberg-Witten equations become vortex equations
\begin{eqnarray}
& & * F_A = i (|\phi|^2 - |\rho|^2) \nonumber \\
& & \bar \partial_B \phi =0  \quad,\quad \bar \partial_B^* \rho = 0 \nonumber \\
& & \phi \bar \rho =0 \nonumber
\end{eqnarray}
A solution to these equations with $\phi=0$ (resp. $\rho=0$)
will be called a positive (resp. negative) vortex.
Indeed, according to the first equation,
the vortex number is given by
$$
\deg (L) = {i \over 2\pi} \int_{\Sigma'} F_A
= {1 \over 2\pi} \int_{\Sigma'} (|\rho|^2 - |\phi|^2)
$$
so that positive and negative vortices have $\deg (L) \ge 0$
and $\deg (L) \le 0$, respectively.
Solutions that have $\phi=0$ and $\rho =0$ are reducible;
these are flat connections on $\Sigma'$.
In what follows, we consider only irreducible
solutions with either $\phi \ne 0$ or $\rho \ne 0$.

Excluding reducible solutions,
we denote by $\CM^+ (E)$ and $\CM^- (E)$
the moduli spaces of positive and negative vortices
in a line bundle $E$ over $\Sigma'$.
Since these moduli spaces are related,
$$
\CM^+ (E) = \CM^- (K_{\Sigma'} \otimes E^{-1})
$$
without loss of generality we can consider only one of them,
say, the moduli space of negative vortices with $\rho=0$.
(To simplify notations, sometimes we denote this moduli
space as $\CM (E)$ or simply as $\CM$.)
Then, using $F_B = \frac{1}{2} (F_A + F_{K_{\Sigma'}})$,
we can write the vortex equations in the familiar form
\begin{eqnarray}
& & 2 \Lambda F_B - \Lambda F_{K_{\Sigma'}} = i |\phi|^2 \nonumber \\
& & \bar \partial_B \phi =0
\end{eqnarray}
where $\Lambda$ denotes the contraction with the K\"ahler form on $\Sigma'$.
These equations are a special case of the $\tau$-vortex
equations (with $\tau = i \Lambda F_{K_{\Sigma'}}$):
\begin{eqnarray}
& \Lambda F_A = {i \over 2} (|\phi|^2 - \tau) \nonumber \\
& \bar \partial_A \phi =0
\end{eqnarray}
which minimize the action of the abelian Higgs model
$$
S = \int_{\Sigma'} |F_A|^2 + |d_A \phi|^2 + {1 \over 4} (|\phi|^2 - \tau)^2
$$
The vortex in this theory is a generalization of
the Nielsen-Olesen vortex in the abelian Higgs model,
which can be recovered by taking $\Sigma' = \R^2$ and $\tau=1$.

Now, let us consider the vortex moduli space $\CM$.
The second equation says that $\phi$ must be
a holomorphic section of $E$. For example, on $\Sigma' = \C$
it is solved with $\phi = \prod_{i=1}^d (z - z_i)$,
so that $\CM = {\rm Sym}^d (\C)$.
More generally,
the vortex moduli space $\CM$ can be identified with
the space of divisors of degree $d = \deg (E)$ on $\Sigma'$
via zeroes of $\phi$. It is a K\"ahler manifold, isomorphic
to the $d$-th symmetric product
\be
\CM = {\rm Sym}^d (\Sigma')
\ee
when
\be
0 \le \deg (E) \le \frac{\deg (K_{\Sigma'})}{2}
\ee
and is empty otherwise.
In particular, we conclude that if $h = \deg (L)$ satisfies
$$
-2g + 2 \le h \le 2g - 2
$$
then the vortex moduli space $\CM$ is non-empty and is isomorphic
to the $d$-th symmetric product with $d=g-1 - {|h| \over 2}$.
This is precisely the moduli space that appears in the construction
of the Heegaard Floer homology.


\section{$S^2\times S^1$ topologically twisted index of 3d $\CN=2$ theories}
\label{app:index}

In this section we give a brief summary of rules for computing topologically twisted $S^2\times S^1$ index of 3d $\CN=2$ theories from \cite{Cecotti:2013mba,Benini:2015noa,Gukov:2015sna}.

Let us denote the collection of flavor and gauge fugacities by $\{x_i\in \C\}$. They parametrize the maximal torus of the total gauge ($G$) and flavor ($F$) symmetry group $G\times F$. Let $\{m_i\in \Z\}$ denote the fluxes through $S^2$ of the corresponding $U(1)$ subgroups. The vector $\bm$ belongs to the weight lattice of $G\times F$. The contribution of a chiral multiplet in represetation $\mathfrak{R}$ of $G\times F$ and R-charge $r$ (w.r.t. the $U(1)$ R-symmetry used to perform the topological twist along $S^2$) to the index reads:
\begin{equation}
	\prod_{\rho\in \mathfrak{R}}\left(\frac{x^{\rho/2}}{1-x^\rho}\right)^{\rho(\bm)+1-r}
\end{equation}
where $\rho$ denotes a weight of the representation $\CR$ and $x$ is treated as the element of the maximal torus of $G\times F$. The gauging operation together with the contribution of the vector multplets is given by
\begin{equation}
	\int\limits_\text{JK} \prod_{i\in G\text{ fugacities}}\frac{dx_i}{2\pi ix_i} \prod_{\alpha \in \text{roots of }G}(1-x^\alpha) \; \cdots
\end{equation}
The integral should be performed according to Jeffrey-Kirwan residue prescription.
For each simple or $U(1)$ subgroup $G_s\subset G\times F$ one can introduce Chern-Simons coupling $k_s$. The index will have the following classical contribution from CS action:
\begin{equation}
	\prod_{i\in G_s\text{ fugacities}}x_i^{k_sm_i}.
\end{equation}
For a pair of $U(1)$ groups $G_{a}$ and $G_{b}$ one can also introduce mixed CS coupling $k_{ab}$:
\begin{equation}
	x_a^{k_{ab}m_b}x_b^{k_{ab}m_a}.
\end{equation}

\section{3d/3d correspondence for plumbed 3-manifolds: extras}

\subsection{$S^2\times S^1$ topologically twisted index of $T{[}M_3{]}$: $G=SU(2)$ case}
\label{app:SU2index}
In this section we write $SU(2)$ analogs of vertex and edge contributions (\ref{index-vertex-u1})-(\ref{index-edge-u1}) to the topologically twisted index of $T[M_3]$ with $M_3$ given by a plumbing graph.
\begin{multline}
 \CI_{SU(2)_a}(x,m)=\frac{1}{2}(1-x^2)(1-1/x^2)\,x^{2am}\,\times\\
\left(\frac{xy^{-1}}{1-x^2y^{-2}}\right)^{2m-2s-1}\left(\frac{x^{-1}y^{-1}}{1-x^{-2}y^{-2}}\right)^{-2m-2s-1}\left(\frac{y^{-1}}{1-y^{-2}}\right)^{-2s-1},
\end{multline}
\begin{multline}
 \CI_{T[SU(2)]}(x_1,m_1;x_2,m_2)=
\left(\frac{y^{-1}}{1-y^{-2}}\right)^{-2s-1}\sum_{n\in\mathbb{Z}}\int\frac{dz}{2\pi i z}\,z^{m_1}x_1^n\; \times\\
\left(\frac{z^{1/2}x_2^{1/2} y^{1/2}}{1-zx_2t}\right)^{n+m_2+s+1}
\left(\frac{z^{1/2}x_2^{-1/2} y^{1/2}}{1-zx_2^{-1}y}\right)^{n-n_2+s+1}
\times\\
\left(\frac{z^{-1/2}x_2^{1/2} y^{1/2}}{1-z^{-1}x_2y}\right)^{-n+n_2+s+1}
\left(\frac{z^{-1/2}x_2^{-1/2} y^{1/2}}{1-z^{-1}x_2^{-1}y}\right)^{-n-n_2+s+1}
\label{ITSU2}
\end{multline}
where $y$ is the fugacity for $U(1)_\beta$ flavor symmetry of adjoint chiral multiplet. The integer parameter $s$ is the flux of $U(1)_\beta$ through $S^2$. It can be also understood as the $U(1)$ fugacity associated to the puncture on the $T^2$ torus on which $SL(2,\mathbb{Z})$ elements $T^a$ and $S$ act.  In the $\mathcal{N}=4$ 3d language $U(1)_\beta$ is the anti-diagonal of $SU(2)_R\times SU(2)_N$ R-symmetry. The supersymmetry is actually broken to $\mathcal{N}=2$ by Chern-Simons terms, but $U(1)_y$ flavor symmetry still survives and $\mathcal{N}=2$ $U(1)$ R-symmetry can mix with it. In the formulas above the $U(1)$ R-symmetry which is used to make the topological twist is the diagonal of $SU(2)_R$, that is the R-charge of the adjoint chiral is $2$ while the R-charge of hypers is zero. Twisting the diagonal of $SU(2)_N$ instead is equivalent to changing $s\rightarrow s-1$.
 The expression (\ref{ITSU2}) has self-mirror property:

\begin{equation}
 \CI_{T[SU(2)]}(x_1,m_1;x_2,m_2)=\CI_{T[SU(2)]}(x_2,m_2;x_1,m_1)|_{y\rightarrow 1/y,\; s\rightarrow -s-1}
\end{equation}

\subsection{$S^3$ partition function}
The aim of this section is to show that $S^3$ parition function of $T[M_3]$, where $M_3$ is given by a plumbing graph share many structural properties with $S^2\times S^1$ topologically twisted index. Consider the case $G=SU(2)$ and $S^3$ without squashing for simplicity. In the case $M_3=L(p,1)$ squashed 2-sphere partition function of $T[M_3]$ was studied in detail in \cite{Pei:2015jsa}. For the vector multiplet with level $p$ CS term the contribution is the folowing:
\begin{equation}
 Z_{SU(2)_a}[S^3](u)=2(\sinh\pi u)^2\,e^{\frac{\pi i a u^2}{2}}
\end{equation}
The contribution from $T[SU(2)]$ reads \cite{Benvenuti:2011ga}
\begin{equation}
 Z_{T[SU(2)]}[S^3](u,v)=\frac{\sin\pi uv}{\sqrt{8i}\sinh\pi u\sinh\pi v}.
\end{equation}
Then
\begin{equation}
 Z_{T[M_3]}=\int \prod_{i\in\text{vertices}} du_i  Z_{SU(2)_{a_i}}[S^3](u_i)
\prod_{\alpha\in\text{edges}}  Z_{T[SU(2)]}[S^3](u_{\alpha_1},u_{\alpha_2})
\end{equation}
where $\alpha_{1,2}$ are two different vertices in $\d \alpha$. Then one can check if from the different Seifert fibration data realizing homeomorphic 3-manifolds (preserving orientation) we get the same answer. For example the following linear plumbings all give the same lens space $L(7,2)\cong L(7,4)$:
\begin{equation}
 \begin{array}{c}
  (4,2)\\
(1,5,2)\\
(3,1,5)\\
\cdots\\
 \end{array}
\end{equation}
The answer is indeed the same up to $i^{n\in \mathbb{Z}}$ which is most likely related to the framing dependence of the complex Chern-Simons theory on $M_3$.


\newpage

\bibliographystyle{JHEP_TD}
\bibliography{classH}

\end{document}